\documentclass[longauth]{aa}
\usepackage{graphicx}
\usepackage{txfonts}
\usepackage[dvipsnames]{xcolor}
\usepackage[normalem]{ulem}
\usepackage{array, multirow, boldline}
\usepackage{multirow}
\usepackage{caption}
\usepackage{threeparttable} 
\usepackage{booktabs} 

\usepackage[colorlinks,linkcolor=MidnightBlue,citecolor=MidnightBlue,linktocpage=true,breaklinks, 
plainpages=false,urlcolor=MidnightBlue]{hyperref}
\newcommand{\newchi}{\raisebox{0pt}[1ex][1ex]{$\chi$}}
\def\thethree{{\sc Three Hundred Project }}
\def\xmm{XMM-{\it Newton}}

\usepackage[]{subfig}
\begin{document}

   \title{CHEX-MATE: A non-parametric deep learning technique to deproject and deconvolve galaxy cluster X-ray temperature profiles
   }

   \subtitle{}

   \author{A. Iqbal
          \inst{\ref{inst1}}
          \and 
          G.W. Pratt \inst{\ref{inst1}}  
          \and
          J. Bobin\inst{\ref{inst2}} 
          \and
          M. Arnaud\inst{\ref{inst1}} 
          \and
          E. Rasia\inst{\ref{inst3},\ref{inst4}} 
          \and
          M. Rossetti\inst{\ref{inst5}} 
          \and
          R.T. Duffy\inst{\ref{inst1}}  \and
          I. Bartalucci \inst{\ref{inst5}}     \and
          H.  Bourdin\inst{\ref{inst6}}          \and
          F. De Luca\inst{\ref{inst6}}          \and
          M. De Petris\inst{\ref{inst17}}          \and
          M.  Donahue\inst{\ref{inst7}}          \and
          D. Eckert \inst{\ref{inst7B}}  \and
          S. Ettori\inst{\ref{inst8},\ref{inst9}}          \and
          A. Ferragamo\inst{\ref{inst10},\ref{inst11}}          \and
          M. Gaspari\inst{\ref{inst12}}          \and
          F. Gastaldello\inst{\ref{inst5}}     \and
          R. Gavazzi\inst{\ref{inst21}, \ref{inst22}}
          \and
          S. Ghizzardi\inst{\ref{inst5}}          \and
           L. Lovisari\inst{\ref{inst14},\ref{inst15}}          \and
           P. Mazzotta \inst{\ref{inst6}}          \and
           B.J. Maughan \inst{\ref{inst16}}  
           \and
           E. Pointecouteau\inst{\ref{inst18}}          \and
          M. Sereno\inst{\ref{inst19},\ref{inst20}}  
          }
   \institute{Universit\'e Paris-Saclay, Universit\'e Paris Cit\'e CEA, CNRS, AIM, 91191, Gif-sur-Yvette, France \label{inst1}
   \and
   CEA IRFU/DEDIP, 91191 Gif-sur-Yvette, France \label{inst2}
   \and
   INAF - Osservatorio Astronomico di Trieste, via Tiepolo 11, I-34131 Trieste,  Italy \label{inst3}
   \and
   IFPU, Via Beirut, 2, 3I-4151 Trieste, Italy \label{inst4}
 \and   
INAF, Istituto di Astrofisica Spaziale e Fisica Cosmica di Milano, via A. Corti 12, 20133 Milano, Italy \label{inst5}
 \and
Dipartimento di Fisica, Universita' di Roma "Tor Vergata", Via Della Ricerca Scientifica, 1, 00133 Roma Italy \label{inst6}
\and
Dipartimento di Fisica, Sapienza Università di Roma, Piazzale Aldo Moro 5, I-00185 Roma, Italy \label{inst17}
 \and
Department of Physics and Astronomy, Michigan State University, 567 Wilson Rd, East Lansing, MI 48864 USA \label{inst7}
 \and
Department of Astronomy, University of Geneva, ch. d’\'Ecogia 16, CH-1290 Versoix Switzerland \label{inst7B}
 \and
INAF, Osservatorio di Astrofisica e Scienza dello Spazio, via Piero Gobetti 93/3, 40129 Bologna, Italy  \label{inst8}\and
INFN, Sezione di Bologna, viale Berti Pichat 6/2, 40127 Bologna, Italy \label{inst9}
 \and
Instituto de Astrofísica de Canarias (IAC), C/ Vía Láctea s/n, E-38205 La Laguna, Tenerife, Spain \label{inst10}\and
Dipartimento di Fisica, Sapienza Università di Roma, Piazzale Aldo Moro 5, I-00185 Roma, Italy\label{inst11}
 \and
Department of Astrophysical Sciences, Princeton University, Princeton, NJ 08544, USA \label{inst12}
 \and
Laboratoire d’Astrophysique de Marseille, Aix-Marseille Université, CNRS, CNES, Marseille, France \label{inst21}
\and
Institut d’Astrophysique de Paris, CNRS, Sorbonne Université, Paris, France \label{inst22}
\and
INAF, Istituto di Astrofisica Spaziale e Fisica Cosmica di Milano, via A. Corti 12, 20133 Milano, Italy \label{inst14}
\and
Center for Astrophysics $|$ Harvard $\&$ Smithsonian, 60 Garden Street, Cambridge, MA 02138, USA \label{inst15}
 \and
 HH Wills Physics Laboratory, University of Bristol, Tyndall Ave, Bristol, BS8 1TL, UK \label{inst16}
 \and
IRAP, Université de Toulouse, CNRS, CNES, UT3-UPS, (Toulouse), France \label{inst18}
\and
INAF, Osservatorio di Astrofisica e Scienza dello Spazio, via Piero Gobetti 93/3, 40129 Bologna, Italy \label{inst19} \and
INFN, Sezione di Bologna, viale Berti Pichat 6/2, 40127 Bologna, Italy \label{inst20}
            }
   \date{}

 
  \abstract{
   {Temperature profiles of the hot galaxy cluster intracluster medium (ICM) have a complex non-linear  structure that traditional parametric modelling may fail to fully approximate. For this study, we made use of neural networks, for the first time, to construct a data-driven non-parametric model of ICM temperature profiles. A new deconvolution algorithm was then introduced to uncover the true (3D) temperature profiles from the observed projected (2D) temperature profiles.}
   {An auto-encoder-inspired neural network was first trained by learning a non-linear interpolatory scheme to build the underlying model of 3D temperature profiles in the radial range of [0.02-2]\,R$_{500}$, using a sparse set of hydrodynamical simulations from the {\sc three hundred project}. A deconvolution algorithm using a learning-based regularisation scheme was then developed. The model was tested using high and low resolution input temperature profiles, such as those expected from simulations and observations, respectively.
}
   {We find that the proposed deconvolution and deprojection algorithm is robust with respect to the quality of the data, the morphology of the cluster, and the deprojection scheme used. The algorithm can recover unbiased 3D radial temperature profiles with a precision of around 5\% over most of the fitting range. We apply the method to the first sample of temperature profiles obtained with XMM{\it -Newton} for the CHEX-MATE project and compared it to parametric deprojection and deconvolution techniques. Our work sets the stage for future studies that focus on the deconvolution of the thermal profiles (temperature, density, pressure) of the ICM and the dark matter profiles in galaxy clusters, using deep learning techniques in conjunction with X-ray, Sunyaev Zel'Dovich (SZ) and optical datasets.}

}   
   
   \titlerunning{Deprojection and deconvolution of temperature profiles with deep learning}
   \maketitle
%
\section{Introduction}

Galaxy clusters are ideal probes of the large-scale structure of the Universe \citep{Holder2001,2016A&A...594A..24P,2015ApJ...799..214B,2017MNRAS.472.1946S,2022PhRvD.105b3520A}. X-ray observations of the hot gas in the ICM, which constitutes the dominant baryonic component in galaxy clusters, provide us with a useful tool for identifying and studying these objects. Shallow, wide-field X-ray surveys by ROSAT (ROentgen SATellite) and eROSITA (extended ROentgen Survey with an Imaging Telescope Array) have now discovered thousands of clusters \citep[e.g.][and references therein]{mcxc,eFEDS}. In recent years, the detailed X-ray follow-up of samples extracted from these surveys has exploited the high spatial resolution of {\it Chandra} and the large field of view and sensitivity of X-ray Multi-Mirror Mission (XMM-{\it Newton})  to investigate the morphological, structural, and scaling properties of the cluster population \citep[e.g.][and references therein]{lov22,kay22}.

The X-ray-derived radial temperatures and density profiles are key ingredients to derive the thermodynamic properties of the ICM, and, under the assumption of hydrostatic equilibrium, the total mass profile in galaxy clusters \citep{Bohringer2007,Pratt2010,2010A&A...524A..68E,ett13,2022A&A...662A.123E}. 
These X-ray studies have revealed the presence of two distinct types of clusters: cool cores (CCs), characterised by dense and the low-temperature cores, and non-cool core (NCCs), which exhibit a relatively flat central density and temperature. Various morphological parameters have been introduced to analyse X-ray images and to link these to the dynamical behavior of galaxy clusters and to the presence or absence of a low temperature cores, providing insights into their structural characteristics, internal dynamics, and evolutionary stages \citep{Rasia2013,2022A&A...665A.117C}. Although it is now well-established that  Active Galactic Nuclei  (AGN) feedback plays a major role in suppressing the ICM cooling in cluster cores, the reason for the CC and NCC dichotomy is still not fully understood \citep{2015ApJ...813L..17R,2018MNRAS.481.1809B}. 

X-ray observations give access to the projected (2D) density and temperature profiles of the ICM. The latter is obtained from fitting a thermal model to the spectra extracted in concentric annuli about a given centre (usually the X-ray peak or centroid). For further scientific applications, these must then be deprojected to obtain the 3D profiles. If needed, the effect of the instrumental point spread function (PSF) can be taken into account in the deprojection step. While the deprojected (3D) gas density in shells can be easily estimated from the X-ray surface brightness \citep{Croston2006,Bartalucci2017,Ghirardini2019}, the deprojection of ICM temperature profiles is not trivial. This is partly due to the need for sufficient photon counts to build and fit the spectrum, leading to the temperature profiles having significantly coarser angular resolution than the density.

The relationship between the observed 2D temperature profile, $\bf{T}_{\rm 2D}$, and the originating 3D temperature profile, $\bf{T}_{\rm 3D}$, can be expressed in matrix form as
\begin{equation}
{\bf T}_{\rm 2D}  ={\bf C}_{\rm PSF}\otimes\bf{C}_{\rm proj}\otimes\bf{T}_{\rm 3D}=\bf{C}\otimes \bf{T}_{\rm 3D}\label{eqn:matrix},
\end{equation}
where $\otimes$ denotes the matrix product. Assuming a cluster is spherically symmetric and that the 3D temperature profile is defined in concentric spherical shells, the $(i,j)^{\rm th}$ element of the matrix $\bf{C}_{\rm proj}$ encodes the projection effect of the $j^{\rm th}$ 3D shell onto the $i^{\rm th}$ 2D annulus on the plane of the sky. The 2D annuli may have the same or different radii to the 3D shells. We note that
$\bf{C}_{\rm PSF}$ is a second matrix that describes the effect of the finite instrumental PSF. Its $(k,i)^{\rm th}$ element contains the fraction of counts from the $i^{\rm th}$ 2D annulus that are redistributed by the telescope into the $k^{\rm th}$ observed 2D annulus. If there are $n$-model 3D shells, and correspondingly $n$-model 2D  annuli, plus $m$-observed annuli, then the dimensions of $\bf{C}_{\rm proj}$ and $\bf{C}_{\rm PSF}$ are $n\times n$ and $m\times n$, respectively. If the PSF is ignored, then the dimensions of $\bf{C}_{\rm proj}$ should change to $m\times n$.

The fitting of projected parametric models of the 3D temperature profiles to both observed and simulated 2D data has been widely used in the literature  \citep{De_Grandi_2002,2003ApJ...592...62P,2008MNRAS.383..369A,2010ApJ...720.1038B,2012MNRAS.424..190G,2019A&A...627A..19G}.
Initially, these were polytropic models that assumed a simple relationship between the density and the temperature distribution ($T \propto \rho^{\gamma -1})$, but this does not fully capture all the complexities of real galaxy clusters, especially the central regions of CCs clusters.
The quality of recent data has necessitated more complicated models to be proposed, perhaps the most widely used being that proposed by  \citet[][]{Vikhlinin2006}:
\begin{equation}\label{eq:tprof}
  {\rm T}_{\rm{3D}}(r) =  T_0\times (x+\tau)/(x+1)\times \frac{(r/R_t)^{-a}}{(1+(r/R_t)^b)^{c/b}}, 
\end{equation}
where $x=(r/R_{\text{cool}})^{a_{\text{cool}}}$ and $\{T_0,\tau,R_t,R_{\text{cool}},a_{\text{cool}},a,b,c\}$ are the model parameters. In the framework of the {\it Representative XMM-Newton cluster structure survey} \citep[REXCESS;][]{Bohringer2007} and {\it Following the most massive galaxy clusters over cosmic time} \citep[M2C;][]{bar19} projects, \cite{Democles2010} and \citet{Bartalucci2018} developed a non-parametric-like deconvolution approach. For this study the \cite{Vikhlinin2006} parametric model was used to perform the PSF correction and deprojection in order to estimate the temperature at the weighted radii of the 2D annular binning scheme.
The 3D uncertainties were then computed consistently from the 2D errors, and random temperatures were drawn within these uncertainties to compute the temperature derivatives which were used in the hydrostatic equilibrium total mass computations.

However, parametric approaches are not fully satisfactory since, with a limited number of parameters, they could fail to capture features in the temperature profile due to shock fronts, edges, mergers and the presence of cool cores with one single model. Moreover, a high degree of degeneracy between the parameters could be present. The  \cite{Vikhlinin2006} parametric temperature model, which was developed for cool core systems, is a complex eight-parameter model, four of which correspond to the cool-core component. It is therefore not well-suited to highly disturbed NCC clusters, which have flatter central temperature profiles instead of declining cool cores.
Furthermore, for typical X-ray data quality, it exhibits a high degree of degeneracy between its parameters, leading to poorly constrained model parameters and the results that depend on the prior choices in MCMC fitting schemes. Recently, \cite{2021MNRAS.502.5115G}, using a sample drawn from high resolution numerical simulations, found that the \cite{Vikhlinin2006} parametric model could only fit well to 50\% of their sample in the range [0.1-1]\,R$_{500}\footnote{The scaled radius $R_{\Delta}$ is defined such that $R_{\Delta}$ is the radius at which the mean matter density is $\Delta \rho_c$, where $\rho_c=3H^2(z)/8\pi G$ is the critical density of the universe at redshift $z$.}$. 

Model-independent direct spectral deprojection methods offer an alternative and are commonly used to deconvolve the 3D temperature profiles. This can involve the onion-skin technique \citep{fab81,2001ApJ...557..546D,2005MNRAS.356..237J,2008MNRAS.390.1207R,2016MNRAS.460.2625L}, where the 3D layers are successively built up from the outside in. However, this approach is strongly dependent on the choice  of the outermost bin because it is necessary to take into account the contribution to the emission from the shells outside the outermost annulus used for the analysis. 
Alternatively, isothermal models can be fitted  to each annular spectrum and then the matrix method (i.e. Eq. \ref{eqn:matrix}) can be used to deproject \citep[e.g.][]{2002MNRAS.331..635E}. Ignoring the PSF effect, the equation for temperature profiles assuming that the observed projected spectra consist of a linear combination of isothermal emission models weighted by the projected emission measure simplifies to
\begin{equation}
T_{2D,k} = \sum_{j=1}^{n} \frac{w_{k,j}}{\sum_{j=1}^{n} w_{k,j}}T_{3D,j}.\label{eqn:matrix2}
\end{equation}
Here, $T_{3D,j}$ and $T_{2D,k}$ are the 3D and 2D temperatures at the $j$th 3D spherical shell and $k$th 2D  observed annulus, respectively, and the weights, $w_{k,j}$, consist of the emission measure contribution of the spherical shells onto the observed annuli \citep[e.g.][]{2001ApJ...546..100M}.

However, such model-independent approaches are often unstable if the data are noisy because Eqn.\ref{eqn:matrix} is an inverse problem, meaning that any noise becomes greatly amplified by the deconvolution procedure. 
In addition, the simplistic emission measure weighting has been found to be inaccurate when applied to X-ray observations. In particular, it has been demonstrated that in the presence of a multi-temperature components gas, $w$ is more appropriately expressed as a non-linear combination of density and temperature \citep{Mazzotta2004,Vikhlinin2006b}, further complicating the deconvolution procedure.
\begin{figure*}
	\centering	
	\includegraphics[width=0.98\textwidth]{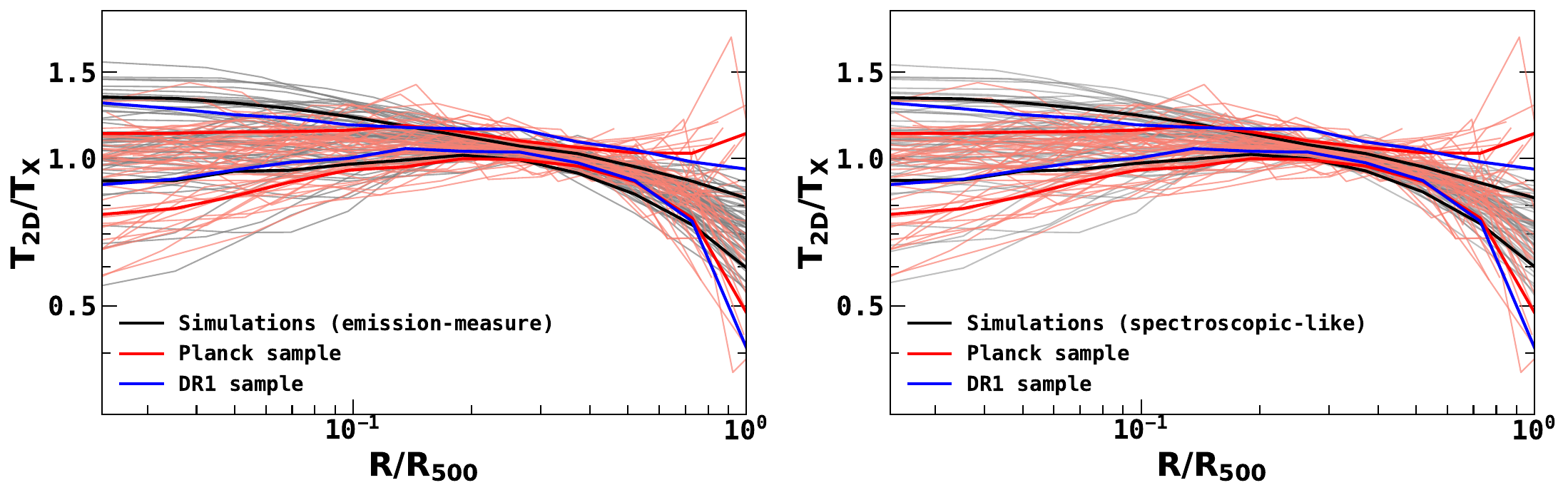}   
	\caption{\footnotesize Comparison of the observed 2D temperature profiles, scaled as a function of $R_{500}$ and T$_{X}$, the temperature in the [0.15-0.75]\,$R_{500}$ region. 
 The thin grey lines show 50 randomly selected simulated 2D temperature profiles from the {\sc Three Hundred Project}, extracted with an observation-like annular binning resolution, derived using emission measure (left panel) and spectroscopic-like (right panel) weighting schemes. The thin red lines show individual profiles in the \cite{2011A&A...536A..11P} sample. For better visibility, the error bars corresponding to the observed profiles are not shown. The regions enclosing thick black and red lines show the 1-$\sigma$ dispersion (16th–84th percentile range) of the temperature profiles of the full simulated sample and the {\it Planck} sample respectively. The regions enclosing the thick blue lines show the  1-$\sigma$ dispersion of the CHEX-MATE DR1 sample. Scaled by R$_{500}$ and T$_X$, both the emission measure and spectroscopic-like derived 2D simulated temperature profiles become somewhat self-similar.}   
	\label{figcom}
\end{figure*}
\begin{figure}
    \includegraphics[width=0.49\textwidth]{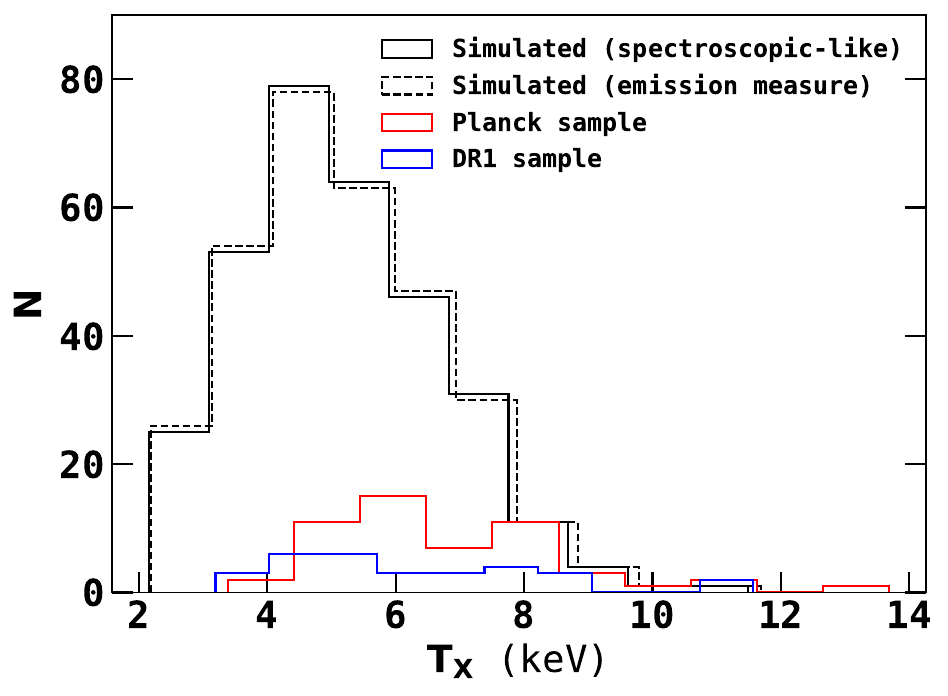}
    \caption{\footnotesize Number of clusters as a function of T$_{X}$ in the \thethree sample, the {\it Planck} SZ sample and the DR1 sample.  }
    \label{fig_dis_Tx}
\end{figure}

Machine Learning (ML) techniques have emerged as a powerful technique for predicting key features of data and for solving inverse problems to reconstruct (deconvolve) signals, images, etc, from observations. ML techniques have been applied to study galaxy clusters too. \cite{Ntampaka} developed an ML algorithm based on Support Distribution Machines to reconstruct dynamical cluster masses  using the velocity distribution of cluster members from simulations, achieving a reduction in the scatter between the predicted and true mass by a factor of two compared to standard methods. More complex ML approaches have  led to similar significant improvements in the mass estimates \citep{Armitage2019,Calderon2019}. 

Using deep learning techniques, Convolutional Neural Network (CNN) models have also been used to infer the dynamical mass of galaxy clusters \citep{2019Ho,2020Ramanah,2021ApJ...908..204H,2022NatAs...6.1325D}. In particular, \cite{2020MNRAS.499.3445Y} used  mock datasets of stellar mass, soft X-ray flux, bolometric X-ray flux, and Compton y-parameter images as input to train a CNN model to infer the mass of galaxy clusters, and \cite{2020ApJ...900..110G,2021ApJ...923...96G}  trained CNN models to estimate cluster masses used mock SZ, cosmic micro-wave background (CMB) lensing maps. \cite{2022arXiv220712337F}, using a combination of an auto-encoder and a random forest regression technique on a sample of 73,138 mock Compton-$y$ parameter maps from the hydrodynamical simulations of the \thethree \citep{Cui2018}, and were able to reconstruct the 3D gas mass profile and total mass in galaxy clusters with a scatter of about 10\% with respect to the true values. \cite{2022NatAs...6.1325D} and \cite{2022NatAs...6..936H} have used real observations to estimate the total mass profiles of galaxy clusters using deep learning models trained on mock simulations. While \cite{2022NatAs...6.1325D} used the {\it Planck} SZ maps \citep{2016A&A...594A..27P} to determine the masses of Planck clusters, \cite{2022NatAs...6..936H} used relative line-of-sight velocities and projected radial distances of galaxy pairs from Sloan Digital Sky Survey (SDSS) data \citep{2015ApJS..219...12A} to determine the mass of the Coma cluster.

In this work, we show the first use of  neural networks, trained on numerical simulations, to deproject the X-ray temperature profiles of galaxy clusters. Our technique is based on that proposed by \citet{9022675,bobin:hal-03265254} where a so-called  Interpolatory Autoencoder ({\bf IAE}) neural network is built to model the 3D temperature profiles by learning a non-linear interpolatory scheme from a limited set of example profiles called `anchor points'. The main advantage of the {\bf IAE} neural network is that it is able to capture the intrinsic low-dimensional, non-linear nature of the profiles even when the training sample is not large in size. This is crucial as a small sample size can otherwise pose several challenges to the effectiveness of a deep learning algorithm. The model is trained and tested with a set of 315 simulated temperature profiles, in the radial range of [0.02-2]\,R$_{500}$, from the {\sc three hundred project} \citep[]{Cui2018}. A robust temperature deconvolution scheme is then introduced to fit the trained {\bf IAE} model, that makes use of an efficient regularisation term in the likelihood, along with Markov chain Monte Carlo (MCMC) sampling.
The technique is then applied to a pilot sample of X-ray temperature profiles from the CHEX-MATE project \citep[Cluster HEritage project with XMM-{\it Newton}: Mass Assembly and Thermodynamics at the Endpoint of structure formation;][]{2021A&A...650A.104C}.

The paper is organised as follows. Section 2 discusses in detail the simulations used in training the {\bf IAE} model for temperature profiles. In Section 3 we present the IAE model, and Section 4 deals with model training and the learning-based deconvolution technique. The performance of the deconvolution algorithm is tested with simulations in Section 5, while in Section 6, we apply our approach for the first time to a representative sample of 28 galaxy clusters from the first data release (DR1 hereafter, \citealt{Rossetti2023}, in prep.) in the CHEX-MATE sample. Finally, in Section 7, we summarise our work. Throughout this work, we adopt a flat $\Lambda$CDM model with $H_{\rm 0}=70$ km s$^{-1}$ Mpc$^{-1}$, $\Omega_{\rm m}=0.3$ and $\Omega_{\rm \Lambda} =0.7$.  Further, $E(z)$ is the ratio of the Hubble constant at redshift $z$ to its present value, $H_{\rm 0}$ and $h_{\rm 70}=H_{\rm 0}/70=1$.
\section{Simulations}
\label{sec:simulations}
In this work, training of the neural network is undertaken using the gas mass-weighted 3D temperature profiles, ${\bf T}_{\rm 3D}$, of galaxy clusters from the \thethree \citep{Cui2018,2020A&A...634A.113A}. 
These  simulations are based on the 324 Lagrangian regions centred on the $z=0$ most massive galaxy clusters selected from the MultiDark dark-matter-only MDPL2 simulation \citep{Klypin2016}, carried out with the cosmological parameters from the {\it Planck} mission \citep{2016A&A...594A..13P}. MDPL2 is a periodic cube of comoving size equal to 1.48\, Gpc containing $3840^3$ dark matter particles. The selected regions were resimulated with the inclusion of baryons and were carried out with the code {\tt{GADGET-X}} \citep{2016MNRAS.455.2110B}. To treat the baryonic physics several processes were included such as: metallicity-dependent radiative cooling, the effect of a uniform time-dependent UV background, a sub-resolution model for star formation from a multi-phase interstellar medium, kinetic
feedback driven by supernovae, metal production from SN-II, SN-Ia and asymptotic-giantbranch stars, and AGN
feedback \citep{2015ApJ...813L..17R}.

In the present work, we ignore the redshift dependence of the profiles, if any, and only consider the simulated sample at a fixed redshift of $z=0.33$, which is the average redshift of the CHEX-MATE sample. However, we consider a mass range of  M$_{500}>10^{14}\,{\textrm M}_\odot$ allowing us to build a library covering the full mass range of the CHEX-MATE sample. This left us with 314 clusters in the simulated sample.

The temperature profiles were derived in 48 fixed  logarithmically spaced radial bins in  
the range  [0.02-2]\,R$_{500}$ \citep{2020A&A...634A.113A}. The lowest radial limit of 0.02\,R$_{500}$ was chosen since it encloses approximately 100 gas particles for the simulated sample, which we call the precision threshold condition, thus ensuring that the analysis is statistically robust and that the results are not affected by numerical fluctuations in the gas properties at small radii \citep{2015ApJ...813L..17R}. The 3D mass-weighted temperature in a given shell $i$, ${T}_{3D,i}$ (i.e the $i^{\rm th}$ element of the ${\bf T}_{\rm 3D}$ vector), was calculated by weighting the temperature of the $p^{\rm th}$ gas particle (${T}_p$) using its gas mass (${m}_p$) as a weighting function $w$,
\begin{equation}
    {T}_{3D,i}=\frac{\sum {{T}}_p {m}_p}{\sum {m}_{p}}\,, 
    \label{eq:mw}
\end{equation}

In this calculation, no attempt was made to exclude low-temperature sub-clumps in the outskirt regions of the clusters, however, only particles with temperature >0.3 keV were considered.

We estimated the projected 2D temperature profiles  (${\bf T}_{\rm 2D}$) along the line of sight ($l$) using the 3D gas density ($\boldsymbol \rho$) and temperature profiles (${\bf T}_{\rm 3D}$). 
The 2D temperature profiles were estimated in pre-defined logarithmically spaced annular bins by first considering the classical emission-measure weights (${\bf C}=\bf{C}_{\rm proj}$, see Eqn.~\ref{eqn:matrix2}):
\begin{equation}
    {\bf T}_{\rm 2D}=\frac{\int {w} {{\bf T}_{\rm 3D}dl}}{\int {w} dl}=\bf{C}\otimes{\bf T}_{\rm 3D}, 
\end{equation}
where $w={\boldsymbol \rho}^2$ \citep[e.g.][]{2001ApJ...546..100M}.
We produced several versions of the ${\bf T}_{\rm 2D}$ profiles: First the ${\bf T}_{\rm 2D}$ profiles were first estimated in the same radial bins as those of the ${\bf T}_{\rm 3D}$ (48 bins) by using a matrix $\bf{C}$ of dimension $48\times48$ (${\bf C}_{48,48}$). We also estimated ${\bf T}_{\rm 2D}$ in a coarser binning scheme to reproduce typical radial sampling from present-day X-ray observatories such as {\it XMM-Newton} and {\it Chandra}. These have either twelve or six logarithmic bins reaching only up to R$_{500}$, corresponding to matrices of dimension $12\times48$  (${\bf C}_{12,48}$) and $6\times48$  (${\bf C}_{6,48}$) respectively.

\begin{figure*}
	\centering	
	\includegraphics[width=0.98\textwidth]{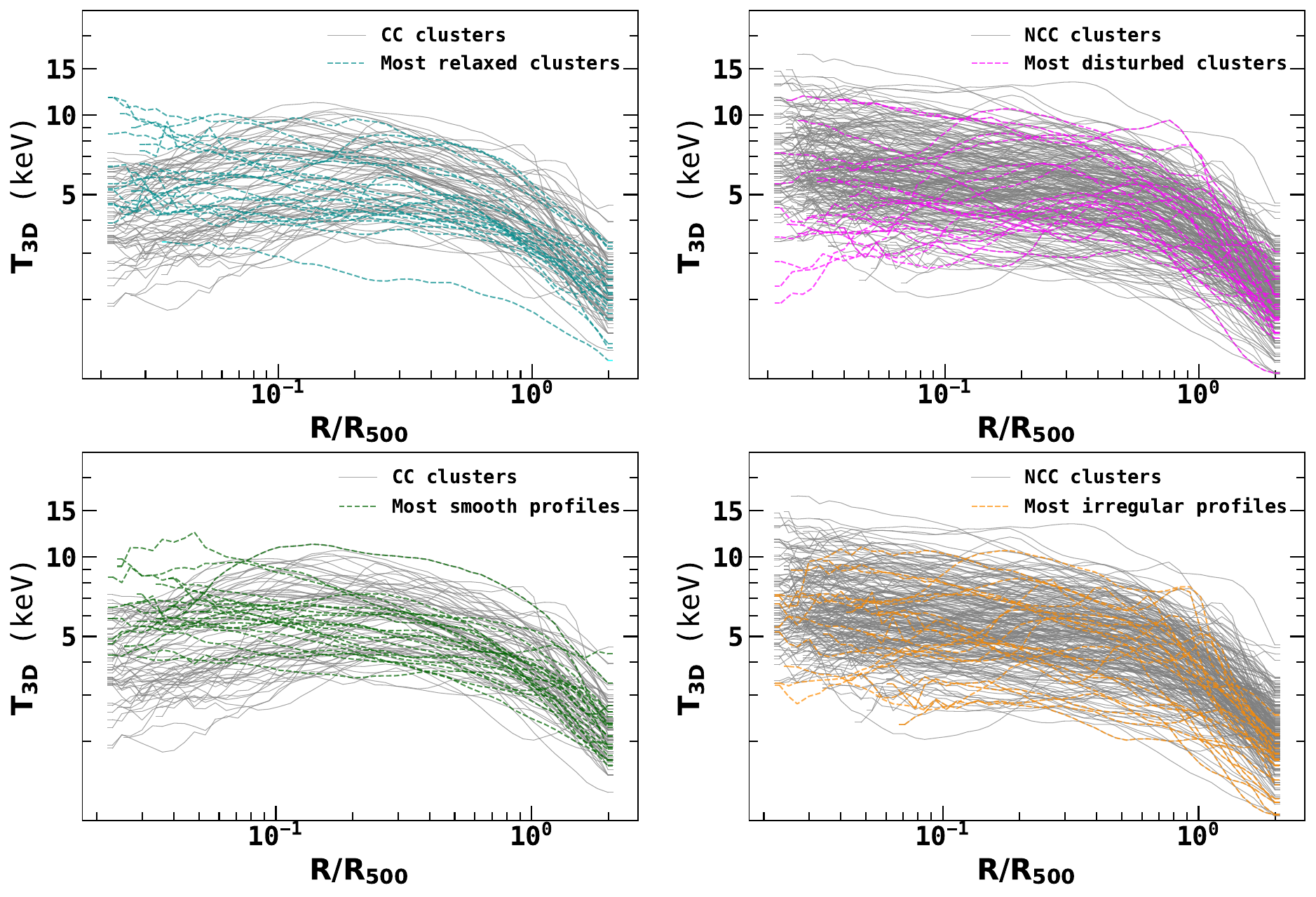}   
	\caption{\footnotesize Classification of temperature profiles in the {\sc Three Hundred Project}. Left panel: Grey line shows the visually classified CC clusters. Cyan and green lines show the 20 most relaxed clusters (top panel) and 20 most smooth profiles (bottom panel). Right panel: Grey line shows the visually classified NCC clusters. Magenta and orange lines show the 20 most disturbed  clusters (top panel) and irregular profiles (bottom panel)}.
	\label{fig2}
\end{figure*}
\begin{figure*}
	\centering	
	\includegraphics[width=0.98\textwidth]{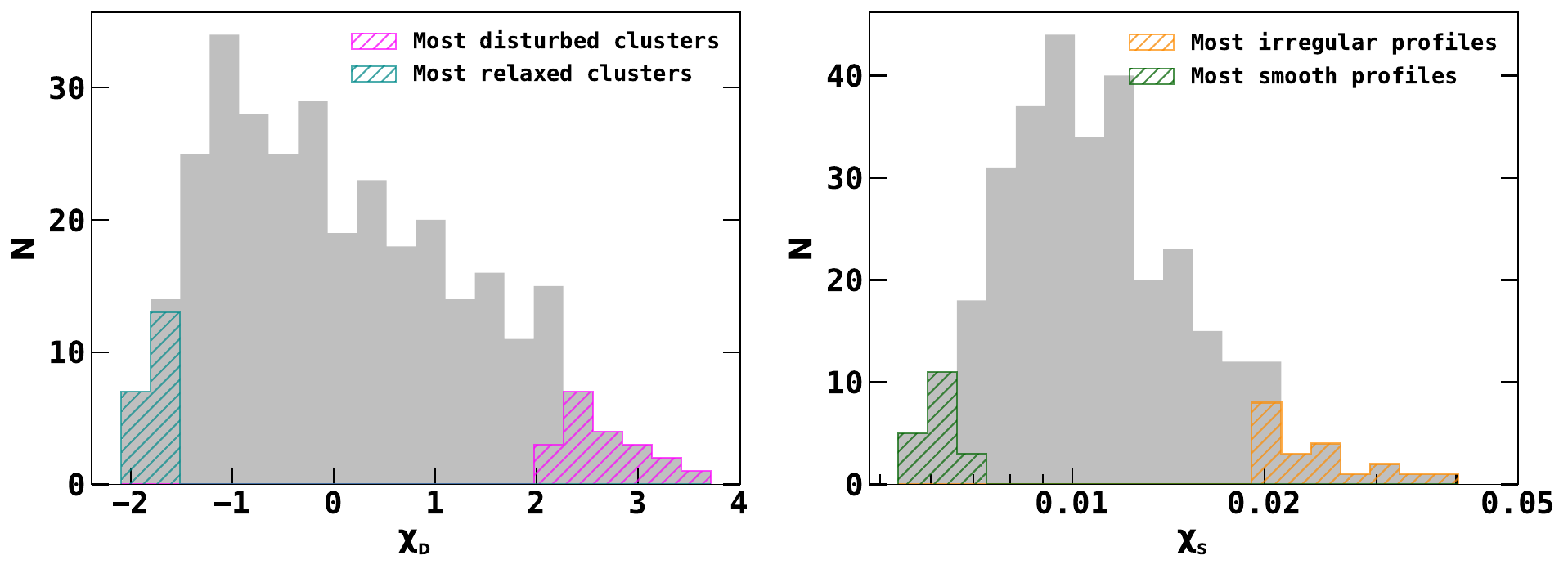}   
	\caption{\footnotesize Distribution of clusters in the \thethree as a function of the $\chi_{\rm D}$ (Eqn.~\ref{dynamical}) and $\chi_{\rm S}$  (Eqn.~\ref{structural}) criteria. The hatched cyan and magenta regions show the 20 most relaxed clusters  and the 20 most disturbed clusters  respectively based on $\chi_{\rm D}$ criterion. The hatched green and orange 20 most show the 20 most regular profiles and the 20 most irregular profiles respectively based on $\chi_{\rm S}$ criteria. }
	\label{fig1}
\end{figure*}
\begin{figure}
    \includegraphics[width=0.49\textwidth]{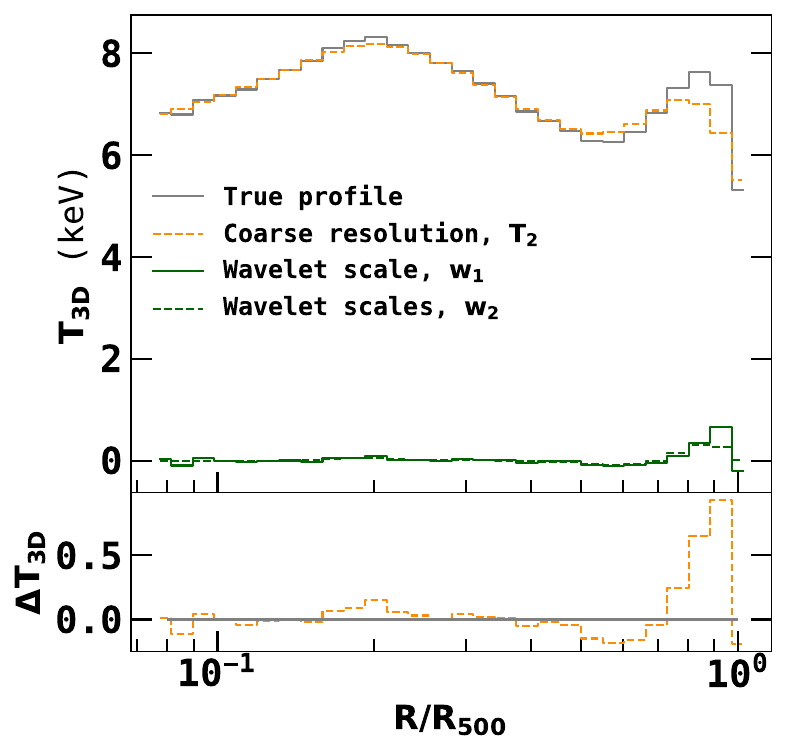}
    \caption{\footnotesize Smooth (coarse) component of a complex temperature profile derived from the application of the Starlet transform with $J =2$. The bottom panel shows the corresponding difference between true and smooth temperature profiles.}
    \label{starlet}
\end{figure}

We also considered a more complex case where we use the spectroscopic-like weighting proposed by \citet{Mazzotta2004} to generate the 2D temperature profiles using the binning schemes discussed above. In this case, apart from the normalisation, the matrix elements of $\bf{C}$ simply change to ${C}_{i,j}={C}_{i,j}T^{3/4}_{3D,j}$ (or equivalently, the weights change to $w = \rho^2 T_{3D}^{-3/4}$), where $T_{3D,j}$ is the mass-weighted 3D temperature profile in the $j^{\rm th}$ bin. 
 
In many clusters in the simulated sample, the temperature profiles in the first few inner bins (typically 0-13 radial bins corresponding to radii between $\approx$ [0.02-0.07]\,R$_{500}$) were noisy (i.e having $< 100$ gas particles). For such systems, the 2D profiles were estimated without considering such bins.

Figure~\ref{figcom} shows the observed scaled 2D temperature profiles of the {\it Planck}  SZ sample \citep{2011A&A...536A..11P} and the XMM-{\it Newton} DR1 sample \citep[][described in detail in Sect.~\ref{sec:DR1desc}]{Rossetti2023}. These are compared to 50 randomly drawn 2D temperature profiles from the \thethree using emission measure (left panel) and spectroscopic-like (right panel)  weighting schemes and an observation-like convolution matrix, ${\bf C}_{12,48}$. Both observational and simulated temperature profiles were scaled by the average 2D temperature (T$_X$) in the radial range of [0.15-0.75]\,R$_{500}$. Figure~\ref{fig_dis_Tx} shows the distribution of the clusters in the simulated sample, {\it Planck} sample and DR1 sample on the basis of T$_X$.  These two figures illustrate three points that will be critical for the following study: 
\begin{enumerate}
\item In common with a number of works over the last 20 years \citep[e.g.][]{deg02,Vikhlinin2006,pra07,lec08,Ghirardini2019}, the structural similarity in the observed temperature profiles are clearly visible in Fig.~\ref{figcom}. The central regions are characterised by a large spread, due to a mixed population of cool core and disturbed systems, while beyond the central 0.15\,R$_{500}$ the profiles all decline in a similar fashion.  \newline
    \item The simulated profiles follow the same general trend as the observed profiles. The average trend and 1-$\sigma$ dispersion of the simulations is very consistent with that of the CHEX-MATE DR1 sample. The simulated temperature profiles on average are slightly hotter in the centre compared to the {\it Planck} SZ sample. This may  be related to the fact that there are more low mass clusters in the simulated sample compared to the {\it Planck} SZ sample. Such low mass clusters are expected to be more strongly affected by AGN feedback, potentially leading to higher temperatures in the central region \citep{2018MNRAS.480L..68I}.  Alternatively, the higher central temperatures  in the simulations may simply be due to the fact that the sample has a large number of NCC clusters.  \newline

    \item Overall, the observed temperature profiles are well represented by the simulated sample. This fact will be key to a successful training stage of the {\bf IAE} model, which relies on identifying underlying trends in the data that would not otherwise be found. We note that the simulated profiles do not have to precisely match the observed data: as we will see, the most important point is that they reproduce the overall structure and diversity of the observed profiles, which is what our {\bf IAE} model learns. 
\end{enumerate}

We further classified the simulated clusters using three schemes. This is important to quantify how well the {\bf IAE} model reconstructs the radial temperature distribution for different types of objects and profile shapes. 

\subsection{CC and NCC classification}

Firstly, we classify the profiles as CC and NCC by visual inspection. The objective here is simply to select simulated profiles that mimic those of observed cool-core-like clusters with a central temperature drop, and non cool-core clusters that display an almost isothermal central temperature profile. The profiles which show a decreasing trend towards the cluster centre (positive temperature gradient) were classified as CC clusters. We identify about one-third of the clusters as belonging to the CC class. 
In Fig.~\ref{fig2}, grey lines in the left panels and right panels show the 3D temperature profiles (${\bf T}_{\rm 3D}$) of CC and NCC clusters respectively. 

\subsection{Dynamical classification}
\label{sec:dyn}
    Clusters in these simulations were classified on the basis of their  intrinsic dynamical state (relaxed or disturbed) using  a variety of  estimators \citep{Rasia2013}. The two important intrinsic estimators are $f_{\rm s}=M_{\rm sub}/M_{\rm tot}$, the fraction of cluster mass ($M_{\rm tot}$) included in substructures ($M_{\rm sub}$), and $\Delta_{\rm r}=|r_{\delta}-r_{\rm cm}|/R_{\rm ap}$, which is the measure of the offset between the central density peak ($r_{\delta}$), and the centre of mass ($r_{\rm cm}$) of the cluster normalised to aperture radius $R_{\rm ap}$. Both of the estimators were computed at R$_{500}$.  Both $f_{\rm s}$ and $\Delta_{\rm r}$ are expected to be lower than 0.1 for relaxed objects \citep{Cialone2018,DeLuca2021}. These two dynamical parameters can be combined \citep{Rasia2013} to give the so-called relaxation parameter $\newchi_D$ 
\begin{equation}
\chi_{_D}=\frac{1}{2}\times \left(\frac{\Delta_{\rm r}-\Delta_{\rm r,med}}{|\Delta_{\rm r,quar}-\Delta_{\rm r,med}|} + \frac{f_{\rm s}-f_{\rm s,med}}{|f_{\rm s,quar}-f_{\rm s,med}|}\right).
\label{dynamical}
\end{equation}
Here $\Delta_{\rm r,med}$ and $f_{\rm s,med}$ are the medians of the $\Delta_{\rm r}$ and $f_{\rm s}$ distributions, respectively, and $\Delta_{\rm r,quar}$ and $f_{\rm s,quar}$ are the first or the third quartiles, depending on whether the parameters of a specific cluster are smaller or larger than the median. According to this definition, clusters with $\chi_{_D}<0$ are classified as relaxed, and clusters $\chi_{_D} > 0$ are classified as disturbed. The left panel of Fig.~\ref{fig1} shows the histogram of $\chi_{_D}$ values. The cyan and magenta hatched regions represent the 20 most relaxed clusters and 20 most disturbed clusters, respectively. We will refer to these sub-samples as MR20 and MD20 hereafter. In the top panel of Fig.~\ref{fig2}, we show the corresponding temperature profiles of the MR20 clusters (left panel) and the MD20 clusters (right panel) with cyan and magenta lines, respectively. It is interesting to note that only a few of the most relaxed objects are also categorised as CC clusters. Visual inspection of emissivity maps shows, as expected, that $\chi_{_D}$ is clearly linked to the overall gas morphology, as also found in \cite{2022A&A...665A.117C}.

\subsection{Structural classification}
\label{sec:str}

To enable a better assessment of the performance  of the {\bf IAE} model for temperature profile reconstruction, we also classified the 3D temperature profiles based directly on their smoothness. Bumps in the temperature profiles are usually associated with complex astrophysical processes such as merger shocks, gas condensation, the presence of cold substructures, sloshing, and turbulence, all of which affect the temperature in a given annulus. To measure the degree of the bumpiness of the 3D temperature profiles, we used the starlet wavelet transform, which is widely used in component separation in astrophysical images \citep{2007ITIP...16..297S}, to split each profile into its smooth and non-smooth components. Using this technique, the 3D temperature profile ${\bf T}_{\rm 3D}(r)$ can be decomposed into a $J+1$ coefficient set
${\bf W} = \{{\bf w}_1,...,{\bf w}_J,{\bf T}_J\}$, as a superposition of the form
\begin{equation}
{\bf T}_{\rm 3D}(r)={\bf T}_J(r)+\overset{J}{\underset{j=1}{\sum}}{\bf w}_j(r)\, ,
\end{equation}
where ${\bf T}_J$ is a smooth (coarse resolution) version of the original temperature profile and ${\bf w}_j$ represents the structure in the temperature profile on scale $2^{-j}$. 
\begin{figure*}
	\centering	
	\includegraphics[width=0.98\textwidth, angle =0 ]{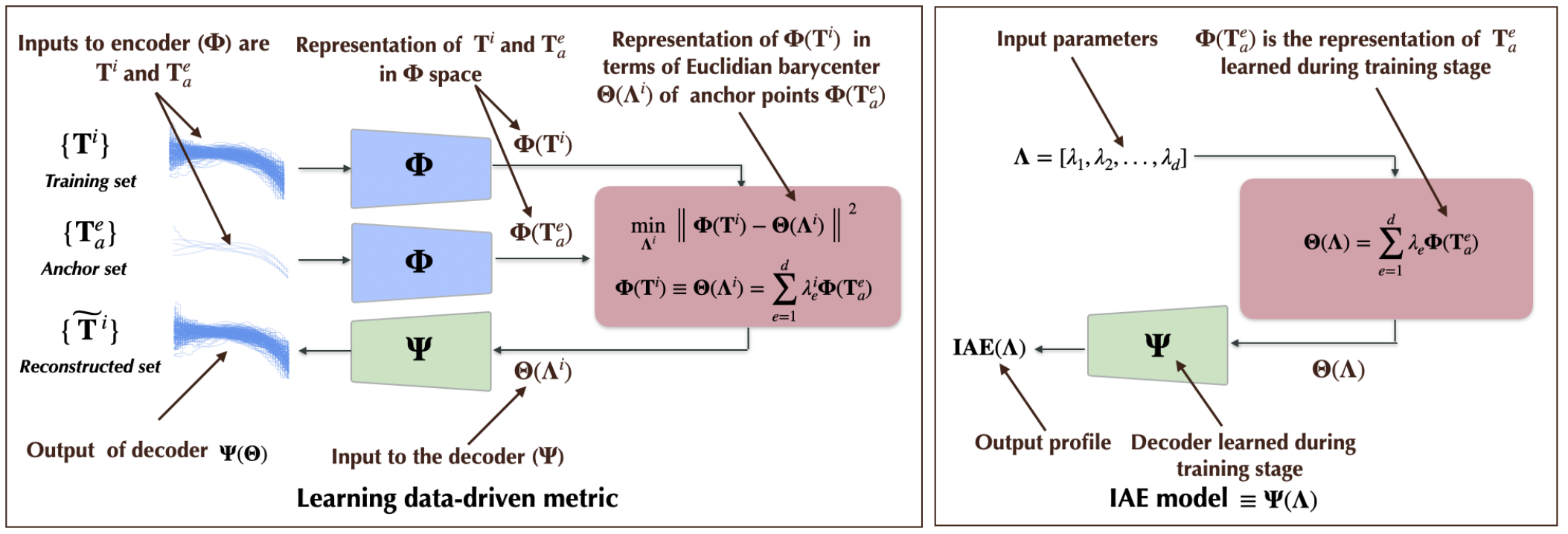}   
        	\caption{\footnotesize Design of the neural network used in this work. Left Panel: Neural network used in the training stage. {$\bf \Phi$} and {$\bf \Psi$} represent the encoder and decoder respectively. ${\bf T}^i$ are the elements of the training set and ${\bf T}^e_a $ are the elements of the anchor set. ${\bf \Phi}({\bf T}^i)$  and ${\bf \Phi}({\bf T}^e_a)$ are the representations of ${\bf T}^i$ and ${\bf T}^e_a $, respectively, in the encoder (feature) space. ${\bf \Theta}({\mathbf \Lambda}^i)= {\bf \Theta}([\lambda_1^{i} ,..., \lambda_d^{i}])$ is the Euclidean barycentric representation of ${\bf \Phi}({\bf T}^i)$ in terms of $d$ anchor points ${\bf \Phi}({\bf T}^e_a)$,  which is fed to the decoder. ${\bf \Psi}({\bf \Theta})$  is the reconstructed output of the decoder. The network is trained by minimising the error between the input ${\bf T}^i$ and output ${\bf \Psi}({\bf \Theta})$ temperature profiles. Right panel: Neural network (IAE model) of temperature profiles, where $\lambda_1,...,\lambda_d $ are the input parameters and IAE($[\lambda_1 ,..., \lambda_d]$) is the output temperature profile. The decoder is not required in any step here.} 
	\label{figA}
\end{figure*}

Figure~\ref{starlet} shows the starlet decomposition for one of the clusters in the \thethree which exhibits a complex shape in the range [0.5-1]\,R$_{500}$. The cluster is experiencing a major merger and there is an enhancement of the temperature due to the propagation of a shock in this region. 
We use the starlet transform with $J=2$, which we have found to be the optimal configuration to measure the non-smoothness, yielding a decomposition into a smooth temperature component and two additional non-smooth components, ${\bf w}_1(r)$ and ${\bf w}_2(r)$. We then define the root mean square deviation, $\chi_{_S}$ of the difference between the true and smooth temperature profiles in the radial range of [0.08-1]\,R$_{500}$ as a measure of the non-smoothness of the temperature profiles. 
\begin{equation}
\chi_{_S}=\sqrt{\frac{1}{u}\sum_{i=1}^{u}(T_{3D,i}-T_{J,i})^{2}}\,,
\label{structural}
\end{equation}
where $u$ is the number of data points in the range of [0.08-1]\,R$_{500}$, and the lower limit of 0.08\,R$_{500}$ corresponds to the radius at which all clusters satisfy the precision threshold condition. The temperature profiles were first scaled (normalised) by the average mass-weighted temperature in the radial range of [0.15-0.75]\,R$_{500}$ before applying the decomposition operator to calculate $\chi_{_S}$. The right hand panel of Fig.~\ref{fig1} shows the distribution of $\chi_{_S}$ for the full sample, which follows an approximately log-normal distribution. The green and orange hatched regions represent the 20 most smooth profiles and 20 most irregular profiles, respectively, based on the $\chi_{_S}$ criterion. We will refer to these sub-samples as MS20 and MI20 henceforth. In the bottom panel of Fig.~\ref{fig2}, we show the corresponding temperature profiles of the MS20 (left panel) and MI20 profiles (right panel) with green and orange lines respectively. Here also, only a few of the clusters with the most smooth profiles are categorised as CC clusters. The correlation between $\chi_{_D}$  and $\chi_{_S}$ is shown in Fig.~\ref{appx1}. They are moderately correlated, with a  Spearman's correlation coefficient of 0.42 and a $P$ value of $5 \times 10^{-15}$.

\section{Neural network model for learning 3D temperature profiles}

The deconvolved temperature profile can in principle be obtained by solving the following classical inverse problem
\begin{equation}
{\bf T}_{\rm 2D}={\bf C}\otimes {\bf T}_{\rm 3D }+\bf N\,,
\label{eq11}
\end{equation}
where {\bf C} is a non-linear operator (matrix) which represents the observational and instrumental effects (projection, PSF, etc) and {\bf N} represents the statistical properties of the noise. The standard way of solving Eqn.~\ref{eq11} is to consider least squares regression with some regularisation $\bf R$
\begin{equation}
 {\bf T}^{\rm fit}_{\rm 3D}=\min_{\bf T} \,\,{\bf R} ({\bf T})+ \left \| {\bf T}_{\rm 2D} - {\bf C}\otimes{\bf T} \right\|^2\, ,
\end{equation}
where ${\bf  T}^{\rm fit}_{\rm 3D}$ is the best-fitting model profile for ${\bf T}_{\rm 3D}$, which is obtained by optimising the above relation with respect to $\bf T$. However, Eqn.~\ref{eq11} is an ill-posed (non-linear) problem, and using standard non-parametric methods does not result in a unique and stable solution. Therefore, one has to resort to advanced deconvolution techniques. In this work, we propose one such algorithm that makes use of neural networks to model the temperature profiles, and whose framework will be explained below. A learning-based regularisation procedure for direct deconvolution using the trained neural network is discussed in Sect.~\ref{sec1}.

Our approach is based on manifold learning, which stems from the manifold hypothesis, that suggests the existence of a lower dimensional manifold on which real-world data lies \citep{2013arXiv1310.0425F}. This is evidently the case for galaxy cluster temperature profiles, which clearly display some degree of regularity, as seen in Fig.~\ref{fig1}. The goal is then to find the lower dimensional manifold by learning the underlying structure of the data. When one has access to a large training set (from observations and/or simulations), it may be possible to make use of machine learning (deep learning) methods to build an underlying manifold. However, this  becomes quite difficult when available training samples are sparse, as is the case for cluster temperature profiles. In such cases, rather than learning the underlying manifold structure, \cite{bobin:hal-03265254} proposed the Interpolator AutoEncoder ({\bf IAE}), that learns to travel on a manifold by way of interpolation between a limited number of {\it anchor points} that belong to it.

We assume that any temperature profile in a training set $\{{\bf T}^i\}^{i=1,...,n}$, where $n$ represents the total number of elements in the set, can be interpolated from a small set of $d$ anchor points $\{{\bf T}^e_a\}^{e=1,...,d}$ using an appropriate metric $\bf \Pi$ 
\begin{equation}
 { \bf \Theta} ({\mathbf \Lambda}^{i})=\min_{{\mathbf\Lambda}^{i}}\sum_{e=1}^d{\mathbf \lambda_{e}^{i}} {\bf \Pi}({\bf T}^i,{\bf T}^e_a)\,,
  \label{eq:3}
\end{equation}
where $\bf \Theta$ is called the {\it barycentre}. The elements of vector ${\mathbf \Lambda}^{i}=[\lambda_{1}^{i},...,\lambda_{d}^{i}]$ are the barycentric weights ($\sum_{e=1}^d\lambda_e^{i}=1$) which are optimised in the above equation.  If we consider the metric $\bf \Pi$ to be Euclidian, then
\begin{equation}
{\bf \Theta} ({\mathbf \Lambda}^{i})=\min_{{\mathbf \Lambda}^{i}} \sum_{e=1}^d \lambda_{e}^{i} ||{\bf T}^i-{\bf T}_a^e||^2.
 \label{eq:4}
\end{equation}
The above equation reduces ${\bf \Theta} ({\mathbf \Lambda}^{i})$ to an orthogonal projection onto the span of anchor points ${\bf T}_{a}^e$, that is
\begin{equation}
{\bf T}^{i} \equiv
{\bf \Theta} ({\mathbf \Lambda}^{i})=\sum_{e=1}^d \lambda_{e}^{i} {\bf T}_{a}^e.
 \label{eq:4b}
\end{equation}

The problem then reduces to finding (optimising) barycentric weights such that the barycentre $\bf \Theta $ accurately reconstructs any input temperature profile in the training sample.

However, if the profiles are non-linear, with varying amplitudes and shapes, as is the case with the temperature profiles in galaxy clusters, the standard metric $\bf \Pi$ may not reconstruct an appropriate barycentric representation. Our method, therefore, uses the approach proposed by \citet{9022675,bobin:hal-03265254}, in which a data-driven metric is constructed using a deep learning neural network that is well adapted to build physically relevant barycentres of anchor points. We introduce an auto-encoder \citep{vincent2010stacked} inspired neural network model which learns to transport points (temperature profiles in our case) onto the underlying manifold using a non-linear interpolation scheme between the anchor points.

The structure of the neural network we are considering is shown in the left hand panel of Fig. ~\ref{figA}. It consists of an encoder (${\bf \Phi}$), that takes an input, and a decoder (${\bf \Psi}$), that generates the desired output. The role of the encoder is to transform the input data into a lower-dimensional representation, while the decoder is responsible for mapping the lower-dimensional data back into the original space. By performing these mappings, auto-encoders are able to learn the underlying structure of the data. In contrast to standard auto-encoders, our model training is performed by minimising the error between the input and the reconstructed training sample according to the Euclidean distance onto the manifold spanned by the anchor points in the encoder (feature space). 

More precisely, for the encoder ${\bf \Phi}$, the representation of the input profile ${\bf T}^i$ (belonging to the training set ${\bf \Phi}({ \bf T}^i)$) is expressed in terms of the barycentre, $\bf \Theta$, in feature space, as an orthogonal projection onto the span of the anchor points ${\bf \Phi}({\bf T}_{a}^e)$ given in  Eqn.~\ref{eq:4b}:
\begin{equation}
{\bf \Phi}({\bf T}^i) \equiv {\bf \Theta} ({\mathbf \Lambda}^{i}) = \sum_{e=1}^d \lambda_e^{i} {\bf \Phi}({\bf T}^e_a).
\label{eq:14}
\end{equation}
The barycentric weights are constrained to sum to one so as to avoid certain scaling indeterminacies, and are not necessarily constrained to be positive like actual barycentric weights, which potentially allows us to extrapolate beyond the affine hull of the encoded anchor points. More precisely, the barycentric weights for the $n$ elements in the training sample are computed as follows:
\begin{equation}
\min_{\substack {\mathbf \Lambda^{i}}} \sum_{i=1}^n\left \| {\bf \Phi}({\bf T}^i)-\sum_{e=1}^d\lambda_e^{i}{\bf \Phi}({\bf T}_a^e)\right \|^2 \mbox{ s.t. } \sum_e \lambda_e^{i}=1,
\end{equation}
which can be approximated by taking the solution to the least-squares problem followed by a rescaling of the barycentric weights in order to make them sum to one.

Once the optimal barycentric weights (${\bf \Lambda}^{i}$) are computed for each element ${\bf T}^i$  of the training sample, the approximations (i.e. the barycenters) go back through the decoder ${\bf \Psi}$ to reproduce the input as {$\widetilde{\bf T}^i$}=${\bf \Psi}({\bf \Theta})={\bf \Psi}(\sum_{e=1}^d\lambda_e^{i}{ \bf \Phi}({\bf T}_e^a))$. 
The learning stage reduces to estimating the weights and biases of layers of $\bf \Phi$ and $\bf \Psi$ using an appropriate cost function  that minimises the error between the input, ${\bf T}^i$ and the output, ${\bf \Psi}(\sum_{e=1}^d\lambda_e^{i}{\bf \Phi}({\bf T}_a^e))$, so that
\begin{equation}
\label{eq:training_loss}
\min_{\substack {\bf \Phi,\Psi}} \mu\sum_{i=1}^n\left \| {\bf T}^i-  {\widetilde{\bf T}^i} \right \|^2 + \sum_{i=1}^n\left \| {\bf \Phi}({\bf T}^i)-  {\bf \Theta} ({\bf T}^i) \right \|^2.
\end{equation}
In the training stage, we thus learn the non-linear interpolation scheme that best approximates the training samples in feature space, and the mapping between the barycentres and real space. The parameter $\mu$ controls the trade-off between these two objectives. 
In the evaluation phase 
only the decoder ${\bf \Psi}(\bf \Theta)$, which embeds the mapping between the barycentric weights and 3D temperature profiles, is used. As shown in the right panel of Fig.~\ref{figA}, the decoder is used as a generative model that is parameterised by the barycentric weights, ${\bf \Lambda}$ (for convenience we drop the subscript `$i$' from now on). This model can  easily be convolved to fit the observed 2D temperature profile so as to recover the true (3D) temperature profile. From now on, we refer to the decoder as the {\bf IAE} model. 
The number and choice of anchor points and the model training will be discussed in the following Section.

\begin{figure}
    \includegraphics[width=0.49\textwidth]{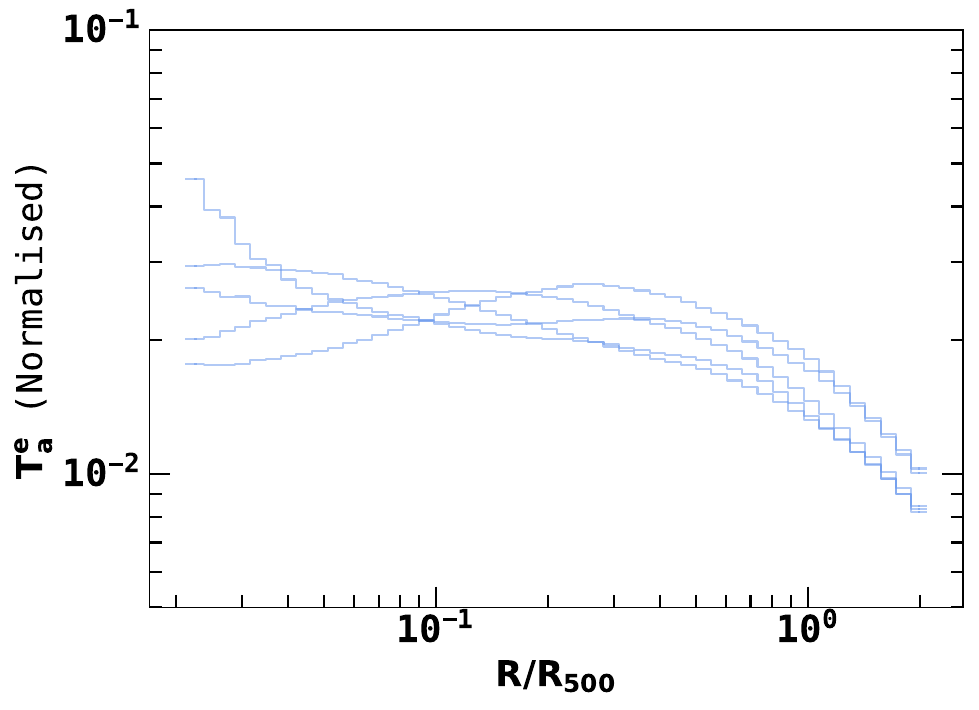}
    \caption{Five anchor points (example profiles), ${\bf T}_a^e$, where $e$ runs from 1 to 5 used in the IAE model.}
    \label{fig_anchor_points}
\end{figure}

\section{Model training and fitting}
\label{sec1}
\subsection{Model training}
We use a JAX \citep{jax2018github} implementation to develop and train the {\bf IAE} model. As a training sample, we use 200 randomly drawn ${\bf T}_{\rm 3D}$ profiles from the full sample of 315 extracted from the {\sc Three Hundred Project} simulations.

Each profile in the training sample is first normalised to entries that sum to $1$. The model is trained at the same fixed radial binning as that of ${\bf T}_{\rm 3D}$ profiles in the [0.02-2]\,R$_{500}$ radial range. 
For the training stage, several configurations were tested, among which the following choices were found to perform the best:
\begin{table*}[!h]   
	 	\caption{\footnotesize Details on the neural network architecture and hyper-parameters used in this work.}
	 	\begin{center}
	\begin{minipage}[t]{.45\linewidth} 
		\centering
		\begin{tabular}{ccccc}
			\toprule
                \toprule
			Layer type & 	Layer &  Activation & Neurons \\
			\midrule
			\multirow{4}{*}{Encoder}  &	Input &  - & 48 \\
			&	Layer 1 & Mish & 48\\
			&	Layer 2 & Mish & 48 \\
			&	Output &  - & 48\\
			\midrule
			\multicolumn{4}{c}{Baycenter representation in $\Lambda$} \\
			\midrule
			\multirow{4}{*}{Decoder}  & 	Input & - & 48 \\
			& 	Layer 1 & Mish & 48 \\
			&	Layer 2 & Mish & 48 \\
			&	Output &  - & 48 \\
			\bottomrule
		\end{tabular}
	\end{minipage}%
	\hspace{0.2cm}  
	\begin{minipage}[t]{.5\linewidth}  
		\centering
		\begin{tabular}{cc}
			\toprule
                \toprule
			Parameters & Values \\
			\midrule
			Optimiser &  Adam \\
			Step size & $10^{-3}$ \\
			Batch size & 32 \\
			Batch normalisation & Yes \\
			Iterations & 25000 \\
                Residual parameter ($\varepsilon_{0}$) & 0.1\\
			Noise & Gaussian \\
			Cost function & Mean squared error \\
			Dropout & None \\
			\bottomrule
		\end{tabular}
	\end{minipage} 
\end{center}
\footnotesize{\textbf{Notes:} The output of the encoder is first transformed into the barycenter of the anchor points using Eqn. \ref{eq:14}.}
  \label{network}
\end{table*}

\begin{itemize}
    \item {\bf Network architecture:} Both the encoder and the decoder are multi-layer perceptron (MLP) neural networks, which are composed of $2$ layers, each of which has a number of hidden units equal to the input signal dimension (i.e. 48). We employ a smooth and non-monotonic Mish\footnote{$\text{Mish}(x) = x \times \tanh(\ln(1 + e^{x}))$} activation function to introduce non-linearity and enhance the learning capacity of our deep neural network model. Since the {\bf IAE} model employs a barycenter transformation of the training sample in encoder space to achieve dimensionality reduction, in this work, we only focus on a specific architecture with a fixed number of neurons per layer, corresponding to the dimension of the input samples. Further exploration of more general architectures is left for future work. For both encoder and decoder, the output $\mathbf{Z}^{l+1}$ of layer $l$ can be expressed as
    \begin{equation}
    \mathbf{Z}^{l+1} = \text{Mish}(\mathbf{W}^{l} \otimes \mathbf{Z}^{l} + \mathbf{b}^{l})+ \varepsilon^{l} \mathbf{Z}^{l}.
    \end{equation}
    Here, the first term represents the standard output of the neural network, with $\mathbf{W}$ and $\mathbf{b}$ defined as weight matrix and bias vector respectively. The second term represents skip connections \citep{2015arXiv151203385H,2016arXiv160806993H}, also known as residual connections. The skip connection acts by partially re-injecting $\mathbf{Z}$ up to a layer-dependent scalar factor $\varepsilon$. In general, the residual injection factors are typically chosen to be small for low-level layers and larger for deeper layers. This approach helps mitigate the vanishing gradient phenomenon, which is commonly encountered during the training of deep networks. For each layer $l$ of encoder and decoder, we consider following the functional form of $\varepsilon^{l}$ as used in \cite{bobin:hal-03265254}
    \begin{equation}
     \varepsilon^{l}   = \varepsilon_{0} \left(2^{1/l} - 1\right),
    \end{equation}
    where $\varepsilon_{0}$ is a constant factor. By using skip connections with re-injection and layer-dependent scaling, the model can leverage both the direct information flow from earlier layers and the higher-level abstractions learned by the deeper layers, which can lead to improved performance and better training in deep neural networks. 
    \newline

    \item {\bf Cost function:} The cost function defined in Eqn.~\ref{eq:training_loss} is composed of two terms. The first term measures the reconstruction error in real space, and the second term defines the error in feature space. The parameter $\mu$ allows one to tune the trade-off between these two terms. An accurate {\bf IAE} model relies on both a low reconstruction error (i.e. first term of the training loss), and an efficient interpolation scheme in  feature space. It has been emphasised in \cite{bobin:hal-03265254} that the second term helps improve the training process by constraining the feature space. In addition, depending on the problem and data at stake, it can help to increase the model accuracy by reducing the interpolation error in feature space, which in turn can reduce  noise propagation at inference. In the present case, we noted that the trained model is not particularly sensitive to $\mu$, which we set to $10\,000$ to minimise the reconstruction error in  real space.   \newline
    \item {\bf Training hyper-parameters:} The batch size (the number of training profiles processed together before updating the neural network weights) is fixed to $32$. The optimisation is performed by back-propagation using the standard Adam solver \citep{2014arXiv1412.6980K} with a step size of $10^{-3}$ and a number of epochs equal to $25000$. It is customary to further regularise the model by adding noise to the training samples, which limits over-fitting effects. To do that, Gaussian noise with mean zero and standard deviation of $2\times10^{-3}$ is added to the samples at the training stage. The batch normalisation was achieved by normalising the input batch using a global mean of 0 and a standard deviation of 1. Finally, we fix the residual parameter ($\varepsilon_{0}$) to 0.1. \newline
    \item {\bf Number of anchor points:} Anchor points serve as the basis on which temperature profiles are reconstructed using barycentric weights. Training with a small number of anchor points results in smoother (more regular) profiles; conversely, a large number of anchor points increases the model-to-data fidelity. Thus, the choice of the number of anchor points used during the training stage is essentially equivalent to choosing a regularisation parameter. For our study, the number of anchor points is fixed at five. These are generated by first dividing the training sample into five groups using a $k$-means clustering algorithm. The anchor points are then assumed to be the central points (centroids) of these five groups. By using five anchor points, we can ensure that the model-to-data residual remains below 10\% over the observable radial range of $\approx$ [0.02-1]\,$R_{500}$, as shown in Sect.~\ref{main_sec_sim}, and at the same time, we can avoid any  possible biases that could be introduced if the observations were shallow (bias-variance problem).
    Figure~\ref{fig_anchor_points} shows the anchor points used in the neural network model.  In Sect.~\ref{sec:apw}, we will discuss the effect of increasing the number of anchor points.

\end{itemize}
   \begin{table}
           \caption{\footnotesize Flat priors used for the IAE model parameters.}
        \begin{center}   
        \begin{tabular}{c c } 
            \toprule
            \toprule
            Parameter     &  Range \\
                \midrule
            $\lambda_1$   & $-14 ; +22$    \\
            $\lambda_2$    & $-10 ; +15$ \\
            $\lambda_3$   & $-30 ; +17 $      \\
            $\lambda_4$   & $-3 ; +8 $          \\
            $\lambda_5$   & $-15 ; +10  $       \\
            $\alpha$   & $100 ; +530 $           \\
            \bottomrule
       \end{tabular}
       \end{center}
      \label{priors}
   \end{table}

Table~\ref{network} provides a comprehensive summary of our neural network architecture, along with the optimal hyper-parameters used in the study. For our implementation, we used publicly available source code hosted on a GitHub repository\footnote{\url{https://github.com/jbobin/IAE}} \citep{9022675,bobin:hal-03265254}. 

Since our simulated sample is small, we 
use the term `validation' to refer to testing of the model performance on simulated data (Sec.~ \ref{main_sec_sim}) before using it on real-world data, where the 3D temperature profiles are not directly available. We therefore used the training sample itself to evaluate the convergence of the cost function. Specifically, we monitored the cost function during training and found that after approximately 25000 iterations, the cost function reached a point where it became flat. At this stage, we considered the training process to be sufficiently converged, and we terminated the training.

\subsection{Model fitting}
The {\bf IAE} model is tested/fitted on the validation sample consisting of the remaining 115 galaxy clusters in the sample which were not used in the training stage. We have verified that the validation sample is representative of the full sample: about one-third of the validation clusters have cool cores, and the fractions of relaxed/disturbed clusters and smooth/irregular profiles are similarly distributed in the training and validation samples.

We employ Markov Chain Monte Carlo (MCMC) analysis to estimate the parameters of the IAE models and use the publicly available {\tt emcee} python package \citep{2013PASP..125..306F} for this purpose. The parameter estimation is undertaken on all the IAE parameters: the five anchor point weights $ {\mathbf \Lambda}=[\lambda_1,\lambda_2, \lambda_3, \lambda_4, \lambda_5]$, and the amplitude (normalisation) parameter $\alpha$. 

The deconvolved temperature profile can be obtained from the trained non-parametric IAE model by minimising the following log-likelihood:
\begin{eqnarray}
\mathcal{L}({\mathbf \Lambda},\alpha)=\Gamma \times \mbox{Tr}\left(({ {\mathbf \Lambda}} - \bar{{\mathbf \Lambda}})^\top \otimes
 {\mathbf  \Sigma}^{-1}_{\bf t} \otimes({ {\mathbf \Lambda}} - \bar{ {\mathbf \Lambda}})\right) + \quad \quad\quad \quad \nonumber \\ 
\frac{1}{2}\times \mbox{Tr}\left(({\bf T}^{\bf val} -{\bf T}^{\bf IAE})^\top \otimes
 {\mathbf  \Sigma}^{-1}_{\bf o} \otimes({\bf T}^{\bf val} -{\bf T}^{\bf IAE})\right),
\label{eq:b}
\end{eqnarray}
where $\bf T^{val}$ is a temperature profile (2D or 3D) in the validation sample to be fitted, ${\bf \Sigma}_{\bf o}$ is the error covariance matrix and ${\bf T}^{\bf IAE}={\bf C} \otimes {\bf IAE}({\mathbf \Lambda},\alpha)$ is the corresponding convolved IAE model predicted profile. Tr and $\top$ represent the trace and transpose of the matrix respectively. The first term represents  the mean proximity term, with $\Gamma$ controlling its overall contribution to the likelihood. This  enforces the solution to be a barycentre of the example profiles (i.e. it searches for the best approximation of the input signal with respect to the learned model/network). We find that $\Gamma$ in the range 0.1-1 generally provides good results, and we, therefore, fix it to 1. $\bar{{\mathbf \Lambda}}$ (the mean value of the ${\mathbf \Lambda}$s) and ${\mathbf  \Sigma}_{\bf t}$ (the covariance matrix of the ${\mathbf \Lambda}$s) are computed from the training set by generating 100 Monte Carlo simulations for each cluster with log-normal noise, which are then subsequently fitted to the {\bf IAE} model using the {\tt Adam} optimiser.  This cost-effective regularisation strategy is introduced to avoid model extrapolation (physically unrealistic results), and enables us to have a robust and effective deconvolution algorithm.
The second term is the standard likelihood related to some additive Gaussian noise perturbation.  We have used flat prior distributions and Tab.~\ref{priors} shows the prior ranges of all parameters. We used {\tt Getdist} \citep{Lewis:2019xzd} with the chains generated by {\tt emcee} to produce 2D contours and marginal posteriors.
\begin{figure*}
	\includegraphics[width=0.49\textwidth]{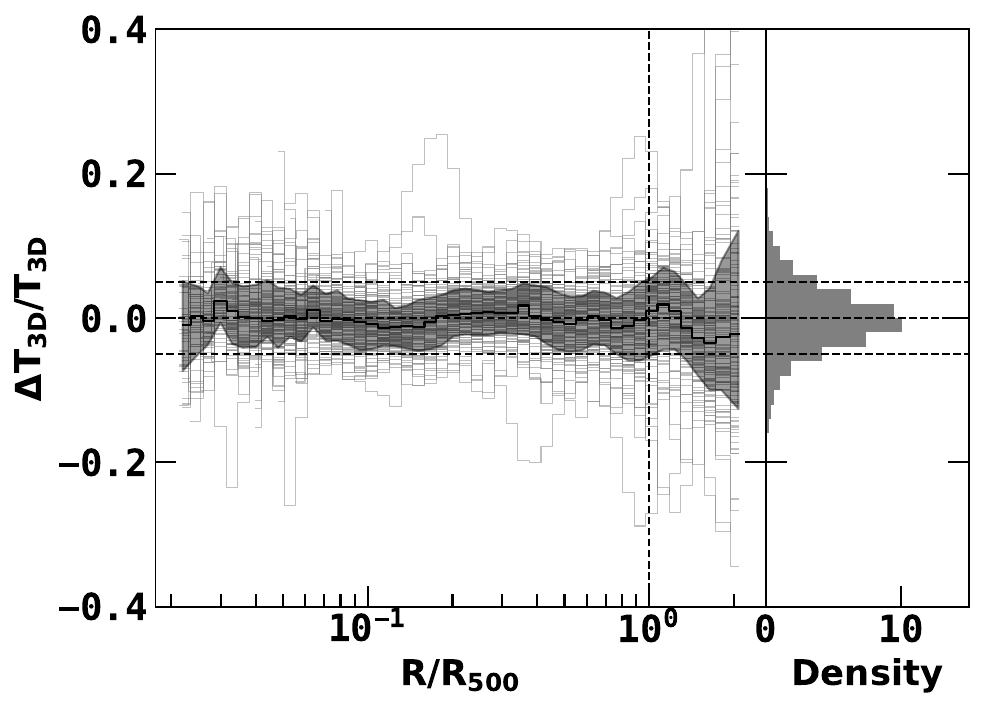}
\hfill
		\includegraphics[width=0.49\textwidth]{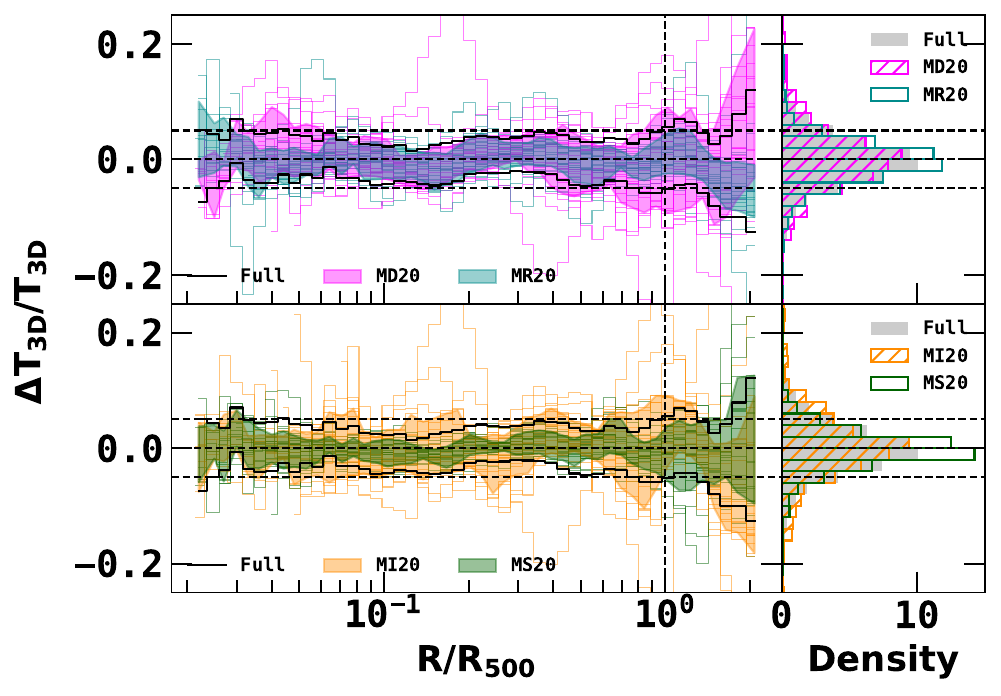}
	\caption{\footnotesize Fractional residuals for 115 clusters in the validation sample with {\bf IAE} for the 3D-3D fit. The three horizontal dashed black lines represent zero and $\pm$5\% fractional residuals; the vertical dashed black lines represent R$_{500}$.
 Left panel: The grey lines show the individual fractional residuals of all the clusters. The solid black line and shaded black region show the median and 1-$\sigma$ dispersion of the fractional residual distribution, respectively.  The histogram shows the distribution of fractional residuals over all radii. Right panel: The cyan and magenta lines in the top panel show the fractional residuals of MR20 and MD20 sub-samples, respectively. The green and orange lines in the bottom panel show the fractional residuals of the MS20 and MR20 sub-samples respectively. Shaded regions show the corresponding 1-$\sigma$ dispersion of the fractional residual distribution. The  histograms show the distribution of fractional residuals over all radii. Regions enclosed by the solid black lines show the 1-$\sigma$ dispersion of 
 the fractional residual of the full validation sample. The {\bf IAE} model can reconstruct 3D temperature profiles with a fractional difference of about 5\% across nearly the full radial range. }
	\label{fig6}
\end{figure*}

%
\begin{figure}[!h]
	\includegraphics[width=0.49\textwidth]{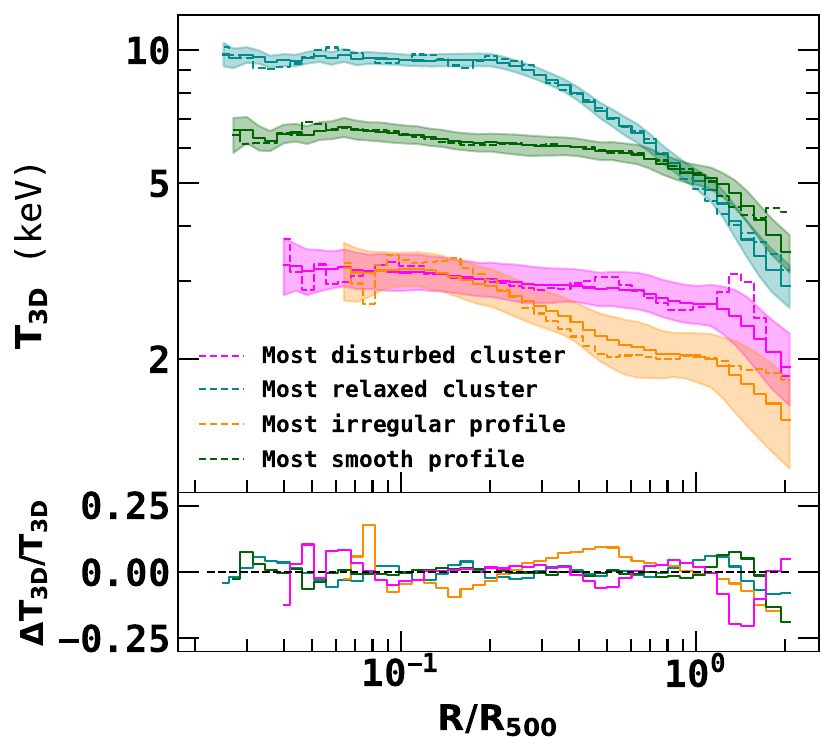}
	\caption{\footnotesize Results for the most relaxed / disturbed clusters and for the most smooth / irregular profiles with {\bf IAE} for the 3D-3D fit.
 Left panel: Dashed cyan and magenta lines show the true 3D temperature profiles of the most relaxed and disturbed clusters respectively in the validation sample. Similarly, dashed green and orange lines show the most smooth and irregular true 3D temperature profiles respectively. The solid lines and the corresponding shaded regions show the median and 1-$\sigma$ dispersion of the reconstructed temperature profile obtained from the IAE model using MCMC.} 
	\label{fig4}
\end{figure}
   
The {\bf IAE} model testing was undertaken by fitting it to the ${\bf T}_{\rm 3D}$ profiles and ${\bf T}_{\rm 2D}$ profiles built in Sect.~\ref{sec:simulations}. For simplicity, we ignore the PSF in the testing phase. We tested and validated our model by considering three fitting cases:

\begin{enumerate}
  \item {\bf 3D-3D fit with fine binning}: the ${\bf T}_{\rm 3D}$ profiles are directly fitted to recover the best-fitting 3D profiles from the {\bf IAE} model. The goal, in this case, is to assess the ability of the {\bf IAE} model to reproduce the input 3D temperature profile shape. In this case, as there is no projection, $\bf C$ in Eqn.~\ref{eq:b} is simply an identity matrix of size $48\times48$ (${\bf C}_{48,48}$).  \newline
  \item {\bf 2D-3D fit with fine binning}: we fitted the 2D projected temperature profiles with the {\bf IAE} model convolved with a projection matrix $\bf C$. In this case, we wish to assess how well the {\bf IAE} model recovers the intrinsic 3D temperature profile when only 2D projected data are available. We used the same 2D radial logarithmic binning as that of the ${\bf T}_{\rm 3D}$ profiles, meaning that $\bf C$ has dimensions of $48\times48$ (${\bf C}_{48,48}$). For this testing phase, we assume standard emission measure weighting to calculate the elements of $\bf C$.  \newline

  \item {\bf 2D-3D fit with coarse binning}: the 2D projected temperature profiles having coarse logarithmic radial binning of twelve or six points up to R$_{500}$ were fitted to the {\bf IAE} model convolved with matrix $\bf C$. Here, the goal is to assess the ability of the IAE model to recover the intrinsic 3D temperature profile when only a coarse 2D projected profile, similar to that obtained from present-day observations, is available. In this case $\bf C$ has a dimensions of $12\times48$ (${\bf C}_{12,48}$) and $6\times48$ (${\bf C}_{6,48}$) for the 2D temperature profiles with twelve and six bins respectively. As above, we use standard emission measure weighting to calculate the elements of $\bf C$. In Sect.~\ref{5p4}, we will also consider the  \cite{Mazzotta2004} temperature-dependent spectroscopic-like weights.    
\end{enumerate}

For case 3, which seeks to mimic the typical characteristics of 2D temperature profiles measured with current X-ray satellites, we assume that the uncertainties increase linearly  with radius.  
Based on our previous experience with XMM-{\it Newton} and {\it Chandra} observations, we assume temperature profile uncertainties that increase from 5\% to 25\% in the [0.02-1]\,R$_{500}$ radial range for the 12-bin profiles, and from  10\% to 30\% for the 6-bin profiles. We built a diagonal error covariance matrix, i.e. ${\bf \Sigma}_{\bf o}$, using this approximation. This was then incorporated in the likelihood and acts as a weighting function, giving more weight to the inner regions in the fit. In general, regardless of whether the errors increase monotonically, the inclusion of errors in the likelihood leads to an overall improvement in the fit. For cases 1 and 2 (fine binning), we do not consider errors in the likelihood and as such ${\bf \Sigma}_{\bf o}$ is a unit matrix. Both model training and fitting a single profile with MCMC can be completed within a few minutes on a 16-core CPU.

In the objects where the temperature profiles in the first few inner bins were not reliable (i.e. having $< 100$ gas particles), these bins were not considered in the fitting.  However, no such constraint was applied during the training stage, as one expects the network to learn only the fundamental structure of the data rather than the noise.

\section{{Model evaluation}}
\label{main_sec_sim}

In this Section, we discuss the robustness of the non-parametric {\bf IAE} model reconstruction using different schemes. We check the performance of our model with respect to the radial binning, which is important since the number of radial bins corresponding to the observations is much lower compared to the resolution of the temperature profiles in the simulated sample. We also consider different weighting schemes in the fit.  The model is tested with the 115 temperature profiles in the validation sample. 

%

\begin{figure*}
\centering
	\includegraphics[width=0.49\textwidth]{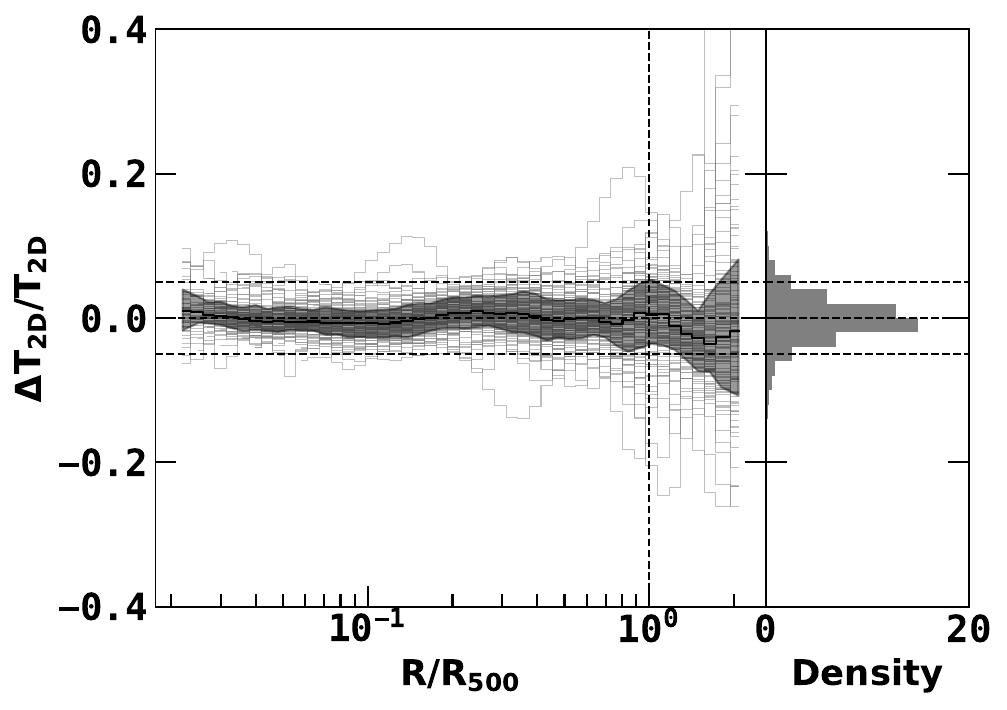}
  \hfill
  \includegraphics[width=0.49\textwidth]{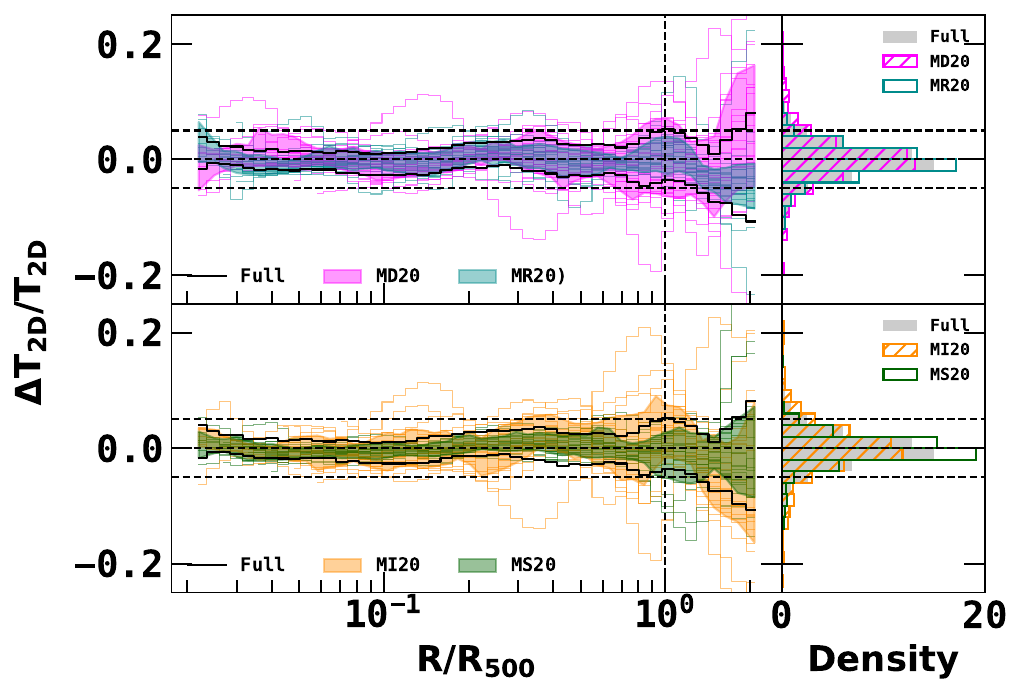} 
    \hfill
	\includegraphics[width=0.49\textwidth]{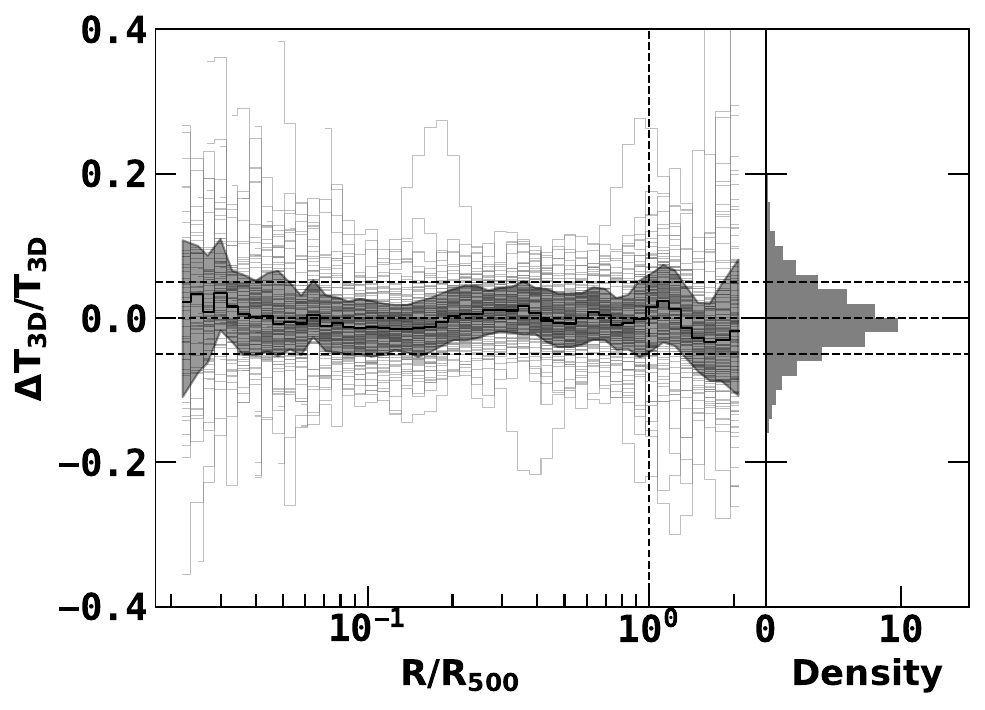}
    \hfill
\includegraphics[width=0.49\textwidth]{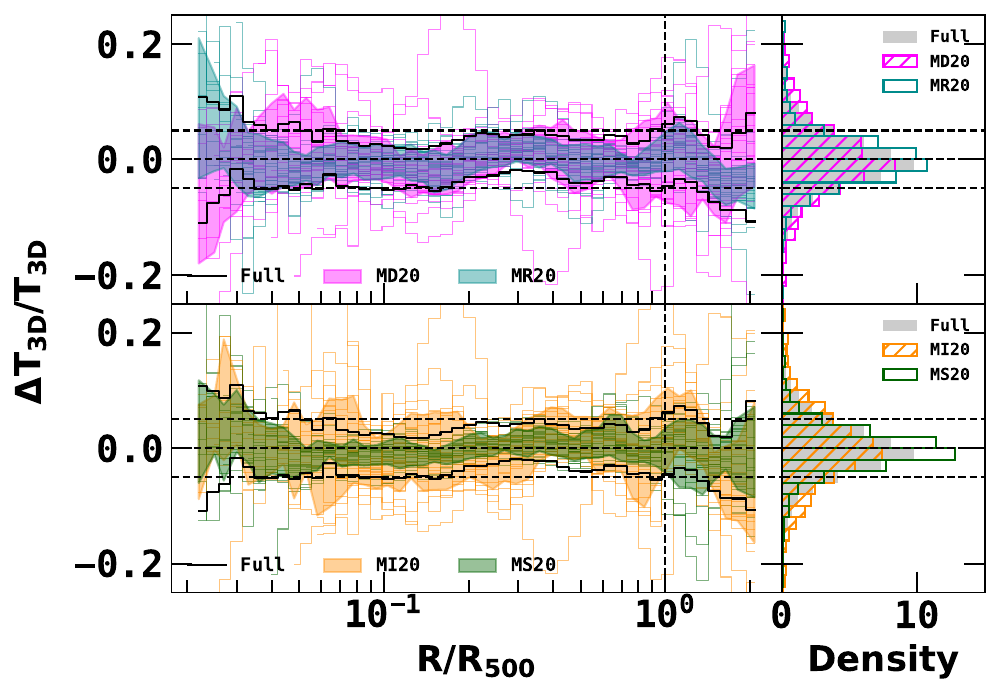}

	\caption{\footnotesize Fractional 2D and 3D residuals for 115 clusters in the validation sample with {\bf IAE} for the 2D-3D fit (fine binning). The three horizontal dashed black lines represent zero and $\pm$5\% fractional residuals; the vertical dashed black lines represent R$_{500}$. Left panel: Grey lines show the individual 2D (top panel) and 3D (bottom panel) residuals of all the clusters. The solid black line and shaded black region in the left panels show the median and 1-$\sigma$ dispersion of the 2D (top panel) and 3D (bottom panel) residual distribution, respectively. The histogram shows the distribution of residuals over all radii. Right panel: The cyan and magenta lines show the 2D (top panel) and 3D (bottom panel) residuals of the MR20 and MD20 sub-samples respectively. Green and orange lines show the 2D (top panel) and 3D (bottom panel) residuals of the MS20 and MI20 sub-samples respectively. Shaded regions show the corresponding 1-$\sigma$ dispersion of the residual distribution. Regions enclosed by the solid black lines show the 1-$\sigma$ dispersion of the median residual of the full validation sample. The histograms show the distribution of residuals over all radii. When given 2D profiles as input, the {\bf IAE} model can reconstruct 3D temperature profiles with a fractional difference of about 5\% across nearly the full radial range.
 }
	\label{fig9}
 \end{figure*}

  \begin{figure}
	\includegraphics[width=0.5\textwidth]{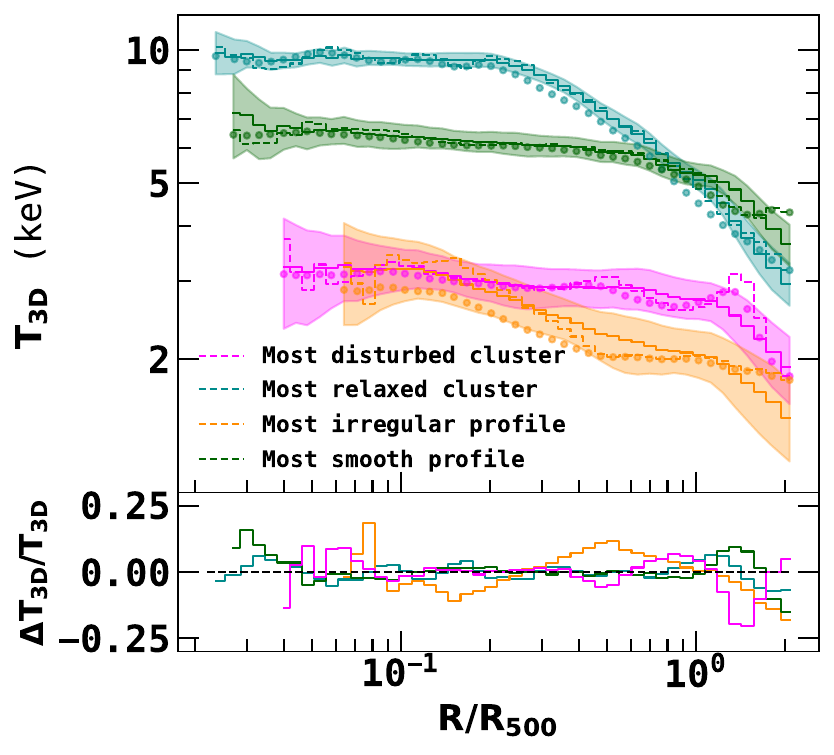}
	\caption{\footnotesize Results for the most relaxed / disturbed clusters and for the most smooth / irregular profiles with {\bf IAE} for the 2D-3D fit (fine binning). Left panel: Dashed cyan and magenta lines show the true 3D temperature profiles of the most relaxed and disturbed clusters in the validation sample, respectively. Dashed green and orange lines show the most smooth and irregular true 3D temperature profiles, respectively. The solid lines and the corresponding shaded regions show the median and 1-$\sigma$ dispersion of the reconstructed temperature profile obtained from the {\bf IAE} model using MCMC. The dotted lines show the 2D temperature profiles actually used in the fitting.}
	\label{fig8}
\end{figure}
The performance of the model was evaluated by comparing the original 3D and 2D temperature profiles with those recovered from the {\bf IAE} model. For each case, we calculated the median fractional residual and its associated  1-$\sigma$ dispersion (16th–84th percentile range) at three scaled radii (0.02\,R$_{500}$, R$_{500}$, and 2\,R$_{500}$), and over the full radial range. These results are presented in Table~\ref{results}, and each case is discussed in more detail below.

\subsection{3D-3D reconstruction of temperature profiles}
\label{secA}

\subsubsection{Overall performance}

We first consider the simplest case, corresponding to the 3D-3D fit with fine binning, where we directly fitted the {\bf IAE} model to the intrinsic 3D gas mass-weighted temperature profiles ($\bf T_{\rm 3D}$), ignoring projection effects. The left hand panel of Fig.~\ref{fig6} shows the fractional residuals 
 ($\Delta \bf{T_{3D}/T_{3D}}$) between the input (true) and recovered temperature profiles for all the individual clusters in the validation sample. The median fractional residual profile along with 1-$\sigma$ dispersion (16th–84th percentile range) are also plotted. 

The median fractional residual profile is found to be close to zero throughout the radial range: at radii,  0.02\,R$_{500}$, R$_{500}$, and 2\,R$_{500}$, the values are $-0.010\pm0.060$, $0.010\pm0.051$ and $-0.020\pm0.120$ respectively. Moreover, the median fractional residual over the full radial range is found to be $-0.001\pm0.042$. The 1-$\sigma$ dispersion in the fractional residuals is nearly constant at around $\pm$5\%, except beyond 1.5\,R$_{500}$.

Within the validation sample, the fractional residuals of the 20 most relaxed / disturbed clusters (MR20 / MD20) are displayed at the top in the right panel of Fig.~\ref{fig6}, while the 20 most smooth / irregular profiles (MS20 / MI20) are shown at the bottom. In all cases, the median fractional residuals are again consistent with zero. The 1-$\sigma$ dispersion in fractional residuals over all radii for the MD20 (MI20) sub-sample is $\pm$0.045 ($\pm$0.053), which is larger, as expected, compared to the dispersion of $\pm$0.032 ($\pm$0.029) found in the MR20 (MS20) sub-sample. This conclusion is supported by the fact that the histogram of the residuals of the MR20 (MS20) sub-sample is more peaked at zero, and hence is narrower compared to the MD20 (MI20) sub-sample. In general, we find that for disturbed clusters and for irregular profiles, the {\bf IAE} model smooths out the sharp small scale variations in the 3D temperature profiles.

\subsubsection{Anchor point weights, $\lambda_i$}\label{sec:apw}

We have shown above that the {\bf IAE} model is able to recover the average shape of the 3D profiles with high accuracy. In this context, it is interesting to consider how the anchor point weights, $\lambda$, change according to the characteristics of the profile under consideration.  Figure~\ref{fig4} shows the temperature profiles of the most relaxed / disturbed clusters in the validation sample, classified according to the $\chi_{\rm D}$ criterion discussed in Sect.~\ref{sec:dyn}, and of the most regular / irregular profiles in the validation sample, classified according to the $\chi_{\rm S}$ criterion introduced in Sect.~\ref{sec:str}. The reconstructed median temperature profile and fractional residuals obtained with the {\bf IAE} using MCMC are also shown. The {\bf IAE} model produces smoother profiles on small scales by ignoring the fluctuations on such scales. At large scales, the {\bf IAE} model is able to reproduce the underlying structure of the input temperature profiles. The bottom left hand panel shows the fractional residuals, which can be seen to be less than 5\% over most of the radial range. 
In Appendix~\ref{app:app1}, the top panel of the Fig.~\ref{L1} shows the corresponding posterior distribution of the parameters of the {\bf IAE} model obtained using MCMC. The parameters are seen to be well-constrained, and as anticipated the relaxed cluster profile (or the most regular profile) has tighter constraints compared to the most disturbed cluster (or the most irregular profile) which has relatively larger contour levels. Figure.~\ref{appx2} of Appendix~\ref{app:app1} shows the comparison of temperature profiles and the reconstructed temperature profiles of 20 example clusters in the validation sample.

We also tested the effect on the {\bf IAE} model of  increasing the number of anchor points. We found that the model fidelity can be improved by increasing the number of anchor points and that the choice of 20 anchor points reduces the residuals significantly. In Appendix~\ref{app:app1}, Fig.~\ref{appx2b} we show the recovered ensemble plot of fractional residuals using the {\bf IAE} model with 20 anchor points for the full validation sample, and for the different sub-samples. There is a significant improvement in the average fractional residual in all the cases. The median of the fractional residuals for the full sample over the entire radial range is found to be $0.002\pm0.030$, about 25\% smaller compared to the fiducial {\bf IAE} model obtained with five anchor points. However, the usefulness of this higher dimensional model is limited to simulations only. The temperature profiles that can be obtained from current X-ray satellites generally have temperature data at around 8-15 points for typical deep observations. Use of the {\bf IAE} model with 20 anchor points in cases such as this would  result in over-fitting and/or large variance.   

\begin{figure*}
	\includegraphics[width=0.49\textwidth]{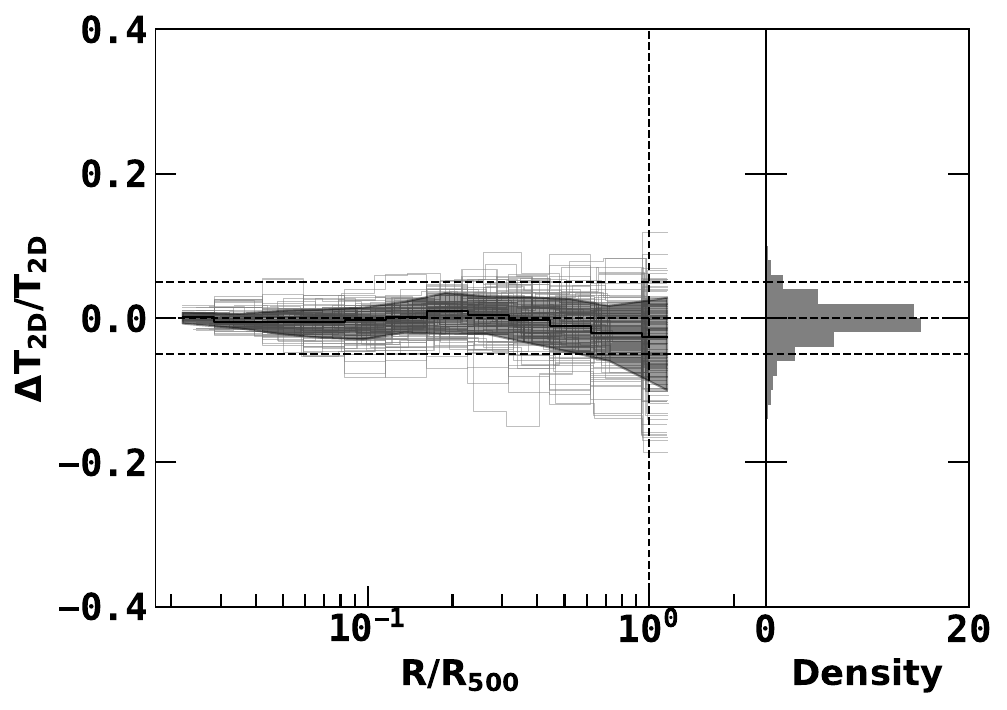}
		\includegraphics[width=0.49\textwidth]{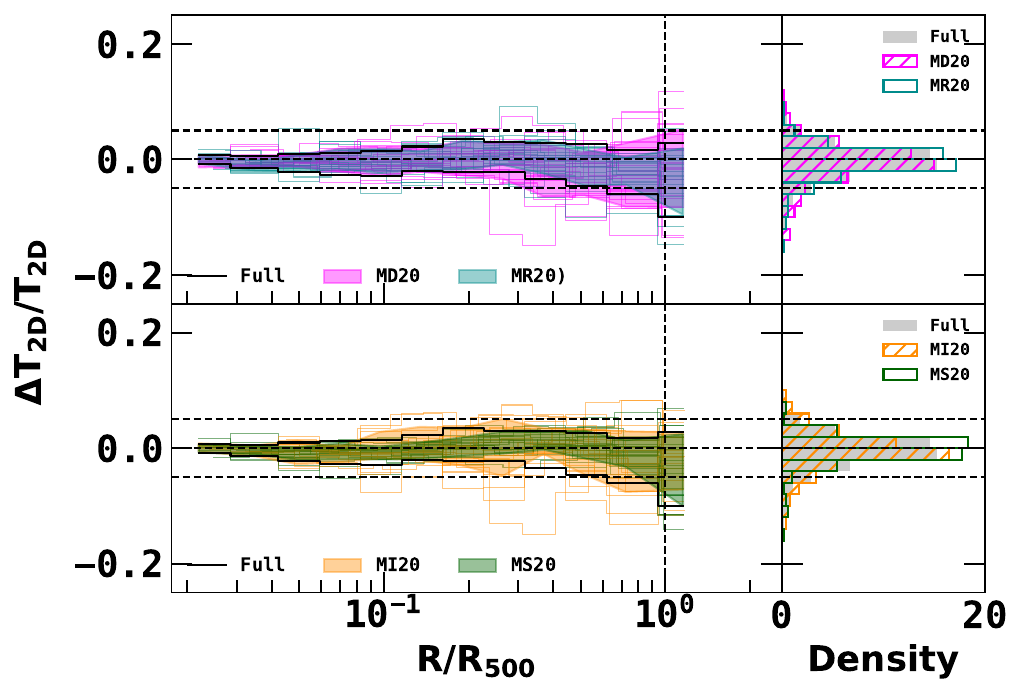}
	\includegraphics[width=0.49\textwidth]{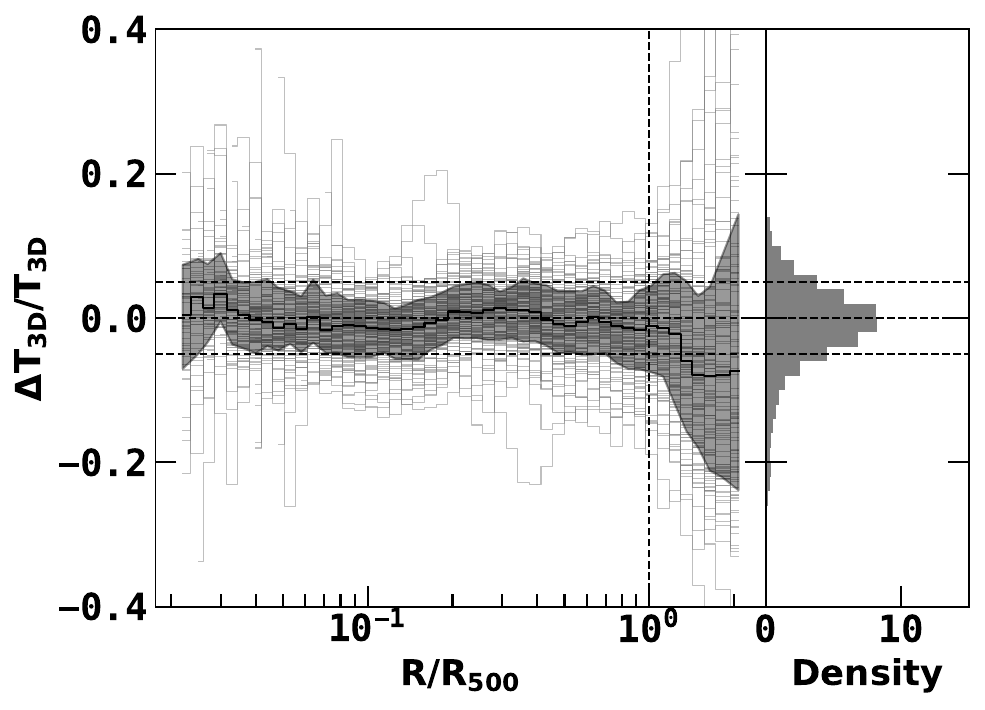}
		\includegraphics[width=0.49\textwidth]{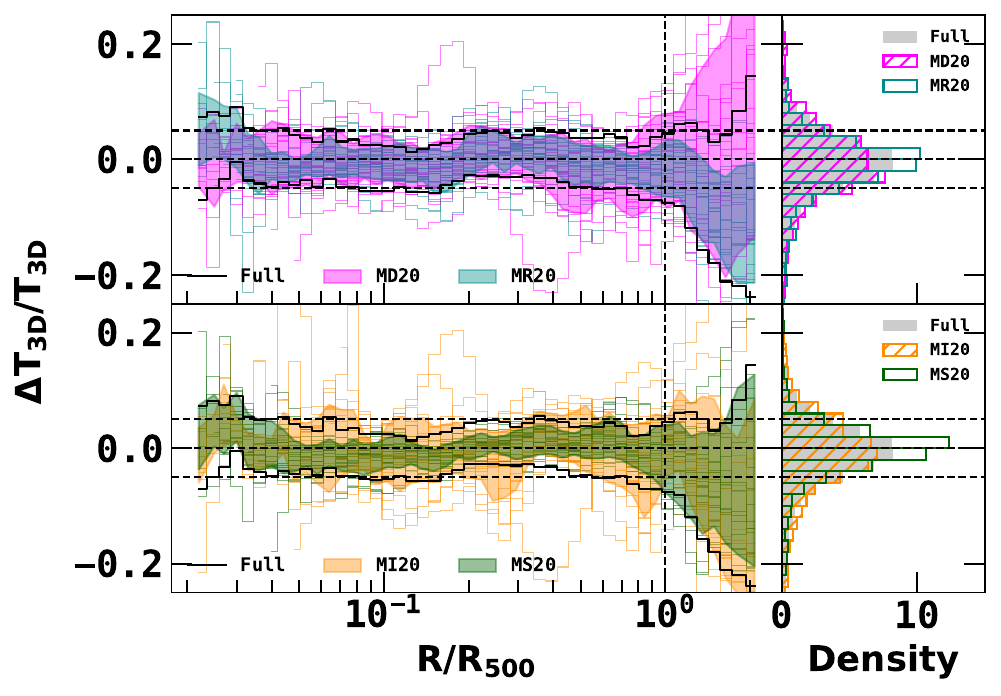}

	\caption{\footnotesize Fractional residuals for 115 clusters in the validation sample with {\bf IAE} for the 2D-3D fit (coarse binning) using 2D temperature profiles defined at twelve radial bins up to R$_{500}$. Colour coding is the same as in Fig.~\ref{fig9}. When given 2D temperature profiles with a binning scheme typical for moderately deep X-ray observations, the {\bf IAE} model can still reconstruct 3D temperature profiles with fractional differences of about 5\% throughout the 2D fitting range (i.e. [0.02-1] R$_{500}$.)}
	\label{fig12}
\end{figure*}

 \begin{figure*}
	\includegraphics[width=0.49\textwidth]{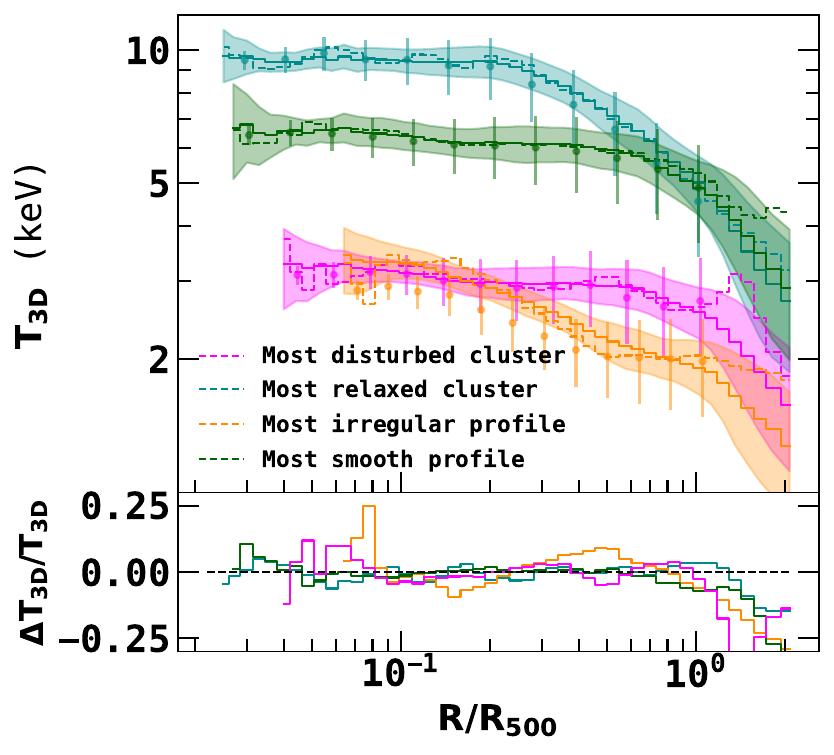}
		\includegraphics[width=0.49\textwidth]{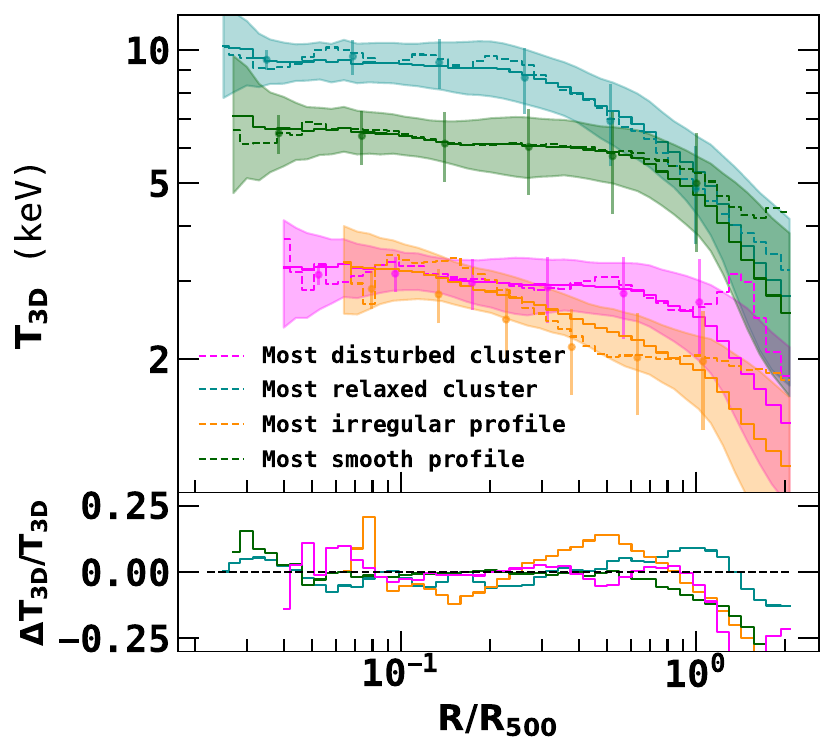}
	\caption{\footnotesize Results for the most relaxed and disturbed clusters and for the most smooth and irregular profile with the 2D-3D fit (coarse binning) using 2D temperature profiles defined at twelve (left panel) and six radial bins (right panel) up to R$_{500}$. Errors in the 2D temperature profiles are assumed to increase linearly with a radius from 5\% (10\%)in the innermost bin to 25\% (30\%) in the outermost bin for the twelve (six) bin case. Colour coding is the same as in Fig.~\ref{fig8}.  }
	\label{fig11}
\end{figure*}

\subsection{2D-3D reconstruction of temperature profiles with fine 2D binning}
\label{secB}

We now discuss the efficiency of the {\bf IAE} model when fitting the 2D (projected) temperature profiles, defined at the same radial grid as in the previous case and at which the {\bf IAE} model is defined (2D-3D fit with fine binning case). Here, the 3D {\bf IAE} model is convolved with the standard emission-measure weighting matrix. The resulting projected model is then fitted to the input 2D temperature profiles, in order to reconstruct the 3D temperature profiles. 

Since projection results in smoother 2D temperature profiles, washing out fluctuations at small scales, one expects the 3D reconstruction obtained from the 2D profile to be more regular compared to what was found in the previous section. It is also important to note that projection effects are dominant in the inner regions (especially in CC clusters), which can introduce degeneracy into the reconstructed 3D temperature profiles in the central region. 
However, both the 2D and 3D profiles of CC clusters will always display a central temperature dip. Thus one can expect a larger scatter in the 3D reconstructed temperature profiles in the central regions, as compared to the 3D-3D fitting case.

In Fig.~\ref{fig9}, we show the ensemble plot of fractional residuals of the 2D (top panel) and 3D (bottom panel) temperature profiles for the validation sample (left panel) and sub-samples (right panel).  The fractional residuals in 2D space (where the fitting is actually performed) are smaller compared to the 3D temperature residuals,  as expected. 

For the 2D fit, we find median fractional residuals at radii  0.02\,R$_{500}$, R$_{500}$, and 2\,R$_{500}$ to be $0.009\pm0.027$, $0.004\pm0.040$ and $-0.018\pm0.095$ respectively. 
The median of fractional residuals for the full sample and over the entire radial range is found to be $-0.002\pm0.027$.
 Unlike in the 3D-3D case, where the dispersion around the median was slightly larger in the outer regions only, here, it also increases towards the centre, as expected from the arguments given above. The dispersion is about $\pm$10\% at the first bin. 
 
 For the 3D reconstruction,  we find median fractional residuals at  0.02\,R$_{500}$, R$_{500}$, and 2\,R$_{500}$ of $0.021\pm0.110$, $0.014\pm0.052$ and $-0.018\pm0.095$, respectively. The median of fractional residuals for the full sample and over the entire radial range is found to be $-0.003\pm0.045$. Moreover, as in the 3D-3D case, here too, the histogram of the fractional residuals over all radii of MR20 and MS20 sub-samples are narrowly peaked compared to the MD20 and MI20 sub-samples, indicating again that the profiles of more relaxed clusters, or intrinsically smoother temperature 2D profiles, are reconstructed with higher fidelity in general. 

In the left panel of Fig.~\ref{fig8}, we show the recovered temperature profiles for the extreme cases of the most relaxed / disturbed cluster and the most smooth / irregular profiles in the validation sample. As in the 3D-3D case, the difference between the input and recovered temperature profiles is less than 5\% over most of the radial range. In Appendix~\ref{app:app1}, the bottom panel of the Fig.~\ref{L1} shows the corresponding posterior distribution of the {\bf IAE} model parameters. Here also, all the parameters are well constrained. The comparison to the equivalent parameters contours for the 3D-3D case, also shown on the plot, show that, understandably, the 2D-3D reconstruction has slightly larger contour intervals compared to the 3D-3D. 

\subsection{2D-3D reconstruction of temperature profiles with an observation-like binning}
\label{secC}

So far we have tested the {\bf IAE} model only with high resolution simulated temperature profiles. However, real observed 2D temperature profiles are of much lower spatial resolution, have fewer data points, and are generally detected   up to R$_{500}$ only. In this Section, we test the accuracy of the {\bf IAE} model to recover simulated temperature profiles with resolutions similar to those found with the current X-ray observations (2D-3D fit with coarse binning case). 

First, we consider a case where we fitted 2D temperature profiles having resolutions similar to that expected with moderately deep X-ray observations. In such  observations, we normally expect around twelve annular data points limited up to R$_{500}$. We also impose more  realistic errors in the 2D temperature profiles: They are assumed to increase linearly with a radius from 5\% in the innermost bin to 25\% in the outermost bin. Later in this Section, we will also consider a fitting case with 2D temperature profiles defined at only six radial points within R$_{500}$, with errors ranging from 10\% to 30\% from the innermost to the outermost radial bin.

\begin{figure*}
	\includegraphics[width=0.49\textwidth]{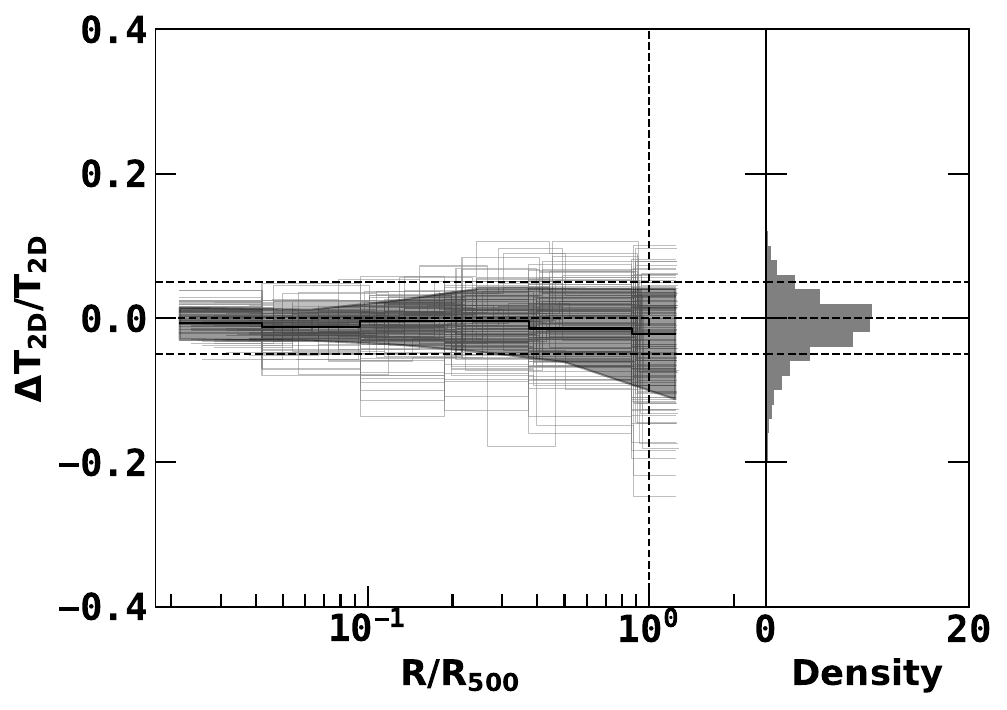}
	\includegraphics[width=0.49\textwidth]{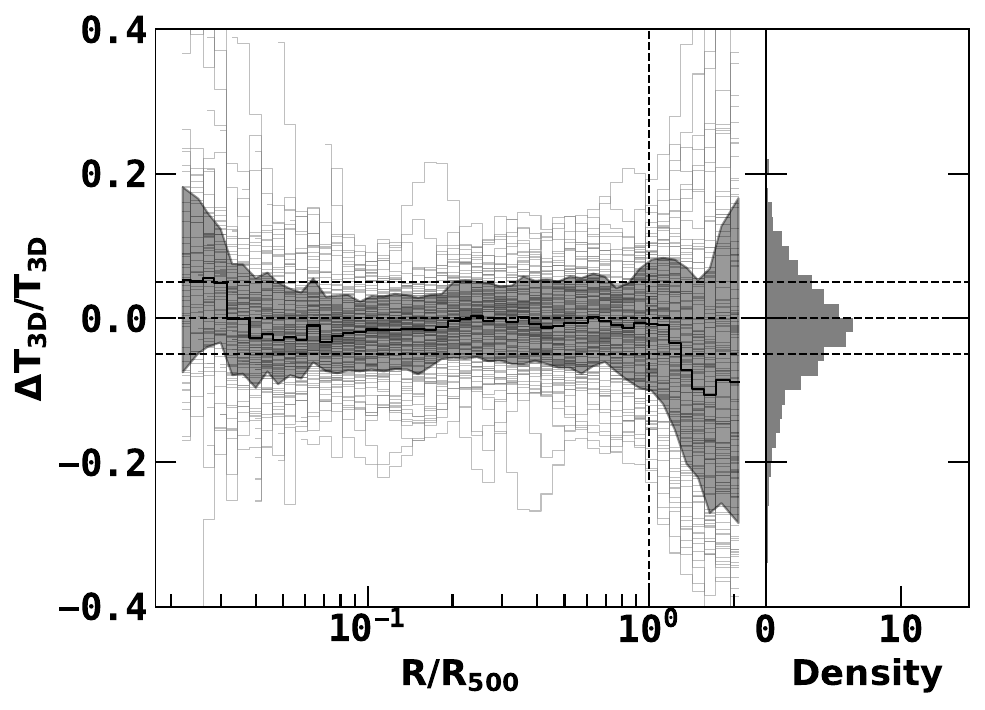}

	\caption{Fractional residuals for 115 clusters in the validation sample with {\bf IAE} for the 2D-3D fit (coarse binning) using 2D temperature profiles defined at six radial bins up to R$_{500}$. Colour coding is the same as in Fig.~(\ref{fig9}). For simplicity, we have not shown the sub-sample cases. Even when input 2D temperature profiles with a binning scheme typical for shallow X-ray observations, the {\bf IAE} model can still reconstruct 3D temperature profiles with fractional differences of about 5\% throughout the 2D fitting range (i.e. [0.02-1] R$_{500}$).}
	\label{fig13}
\end{figure*}

\subsubsection{Twelve bin case}

In Fig.~\ref{fig12}, we show the ensemble plot of the 2D and 3D fractional residuals for the 2D-3D fit with the coarse binning case, by considering twelve 2D temperature data points within R$_{500}$. Even with the lower resolution, we find that within the 2D fitting range (i.e. up to R$_{500}$), the 3D fractional residuals are still close to zero, with a 1-$\sigma$ dispersion of about $\pm$5\%, as in the previous cases. The median 3D fractional residuals at radii  0.02\,R$_{500}$, R$_{500}$, and 2\,R$_{500}$ is found to be $0.003\pm0.071$, $-0.010\pm0.064$ and $-0.070\pm0.185$ respectively. The median of fractional residuals for the full sample and over the entire radial range is found to be $-0.006\pm0.051$. Beyond R$_{500}$, where no 2D temperature data were available to fit, and thus where the constraints on the 3D reconstruction are only due only to projection effects, the scatter increases with radius, reaching a  1-$\sigma$ dispersion of $\pm$20\% at the last bin (2\,R$_{500}$). Moreover, beyond  1.5\,R$_{500}$, 3D temperature profiles are underestimated by about 7\%. However, it is important to mention that the true 3D temperature profiles mainly lie within the 1-$\sigma$ dispersion of reconstructed temperature profiles. As before in the fine binning case, the dispersion in the 2D fractional residual is  much smaller compared to the 3D reconstruction. 

For the 2D fit,  we find median fractional residuals at radii 0.02\,R$_{500}$ and R$_{500}$ to be $ 0.001\pm 0.008$, $-0.026\pm0.073$ respectively. The median of fractional residuals for the full sample and over the entire radial range is found to be $-0.002\pm0.026$ for the 2D profiles, similar to that found in the 2D-3D fit with the fine binning case. Since we assumed that the errors increase radially outwards such as in real observations, putting more weight on the inner regions in the fit, the constraints in the inner region are better compared to the 2D-3D fit with the fine binning case. For comparison, Fig.~\ref{fig:appx4}, in Appendix.~\ref{appx:appx4} shows the 3D fractional residuals for the case where we do not consider error bars in the fit. Here,  we find that the scatter is increased in the inner regions as compared to both 2D-3D fit with fine binning case (previous case) and coarse binning case (present case).  

As in the previous cases, the histogram of the residuals of the MR20 (MS20) sub-sample has a stronger peak around zero and reduced wings compared to the MD20 (MI20) sub-sample.  For example, the 1-$\sigma$ dispersion in 3D fractional residuals over all radii for MD20 (MI20) sub-sample is found to be $\pm$0.055 ($\pm$0.065) and for  MR20 (MS20) sub-sample it is $\pm$0.041 ($\pm$0.036). 

\begin{figure*}
	\includegraphics[width=0.49\textwidth]{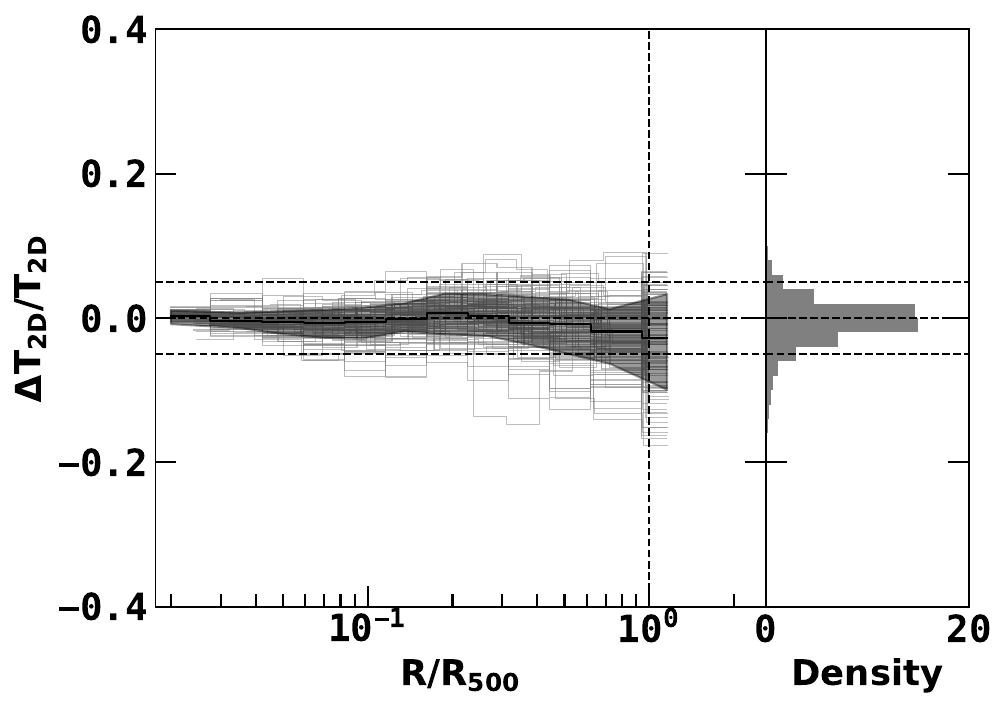}
	\includegraphics[width=0.49\textwidth]{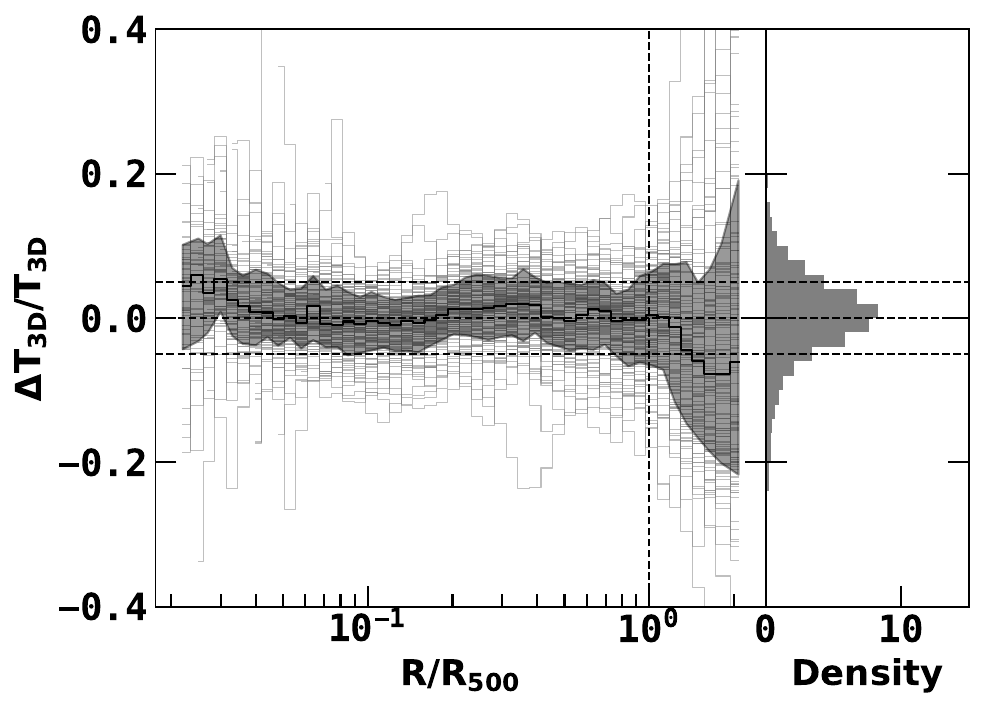}

	\caption{Fractional residuals for 115 clusters in the validation sample with {\bf IAE} for the 2D-3D fit (coarse binning) using spectroscopic-like 2D temperature profiles defined at twelve radial bins up to R$_{500}$. For simplicity, we have not shown the sub-sample cases. }
	\label{fig15}
\end{figure*}

In the left hand panel of Fig.~\ref{fig11}, we show the {\bf IAE} recovered temperature profiles of the most relaxed and disturbed cluster and of the most regular and irregular profile in the validation sample. As in previous cases, here also the difference between the input and recovered temperature profiles is less than 5\% in the 2D fitting range of [0.02-1]\,R$_{500}$. Beyond R$_{500}$, as expected, the residuals can be high. In Appendix~\ref{appx:appx4}, the top panel of Fig.~\ref{L2},  shows the corresponding posterior distribution of the parameter. One finds that the confidence intervals for the {\bf IAE}  model parameters are larger compared to fine binning cases (i.e. cases 1 and 2). However, we were still able to put relatively good bounds on the parameters, which are represented by nearly Gaussian posterior distributions. Figure~\ref{appx5} in Appendix~\ref{appx:appx4} shows the comparison of true 2D and 3D temperature profiles and the reconstructed temperature profiles of 20 example clusters in the validation sample for the twelve bin case.

\subsubsection{Six bin case}

The 2D and 3D fractional residuals for a fit considering only six data points with errors linearly increasing from 10\% in the innermost bin to 30\% in the outermost bin in the range [0.2-1]\,R$_{500}$ are shown in Fig.~\ref{fig13}. We find that the median 2D and 3D fractional residuals are still consistent with zero in the 2D fitting range. However, as expected, the 1-$\sigma$ dispersion is larger compared to the previous cases and temperature profiles are underestimated by about 8\% beyond 1.5\,R$_{500}$ (where there are no 2D data).  We find median 2D fractional residuals at radii 0.02\,R$_{500}$ and R$_{500}$ to be $-0.006\pm0.022$, $-0.022\pm0.070$ respectively. For the 3D reconstruction,  we find median fractional residuals at  0.02\,R$_{500}$, R$_{500}$, and 2\,R$_{500}$ to be  $0.05\pm0.128$, $-0.004\pm0.090$ and $-0.080\pm0.235$ respectively. The median of fractional residuals for the full sample and over the entire radial range is found to be $-0.008\pm0.038$ and $-0.014\pm0.075$ for the 2D and 3D profiles respectively.  In the right panel of Fig.~\ref{fig11}, we show the temperature profiles of the most relaxed and disturbed cluster and of the most regular and irregular profile in the validation sample. We find that even with only six data points in the fit, the {\bf IAE} is still able to recover the 3D temperature profiles with residuals less than 10\% over most of the cluster region. However, the confidence intervals of the reconstructed profiles and {\bf IAE} parameters, shown in the bottom panel of Fig.~\ref{L2} in Appendix~\ref{appx:appx4}, are larger compared  to previous cases. Finally Fig.~\ref{appx5a} in Appendix~\ref{appx:appx4} shows the comparison of true 2D and 3D temperature profiles and the reconstructed temperature profiles of 20 clusters in the validation sample for the six bin case.

For comparison,  Table~\ref{results} provides the median fractional residuals obtained for the different cases of fitting schemes discussed in this Section. Similarly, Table~\ref{results2} shows the best-fitting parameters of {\bf IAE} model for different cases obtained with MCMC. One can see that as we go from the high resolution simulated profiles to lower resolution observational-like profiles, the dispersion in fractional residuals and parameter estimates increases.

We also checked the performance of the model with other binning  schemes and found the performance of the {\bf IAE} model to be robust. In particular, we checked the performance by considering five 2D data points up to 0.5\,R$_{500}$ in the fit. We find that the {\bf IAE} model is able to reproduce the results with an average fractional difference of  about 5\% up to 0.5\,R$_{500}$ which then increases with radius and becomes about 10\% at R$_{500}$ and 25\% at 2\,R$_{500}$. We also considered an IAE model with 20 anchor points, applied to the two observation-like cases, and found that its performance is very similar to that of our fiducial five-parameter IAE model, unlike in the 3D-3D case where it is found to have better performance. This implies that increasing the number of anchor points does not necessarily increase the model fidelity for these cases, as one must also have higher resolution input 2D temperature profiles for the model to be fitted against.

   \begin{table}
   	 \caption{Flat priors used for the  \cite{Vikhlinin2006}  model parameters.}
     \begin{center}
        \begin{tabular}{c c } 
            \toprule
            \toprule

            Parameter     &  Range \\
                \hline
            $T_0$   & $0 ; +4$    \\
            $\tau$    & $+0.2 ; +1$ \\
            $\log(R_{cool}/\rm R_{500})$   & $-2.5 ; 0 $      \\
           $a_{cool}$  & $0 ;+4 $          \\
           $\log( R_{t}/\rm R_{500})$   & $-1 ; +1  $       \\
            $a$   & $0; +0.6 \,(0; +0.1) $           \\
            $b$   & $+1 ; +4 $           \\
            $c$   & $0 ; +4\,(+1 ; +4)$           \\

            \bottomrule
       \end{tabular}
       \end{center}
      \footnotesize{\textbf{Notes}: Numbers in the brackets represent the optimal priors found in this work. }
      \label{tab_vik_priors}
   \end{table}

\begin{figure}
    \includegraphics[width=0.49\textwidth]{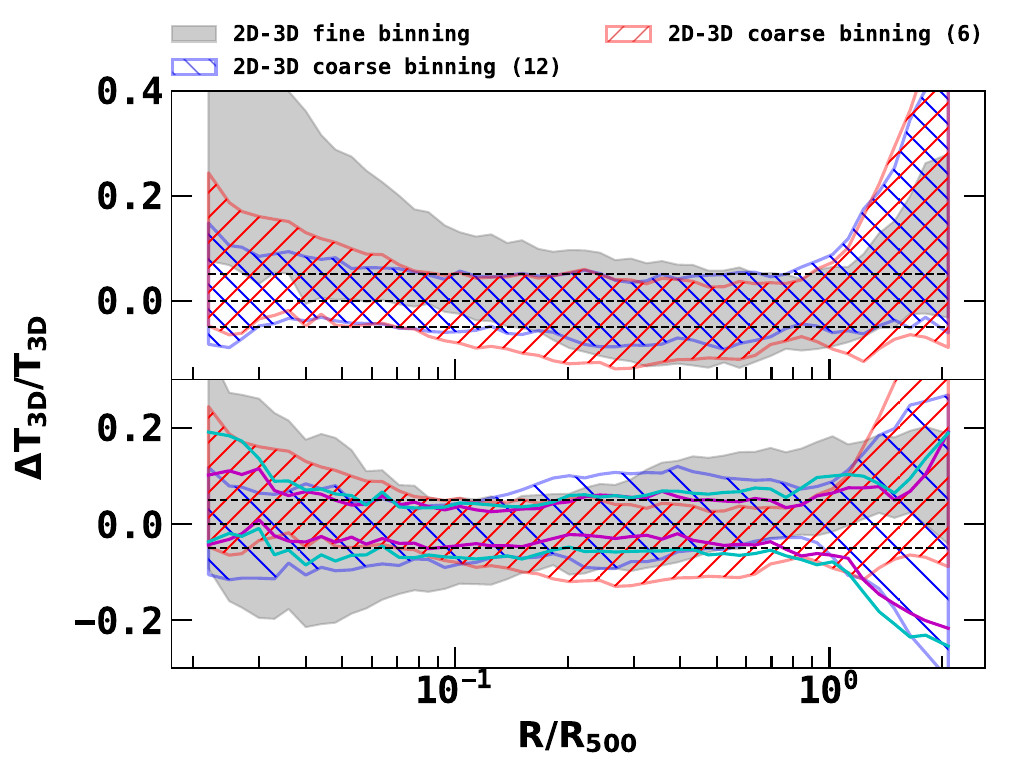}
    \caption{\footnotesize  The 1-$\sigma$ dispersion in the 3D fractional differences obtained with MCMC for priors provided in Table~\ref{tab_vik_priors} for the \cite{Vikhlinin2006} parametric model (Eqn.~\ref{eq:tprof}).  In the  figure, we  consider the 2D-3D fine binning case and 2D-3D observational-like coarse binning cases with twelve and six bins. The top panel shows the results with prior ranges for $a=0-0.6$ and $c=0-4$, while the bottom panel presents the results with priors ranges for $a=0-0.1$ and $c=1-4$. The regions enclosed by cyan and magenta lines in the bottom panel show the corresponding dispersion recovered with the {\bf IAE}  model for the observational-like cases with twelve and  six bins respectively. }
    \label{fig_VIk_profiles}
\end{figure}
   
\begin{figure*}
	\includegraphics[width=0.49\textwidth]{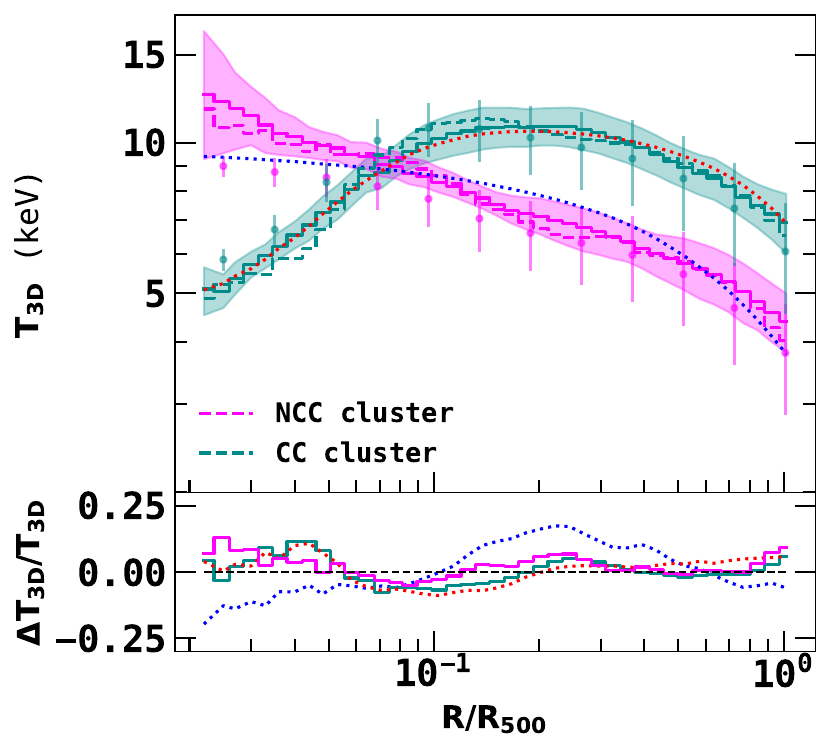}
		\includegraphics[width=0.49\textwidth]{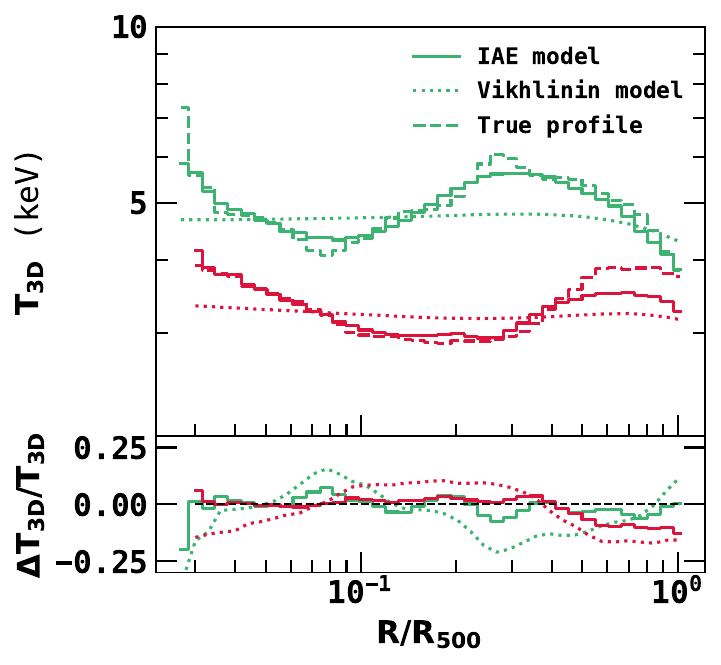}
	\caption{\footnotesize CC and NCC model recover comparison. Left panel: Comparison of the 3D temperature profiles of typical CC and NCC clusters in the \thethree sample recovered with the {\bf IAE} and parametric models using twelve 2D annuli within R$_{500}$ (points with error bars). The dashed line shows the true 3D temperature profiles. The solid lines and shaded regions show the reconstructed 3D temperature profiles with 1-$\sigma$ dispersion obtained with the {\bf IAE}  model. The dotted lines are the 3D temperature profiles recovered with the \cite{Vikhlinin2006} parametric model. For better visibility, the 1-$\sigma$ dispersion for the parametric model is not shown. Right panel: 3D temperature profile reconstruction with the {\bf IAE} and parametric models for two complex cases in the \thethree. For better visibility, 2D profiles and the 1-$\sigma$ dispersion are not shown.  For both figures, the bottom panel shows a fractional difference between the  true and recovered 3D profiles. For NCC and CC clusters, both the {\bf IAE} model and parametric model reconstruction  with optimal priors are comparable, but the former exhibits slightly better performance. For the complex cases, the {\bf IAE} model is more accurate in uncovering the profile shapes.}
	\label{fig_IAE_PAR_sim}
\end{figure*}
    \begin{table*}[!h]
\scriptsize     
\caption{\footnotesize Median fractional 3D and  2D residuals obtained at 0.02\,R$_{500}$ (third column),  R$_{500}$ (fourth column), 2\,R$_{500}$ (fifth column), and over the full radial range (sixth column) for the fitting schemes and samples in Sects.~\ref{secA}, \ref{secB}, and \ref{secC}. } 
\begin{center}
        \begin{tabular}{c c c  c c c } 
            \toprule
            \toprule
             Sample  & Case   &0.02\,R$_{500}$  &   R$_{500}$  &  2\,R$_{500}$  &  Full range\\
                \midrule
               \multicolumn{6}{c}{3D-3D fit with fine binning (48 data points)}\\
              \midrule
            Full sample &3D  & $-0.010\pm0.060$& $0.010\pm0.051$ & $-0.020\pm0.120$ & $-0.001\pm0.042$ \\
            MD20  &3D    & $-0.018\pm0.017$& $0.014\pm0.085$ & $0.093\pm0.128$ & $0.000\pm0.048$ \\
            MR20 &3D    &$0.035\pm0.065$& $0.014\pm0.035$ & $-0.060\pm0.045$ & $-0.002\pm0.032$ \\
            MI20 & 3D   &  $0.002\pm0.060$& $0.027\pm0.061$ & $-0.125\pm0.132$ & $0.001\pm0.052$ \\
            MS20 & 3D   &  $-0.033\pm0.048$& $0.002\pm0.045$ & $-0.037\pm0.110$ & $-0.001\pm0.029$ \\            
               \midrule
            \multicolumn{6}{c}{2D-3D fit with fine binning (48 data points)}\\
                          \midrule  
           Full sample & 2D   & $0.009\pm0.027$ & $0.004\pm0.040$ & $-0.018\pm0.095$ & $-0.002\pm0.027$  \\
            Full sample &3D    &$0.021\pm0.110$ &$0.014\pm0.052$ & $-0.018\pm0.095$ &$-0.003\pm0.045$  \\
            MD20 & 2D    &$-0.001\pm0.037$ &$-0.009\pm0.065$ & $0.067\pm0.105$ &$0.000\pm0.033$  \\
            MD20  &3D   &$-0.021\pm0.120$ &$0.021\pm0.088$ & $0.067\pm0.105$ &$-0.002\pm0.058$  \\
            MR20 & 2D   &  $0.020\pm0.035$ &$0.016\pm0.031$ & $-0.050\pm0.037$ &$-0.003\pm0.023$  \\
            MR20 &3D    & $0.089\pm0.120$ &$0.015\pm0.035$ & $-0.050\pm0.037$ &$-0.003\pm0.035$  \\
            MI20  &2D    & $0.019\pm0.015$ &$0.014\pm0.035$ & $-0.106\pm0.120$ &$0.000\pm0.038$  \\
            MI20 & 3D   &   $0.049\pm0.135$ &$0.025\pm0.070$ & $-0.106\pm0.120$ &$-0.001\pm0.064$  \\
            MS20 &2D    &  $0.010\pm0.013$ &$0.006\pm0.040$ & $-0.036\pm0.065$ &$-0.002\pm0.018$  \\
            MS20 & 3D   &  $0.015\pm0.080$ &$0.012\pm0.043$ & $-0.036\pm0.065$ &$-0.002\pm0.031$  \\

            \midrule 
            \multicolumn{6}{c}{2D-3D fit with coarse binning (12 data points)}\\
            \midrule
           Full sample & 2D   &  $ 0.001\pm 0.008$ & $-0.026\pm0.073$& - &$-0.002\pm0.026$\\
            Full sample &3D    &  $0.003\pm0.071$& $-0.010\pm0.064$ & $-0.07\pm0.185$ & $-0.006\pm0.051$\\
            MD20 & 2D   & $-0.003\pm 0.010$ & $-0.009\pm0.065$& - &$-0.004\pm0.028$ \\
            MD20  &3D    &  $0.006\pm0.045$& $-0.010\pm0.80$ & $0.113\pm0.243$ & $-0.006\pm0.058$\\
            MR20 & 2D   &   $-0.001\pm 0.007$ & $-0.024\pm0.056$& - &$-0.003\pm0.021$ \\
            MR20 &3D    & $0.069\pm0.063$& $0.006\pm0.50$ & $-0.138\pm0.101$ & $-0.006\pm0.042$\\
            MI20  &2D    &  $-0.002\pm 0.007$ & $-0.014\pm0.051$& - &$-0.003\pm0.030$ \\
            MI20 & 3D   & $-0.014\pm0.047$& $0.012\pm0.50$ & $-0.174\pm0.175$ & $-0.007\pm0.065$\\
            MS20 &2D    & $-0.000\pm 0.006$ & $-0.025\pm0.060$& - &$0.000\pm0.018$ \\
            MS20 & 3D   & $0.034\pm0.055$& $0.005\pm0.60$ & $-0.08\pm0.165$ & $0.000\pm0.036$\\
           \midrule  
            \multicolumn{6}{c}{2D-3D fit with coarse binning  (6 data points)}\\
            \midrule
           Full sample & 2D   & $-0.006\pm0.022$& $-0.022\pm0.070$&-&  $-0.008\pm0.038$ \\
            Full sample &3D    &  $0.050\pm0.128$& $-0.004\pm0.090$ & $-0.080\pm0.235$&$-0.014\pm0.075$ \\
            MD20 & 2D   & $-0.008\pm0.027$& $-0.016\pm0.070$&-&  $-0.009\pm0.040$ \\
            MD20  &3D    &$0.005\pm0.060$& $0.002\pm0.115$ & $0.101\pm0.255$&$-0.016\pm0.082$ \\
            MR20 & 2D   & $-0.016\pm0.015$& $-0.018\pm0.062$&-&  $-0.009\pm0.032$ \\
            MR20 &3D    &$0.073\pm0.130$& $0.009\pm0.075$ & $-0.131\pm0.140$&$-0.016\pm0.060$ \\
            MI20  &2D    & $-0.004\pm0.0175$& $-0.040\pm0.055$&-&  $-0.009\pm0.040$ \\
            MI20 & 3D    &$0.024\pm0.035$& $0.000\pm0.070$ & $-0.194\pm0.225$&$-0.017\pm0.090$ \\     
            MS20 &2D    & $-0.008\pm0.0150$& $-0.008\pm0.065$&-&  $-0.003\pm0.025$ \\
            MS20 & 3D   &$0.102\pm0.110$& $0.011\pm0.95$ & $-0.072\pm0.200$&$-0.006\pm0.050$ \\
            \bottomrule
       \end{tabular}
   \end{center}
\footnotesize{\textbf{Notes}: The errors are given at 1-$\sigma$ level. The reconstructed 2D and 3D temperature profiles at the last point (i.e 2\,R$_{500}$) are by construction identical for the 2D-3D fit with fine binning case. The fine binning cases represent high resolution simulated temperature profiles, and the coarse binning cases represent lower resolution observational-like temperature profiles. The constraints on the residuals become weaker with decreasing resolution. Even in the coarse binning cases, the residuals remain consistently below 5\% within the fitting range (i.e. [0.02-1]\,R$_{500}$). In all the cases, MS20 and MR20 sub-samples provide the tightest constraints across the full radial range. }
      \label{results}
   \end{table*}

      \begin{table*}[!h]
\scriptsize 
\caption{\footnotesize Best fit results for the {\bf IAE} parameters derived with the MCMC for the fitting schemes and samples considered in Secs.~\ref{secA}, \ref{secB} and  \ref{secC}. }    
\begin{center}
        \begin{tabular}{c c c c  c c c } 
            \toprule
            \toprule
            Case &  $\lambda_{1}$  &  $\lambda_{2}$  & $\lambda_{3}$  &  $\lambda_{4}$ &$\lambda_{5}$  & $\alpha$\\
                \midrule
               \multicolumn{7}{c}{3D-3D fit with fine binning (48 data points)}\\
              \hline  
            Most disturbed cluster    &$0.0\pm 1.1$ &$0.03\pm 0.77$ & $-0.2\pm 1.4$ &$0.34^{+0.28}_{-0.32}$& $0.91\pm 0.69 $& $380.9\pm 7.0$ \\
            Most relaxed cluster & $3.0\pm 2.2$&  $2.2\pm 1.5 $ & $-2.7\pm 2.9$ & $0.54^{+0.63}_{-0.78}$ &  $-2.0\pm 1.5$& $143.6\pm 9.4$ \\
            Most irregular profile  &$1.8\pm 1.0$  & $1.43\pm 0.71$ &$-1.3\pm 1.4$  & $0.14\pm 0.33$&$-1.08\pm 0.66$&$284.5\pm 7.1$\\            
            Most smooth profile   & $3.2^{+2.6}_{-2.4}$   &  $2.6\pm 1.8$&$-4.0\pm 3.5$& $0.81\pm 0.79$ & $-1.5\pm 1.7$& $129^{+11}_{-12}$ \\
            
               \midrule
            \multicolumn{7}{c}{2D-3D fit with fine binning (48 data points)}\\
                \midrule  
            Most disturbed cluster& $0.2\pm 1.2$  & $0.16\pm 0.82$  & $-0.5\pm 1.6$ & $0.40^{+0.35}_{-0.49}$  & $0.78\pm 0.78$& $381.7\pm 8.0$\\              
            Most relaxed cluster & $2.8^{+2.5}_{-2.2}$ & $2.1^{+1.6}_{-1.4}$  & $-2.4\pm 3.2$ & $0.40^{+0.66}_{-0.81}$ & $-1.8\pm 1.5$ & $144^{+12}_{-15}$\\
            Most irregular profile & $2.4^{+1.3}_{-1.5}$ & $1.78\pm 0.93$ & $-2.0\pm 2.0$ & $0.44^{+0.46}_{-0.66}$ &$-1.58^{+1.1}_{-0.79}$  & $289\pm 11 $\\
            Most smooth profile &$3.0^{+2.8}_{-2.5}$& $2.5^{+2.0}_{-1.7}$& $-3.6^{+3.5}_{-3.9}$& $0.69^{+0.76}_{-0.89}$& $-1.6\pm 1.7$ & $130^{+12}_{-21}$ \\
              
            \midrule  
            \multicolumn{7}{c}{2D-3D fit with coarse binning (12 data points)}\\
            \midrule
            Most disturbed cluster   &$-0.6^{+3.0}_{-2.6}$& $-0.4^{+2.2}_{-1.9}$ &$0.5\pm 3.6$  & $0.20\pm 0.77$  & $1.3^{+1.7}_{-2.0}$  & $378\pm 16$  \\
            Most relaxed cluster    & $0.99^{+3.6}_{-2.9}$& $0.8^{+2.5}_{-2.0}$ & $-0.3^{+3.8}_{-4.5}$ &  $0.15^{+0.81}_{-0.92}$ & $-0.6^{+2.0}_{-2.3}$& $141.2^{+9.8}_{-11}$\\
            Most irregular profile    &$0.5^{+3.2}_{-2.8}$ & $0.4^{+2.3}_{-2.0}$ &$0.3^{+3.7}_{-4.1}$& $0.01\pm 0.87$ & $-0.1\pm 2.1$& $282\pm 13$\\
            Most smooth profile    & $1.9^{+3.4}_{-2.8}$  &  $1.9^{+2.4}_{-1.8}$& $-2.6^{+3.7}_{-4.3}$ & $0.56^{+0.75}_{-0.91}$ & $-0.8\pm 2.0$ & $131^{+11}_{-15}$\\  
               \midrule
            \multicolumn{7}{c}{2D-3D fit with coarse binning  (6 data points)}\\
            \midrule
            Most disturbed cluster &  $-0.6^{+2.9}_{-2.5}$  & $-0.5^{+2.1}_{-1.7}$ & $0.5^{+3.2}_{-3.7}$  & $0.31^{+0.74}_{-0.84}$ & $1.3^{+1.7}_{-2.0}$ & $380\pm 16$ \\
            Most relaxed cluster & $0.6^{+3.9}_{-3.5}$  & $0.5^{+2.7}_{-2.4}$ & $0.0\pm 4.7$  & $0.08\pm 0.96$  & $-0.3^{+2.2}_{-2.5}$ & $141^{+15}_{-22}$\\
            Most irregular profile  &  $-0.3\pm 3.7$ & $-0.1\pm 2.7$ & $1.0\pm 4.7$& $0.0\pm 1.0$ & $0.4\pm 2.5$& $281\pm 24$\\
            Most smooth profile  &$0.8^{+3.8}_{-3.4}$ &$0.97^{+2.8}_{-2.4}$  & $-1.0^{+4.4}_{-4.9}$ & $0.34\pm 0.96$& $-0.1\pm 2.4$ & $129^{+12}_{-20}$ \\              
               \bottomrule
       \end{tabular}
       \end{center}
   \footnotesize{\textbf{Notes}: The errors are given at 1-$\sigma$ level. As in Table.~\ref{results}, the constraints decrease in strength as we go from the fine binning cases to observation-like coarse binning cases. }
      \label{results2}
   \end{table*}
\subsection{2D-3D reconstruction of temperature profiles with spectroscopic-like weighting}
\label{5p4}

In the previous Sections, we have only focused on 3D temperature reconstruction from the {\bf IAE} model using 2D temperature profiles derived using standard emission-measure weights \citep{2001ApJ...546..100M}. In this Section, we consider more complex spectroscopic-like weighting \citep{Mazzotta2004}, which has a stronger dependence on the 3D temperature profiles. This makes deconvolution a more complicated problem and, therefore, it is important to check the accuracy of the {\bf IAE} model in this case. 

In Fig.~\ref{fig15}, we show the fractional residual for 2D and 3D temperature profiles between the input and {\bf IAE} recovered temperature profiles in 2D-3D fit with twelve data points in the range [0.02-1]\,R$_{500}$.  We find the median fractional residuals at radii 0.02\,R$_{500}$ and R$_{500}$ to be $0.002\pm0.008$, $-0.027\pm0.065$ respectively for the 2D profiles. For the 3D reconstruction,  we find median fractional residuals at  0.02\,R$_{500}$, R$_{500}$, and 2\,R$_{500}$ to be  $0.040 \pm0.072$, $-0.003\pm0.065$ and $-0.060\pm0.180$ respectively.
We see that on average there is a small but noticeable 4\% over-estimation in the 3D temperature profiles in the first 4 radial bins. This could be caused by the presence of dense and cold substructures that in the simulated objects could lower the central value of the 3D spectroscopic-like temperature in the innermost region, where the impact of this formulation is the strongest (see e.g. Fig.~3 of \cite{2014ApJ...791...96R}). Similarly, beyond R$_{500}$ the temperature profiles are underestimated by 8\% on average. This effect could also play a role for the central mismatch, since the convolution is temperature dependent, the slight overestimation in the first few innermost bins may be also linked to the underestimation of temperature profiles in the outermost bins. This suggests the importance of  
deriving accurate estimation of the  temperature profiles beyond the 2D fitting range. More detailed treatment in this regard is beyond the scope of this paper and we propose possible explanations as an important future direction. However, we do find the median residual is consistent with zero over all the radial range of [0.02-2]\,R$_{500}$ and as in the previous cases, for the majority of the clusters the true 3D temperature profiles lie within 1-$\sigma$ dispersion of the {\bf IAE} recovered temperature profiles. The median of fractional residuals for the full sample and over the entire radial range is found to be $-0.003\pm0.038$ and $-0.003\pm0.075$ for the 2D and 3D profiles respectively.

\subsection{Comparison of {\bf IAE} model to a parametric model}  

In this Section, we use the validation sample of 115 clusters to compare the non-parametric results from {\bf IAE} model to those obtained from a parametric temperature model. We first obtain the best-fitting 3D temperature from the \cite{Vikhlinin2006} model (Eqn.~\ref{eq:tprof}) considering the prior range on each parameter given in Table~\ref{tab_vik_priors}, and using the same binning schemes as used for the {\bf IAE} model in previous sections, assuming a spectroscopic-like weighting scheme. Temperature profiles were first scaled by ${\rm T}_X$ before fitting them to the parametric model, so as to bring the parameter $T_0$ to a comparable scale. We find that in the 2D-3D (or 3D-3D) fine binning case, the 3D reconstruction is poor compared to the observational-like cases where the fitting is weighted according to the errors, which increase with radius. We also tried to fit the temperature profiles in log space, which could effectively address any heteroscedasticity issues and stabilise the variance over the large radial range. However, this still did not improve the model reconstruction in the 2D-3D (or 3D-3D) fine binning case. This indicates that such a parametric model struggles to accurately capture the true underlying patterns in the noiseless data, or when the noise covariance is negligible. By weighting the fitting according to the errors, which reflect the inherent uncertainties in the data and which increase with radial distance, the model can better adapt to the complexities of the noiseless data, resulting in improved performance. The significant improvement achieved by incorporating error covariance can be visually observed in Fig.~\ref{fig_VIk_profiles}. Even with coarse resolution, as discussed in the next paragraph, the fit shows a remarkable enhancement when realistic error covariance is considered during the fitting process. Another reason for the sub-optimal performance of the parametric model can be attributed to its highly non-linear nature and the strong degeneracy between the parameters. This results in poor constraints on the parameters, and the reconstructed 3D temperature profiles could depend strongly on the choice of fitting priors.

The arguments discussed above can be explained with Fig.~\ref{fig_VIk_profiles}. The top panel of the Fig.~\ref{fig_VIk_profiles} shows the dispersion for the 2D-3D fine and coarse binning cases with prior ranges of parameters $a=0-0.6$ and $c=0-4$, which have a significant effect on the profiles in the central and outer regions respectively. We find, for the 2D-3D fine binning case, that the 3D reconstructed temperature profiles obtained from this parametric fitting have a large bias in both the central and outer regions, with median fractional residuals of values about 30\% and 11\% at the first and last bin respectively. For observational-like binning, having a weighted fitting, the bias in the central regions becomes consistent with zero, however, there is still a bias  beyond the R$_{500}$ which increases with  the median fractional residual of values about 18\%. We find that the optimal priors for parameters $a$ and $c$ are $a=0-0.1$ and $c=1-4$ respectively, leading to a minimal bias in the central and outer regions respectively. This is shown in the bottom panel of  Fig.~\ref{fig_VIk_profiles}, where one finds a median consistent with zero, but with slightly larger dispersion compared to the {\bf IAE} model for the observational-like cases.  In the outer regions, however, the dispersion in the 2D-3D fine binning case is barely consistent with zero for the parametric model. 

Considering the optimal priors for the  $a$ and $c$ parameters discussed above, the left panel of Fig.~\ref{fig_IAE_PAR_sim} shows the  reconstruction of the 3D temperature profiles with the {\bf IAE} and parametric models for typical CC and NCC clusters in the simulated sample with observational-like binning having twelve bins. While the CC profile is recovered well by both models, the reconstruction is poor in the central region for the parametric fit to the NCC case, and would require  larger values of $a$ to improve the fit in the central region. Similarly, in the right panel of Fig.~\ref{fig_IAE_PAR_sim}, we show the 3D reconstruction of two complex profiles. These two clusters are experiencing ongoing merger shocks. Here one sees that, in such scenarios, the parametric model performs poorly compared to the {\bf IAE} model, being unable to capture the true underlying structure of the data. We find that even increasing the priors on $a$ and $c$ did not have any significant improvement in the parametric fit for such complex profiles. The accurate estimation of the shape of the temperature profile is vital since the estimation of total mass profiles depends on it.

\section{First application to CHEX-MATE X-ray data}

\subsection{Modifications to the {\bf IAE} model}

Although the \thethree provide us with one of the highest resolution hydrodynamical simulation samples to date, due to numerical issues, the thermal profiles could only reliably be estimated above 0.02\,R$_{500}$ for most of the galaxy clusters in the sample. 
The number of available 2D annular temperature data points and their radial distribution will depend on the object mass and luminosity, the presence or absence of a cool core, and the depth of the observation\footnote{See \citet[][]{che23} for a discussion of an optimal binning method.}. From our experience of X-ray analysis of typical observations of local ($z<0.5$) massive ($M_{500} > 10^{14}$~M$_{\odot}$) galaxy clusters available in the \xmm\ or {\it Chandra} archives, we find that for many objects, one is generally able to obtain some temperature data points interior to 0.02\,R$_{500}$ (corresponding to $20-40\arcsec$ at $z=0.05$ and $5-10\arcsec$ at $z=0.3$ for typical cluster masses).

Therefore, in order to make the best use of the available data, one needs to look for an optimal extrapolation of the {\bf IAE} model that is able to reconstruct the temperature profiles robustly even in the very central regions.  To build an {\bf IAE} model that  is suitable for application to such observations, we first extrapolated the  simulated temperature profiles to 0.005\,R$_{500}$ by fitting a   \cite{Vikhlinin2006} parametric model in the inner regions (up to 0.5\,R$_{500}$). We then re-trained the {\bf IAE} model in the full radial range of [0.005-2]\,R$_{500}$ with the simulated dataset, augmented by the parametric model extrapolation in the very central regions. 

\begin{figure*}
	\centering	
	\includegraphics[width=0.98\textwidth]{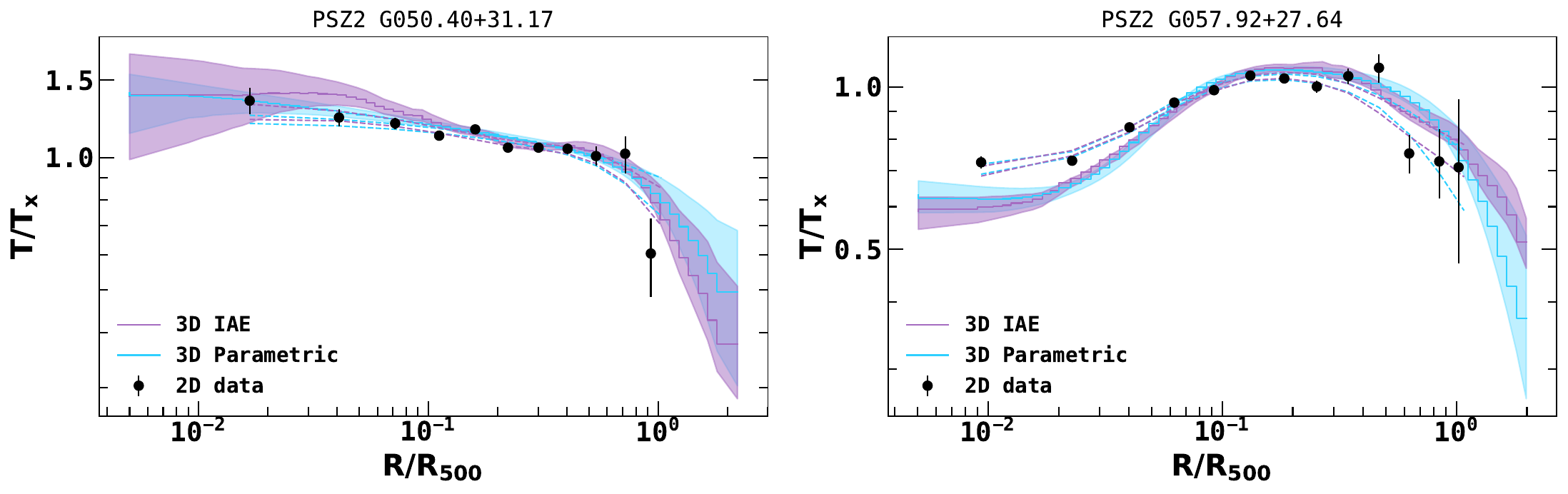}   
	\caption{\footnotesize Comparison of the scaled 2D and 3D temperature profiles of a typical NCC (PSZ2 G050.40+31.17) and CC (PSZ2 G057.92+27.64)  cluster in the DR1 sample recovered with the {\bf IAE} and  parametric models. Solid lines and the associated shaded regions show the median and 1-$\sigma$ dispersion of the reconstructed 3D temperature profile obtained  with MCMC. Regions enclosed by the dashed lines represent the corresponding 1-$\sigma$ dispersion 2D temperature profiles fitted to the observed 2D data (black dots). In line with our results with simulations for observational-like cases, we find that both the IAE model and parametric model with optimal priors generate comparable profiles for NCC and CC clusters.}
	\label{figcom_IAE_PAR}
\end{figure*}
\begin{figure}
    \includegraphics[width=0.49\textwidth]{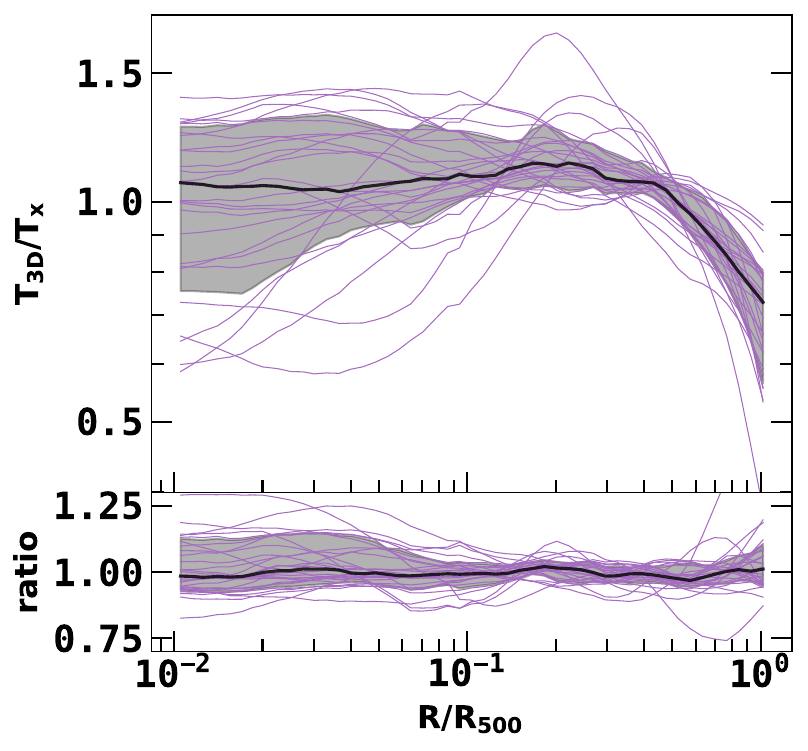}
    \caption{\footnotesize Scaled 3D temperature profiles of the DR1 sample recovered with the {\bf IAE} model. Also shown in the bottom panel is the ratio of 3D temperature profiles recovered with the {\bf IAE} model to the parametric models. For better visibility, the error bars corresponding to the individual profiles are not shown. The black lines and grey shaded grey regions represent the median and 1-$\sigma$ dispersion of the sample. 
    The difference between the {\bf IAE} model and the parametric model can be as high as 20\%, although the average ratio between them remains close to unity. }
    \label{fig_IAE_profiles}
\end{figure}
\begin{figure}
    \includegraphics[width=0.49\textwidth]{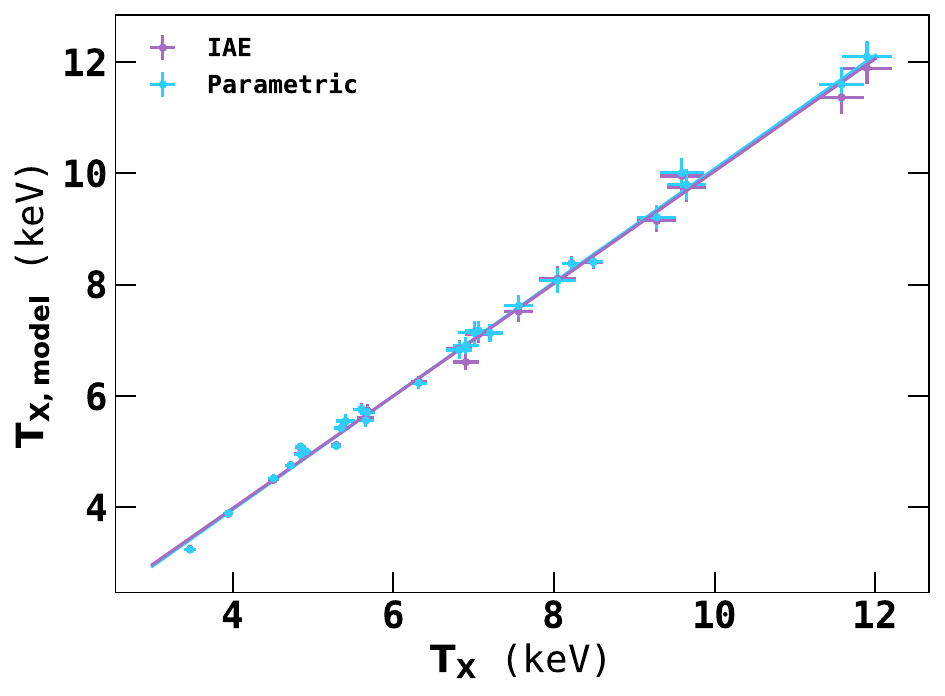}
    \caption{\footnotesize Left Panel: Comparison of the observed T$_X$ and the best-fit T$_{X,\rm{model}}$ obtained with non-parametric {\bf IAE} and parametric \cite{Vikhlinin2006} models. Solid lines show the best fit for the data. We see that both our non-parametric and parametric approaches provide tight and accurate constraints on the average temperature of clusters.}
    \label{fig_Tx_profiles}
\end{figure}

\begin{figure*}
	\centering	
	\includegraphics[width=0.98\textwidth]{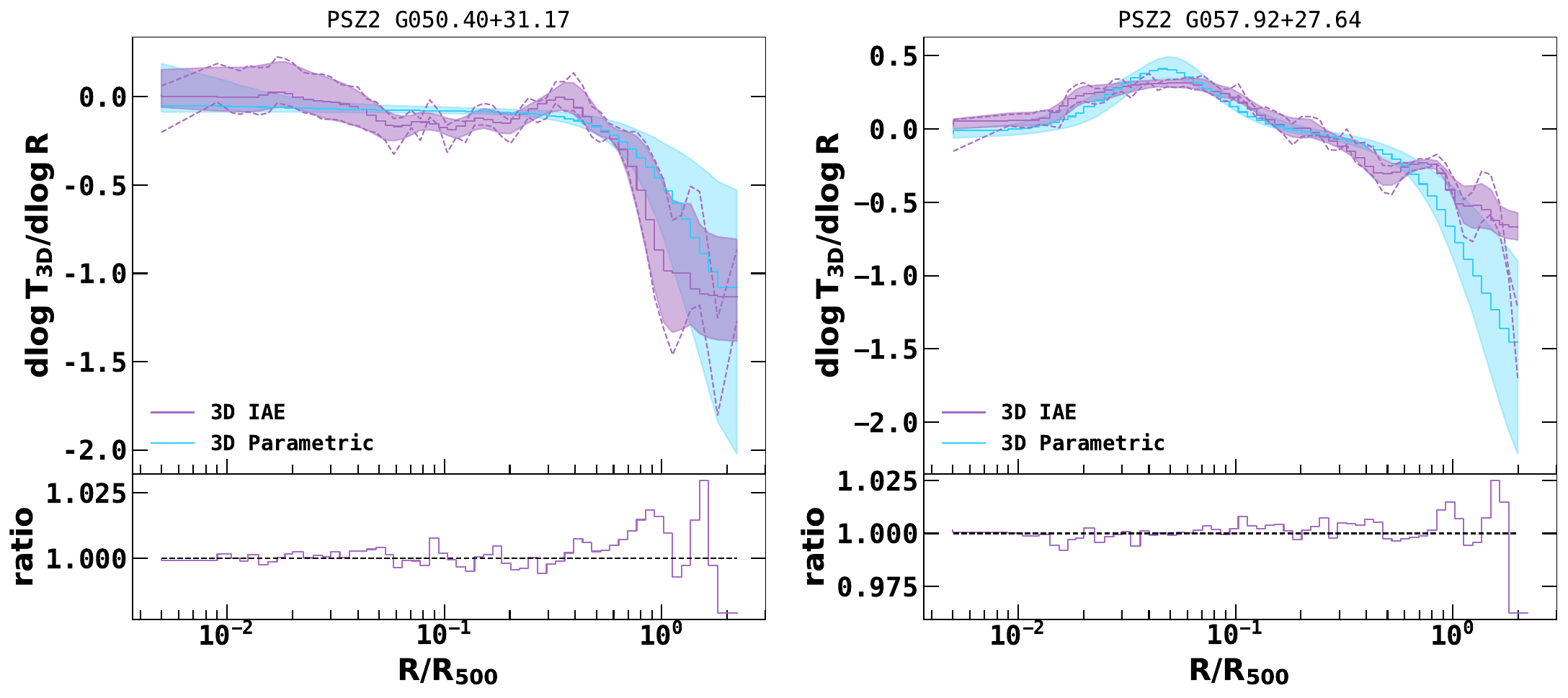}   
	\caption{\footnotesize Comparison of the logarithmic derivatives 3D temperature profiles of a typical NCC (PSZ2 G050.40+31.17) and CC (PSZ2 G057.92+27.64)  cluster in the DR1 sample recovered with the {\bf IAE} and  parametric models. Solid lines and the associated shaded regions show the median and 1-$\sigma$ dispersion obtained  with MCMC. The region enclosed by the dashed lines represents 1-$\sigma$ dispersion, if no smoothing is applied to the profiles derived from the MCMC chain. The bottom panel shows the ratio of the median 3D temperature profiles obtained using {\bf IAE} with and without smoothing.}
	\label{figcom_IAE_der_PAR}
\end{figure*}
\begin{figure}
    \includegraphics[width=0.49\textwidth]{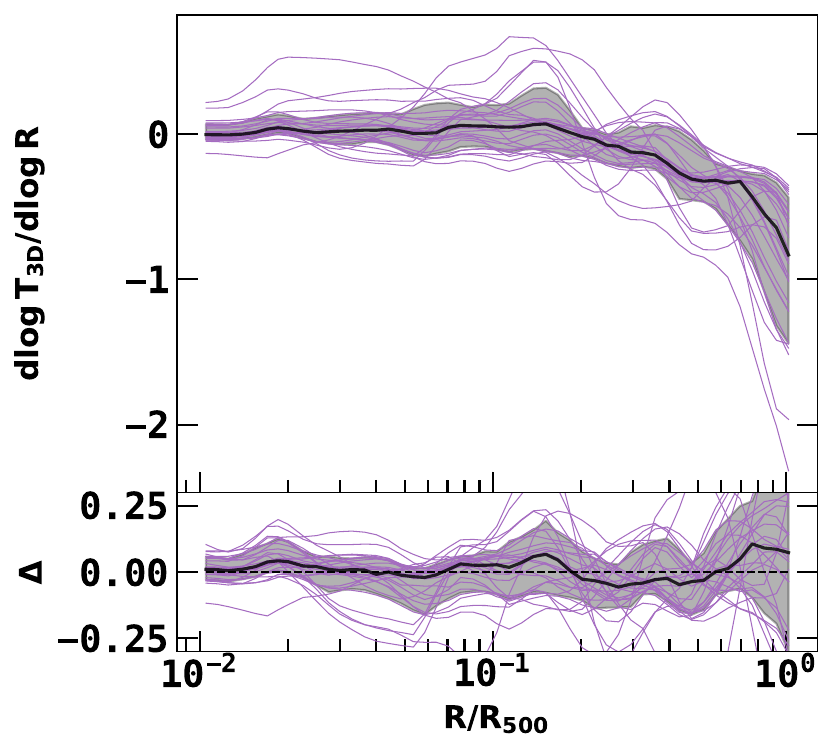}
    \caption{\footnotesize Logarithmic derivatives of 3D temperature profiles of the DR1 sample recovered with the {\bf IAE} model. Also shown in the bottom panel is the difference between  profiles recovered with the {\bf IAE} model and the parametric model. For better visibility, the error bars corresponding to the individual profiles are not shown. The black lines and grey shaded grey regions represent the median and 1-$\sigma$ dispersion of the sample. }
    \label{fig_IAE_der_profiles}
\end{figure}
\subsection{Observed sample}
\label{sec:DR1desc}

We then use this updated {\bf IAE} model on the latest CHEX-MATE Data Release 1, DR1 sample  \citep{Rossetti2023} to deconvolve the temperature profiles. The DR1 sample is a `technical but representative' sub-sample, which was built to test our pipeline for the extraction and reconstruction of the radial temperature and density profiles. It is composed of 30 clusters, whose distribution in mass, redshift, and Planck signal-to-noise-ratio (S/N) reflect the properties of the CHEX-MATE parent sample.  In Appendix~\ref{appx:final}, Table~\ref{allclusters} provides the details of all the clusters in the DR1 sample. For data reduction and analysis, we used the XMM-Newton Science Analysis System (SAS), version $16.1$. We refer to \citet[][]{Bartalucci2023} for details on the data reduction procedures (calibration, standard pattern cleaning, removal of noisy MOS CCDs, and light-curve filtering) and on the detection of contaminating sources. From the EPIC images in the 0.7-1.2 keV band, we extracted both mean and median surface brightness radial profiles, centered on the peak and on the centroid within $R_{500}$. For the temperature profile, we extract spectra in concentric annuli centered on the surface brightness peak, using the MOS-spectra and PN-spectra ESAS tools \citep{2008A&A...478..615S} embedded in SAS. For each region, we perform a joint fit of the MOS1, MOS2, and PN spectra with an adsorbed thermal model, to which we add a model for all the background components (Galactic foregrounds, CXB, Cosmic-ray particle background, residual soft protons). We estimate priors for the parameters of this background model that are allowed to vary within their uncertainty during the joint fit with the cluster parameters, running the Markov Chain Monte Carlo method within XSPEC (see \citealt{Rossetti2023}, for more details). In this work, two clusters (PSZ2 G046.88+56.48 and PSZ2 G057.78+52.32) that require background treatments using off-set observations were not considered in the analysis.  

\subsection{Method}

For deconvolution of these observed profiles, we assume that the 3D temperature profiles can be represented by the {\bf IAE} model, convolved with a response matrix ${\bf C} = {\bf C}_{\rm PSF} \otimes {\bf C}_{\rm proj}$, which simultaneously takes into account projection and PSF redistribution. 

The projection matrix, ${\bf C}_{\rm proj}$, is built by using the DR1  density profiles from \citet[][in prep.]{Duffy2023}, derived using the non-parametric deconvolution algorithm of \citet{Croston2006}. More details of the derivation of the density profiles can be found in \citet{cro08} and \citet{pra22}. 
${\bf C}_{\rm PSF}$ is constructed as in \cite{Croston2006}, which uses the parametric PSF model of \citet{ghi01} as a function of the energy and angular offsets, the parameters of which can be found in EPIC-MCT-TN-011\footnote{\url{http://www.iasf-milano.inaf.it/~simona/pub/EPIC-MCT/EPIC-MCT-TN-011.pdf}} and EPIC-MCT-TN-012\footnote{\url{http://www.iasf-milano.inaf.it/~simona/pub/EPIC-MCT/EPIC-MCT-TN-012.pdf}}. 

The IAE model was then projected, taking into account the spectroscopic-like weighting scheme proposed \citet{Mazzotta2004}, and fitted to the observed 2D profiles. In our future work, we will examine the more  complex \cite{Vikhlinin2006b} weighting scheme, which is more robust for lower temperature clusters/groups, and compare the results to other weighting schemes.

\subsection{Results}
\subsubsection{Estimation of profiles}

In Fig.~\ref{figcom_IAE_PAR}, we show the 3D temperature profiles reconstructed using the {\bf IAE} model and the \cite{Vikhlinin2006} 8-parameter parametric model for a typical NCC and a typical CC cluster in the DR1 sample.  In general,  we find that with the annular resolution of the present 2D profiles, both models produce similar reconstructed 3D temperature profiles. However, the parameters of the \cite{Vikhlinin2006} model are poorly constrained, and the final reconstructed temperature profiles (especially the inner and  outer regions) may depend on the chosen priors.  

Figure~\ref{fig_IAE_profiles} shows the 3D temperature profiles of the clusters in the DR1 sample obtained with the {\bf IAE} model, scaled by the average temperature (T$_X$) in the [0.15-0.75]\,R$_{500}$ region. We find that fractional dispersion is about 22\% in the inner region which first decreases with the radius and attains a minimum value of 3\% at around 0.5\,R$_{500}$. It then starts to increase with radius, achieving a maximum value of 22\% in the outer regions. Also plotted in the sub-panel is the ratio of the 3D temperature profiles recovered with the {\bf IAE} and parametric models. One finds that within the radial range of [0.1-1]\,R$_{500}$, the difference between {\bf IAE} and \cite{Vikhlinin2006} model is less than 10\%. The difference between them can be as high as 25\% in the inner and outer regions. However, on average both models predict very similar profiles with a difference of less than 2\% over the entire radial range of [0.005-2]\,R$_{500}$.

As a consistency check,  we compared the values of the average temperature in the [0.15-0.75]\,R$_{500}$ region. Figure~\ref{fig_Tx_profiles} shows the observed T$_X$ compared to T$_{X,\rm{model}}$, the temperature derived from a projection of the 3D non-parametric {\bf IAE} and parametric models in the same annulus. Fitting a straight line to the  (T$_{X,\rm{model}}$,T$_X$) one finds the slope for the {\bf IAE} and parametric model to be  $1.01\pm0.01$ and $1.01\pm0.02$ respectively. 

\subsubsection{Estimation of derivatives}

While non-parametric models offer greater flexibility in modelling complex patterns and relationships, one requires a large amount of data to accurately estimate derivatives. Small irregularities in the profiles often amplify the noise in the derivatives. Therefore, it is often desirable to apply some degree of smoothing to the profiles to have accurate derivatives in the non-parametric approaches. As can be seen from Fig.~\ref{fig_IAE_profiles}, the reconstructed 3D temperature profiles from the {\bf IAE}  model have a reasonably smooth underlying structure. We find that the direct computation of numerical derivatives of individual profiles derived from the MCMC chains using spline interpolation, without applying any smoothing, usually provided a good estimate of the logarithmic derivatives and corresponding 1-$\sigma$ interval.  Nonetheless, we sometimes found the derivative estimates to be noisy, particularly beyond the 2D fitting range.  This noise can be attributed to logarithmic binning, which can create sparsity in the outer regions. Another potential cause for the noise is small spikes in the temperature profiles between consecutive radii in the profiles inherited by the model from the simulations itself in the inner regions due to the limited resolution there. We, therefore, choose to apply a very minimal smoothing, such that only the sharp discontinuities, if any (usually small in magnitude),  on local scales (2-3 radial bins) are affected/corrected  and the general non-linear structure is preserved. We use the algorithm developed by \cite{Cappellari2013b} which implements the one-dimensional locally linear weighted regression \cite{doi:10.1080/01621459.1979.10481038}\footnote{\url{https://pypi.org/project/loess/}}. It uses a tri-cube weighting function with weights $(1-u^3)^3$ where $u$ is a distance from the local point R under consideration and a smoothing parameter $f$ which is the fraction of neighborhood points to be considered in the local fit around R.  Increasing the value of $f$ increases the neighborhood of influential points leading the smoother profiles. For our case, we apply modest smoothing with $f=0.15$.

Figure~\ref{figcom_IAE_der_PAR} shows the corresponding  logarithmic derivatives of the temperature profiles of the two clusters discussed in the previous sub-section. Here also, both the {\bf IAE} and the parametric models produce consistent profiles. Furthermore, for the {\bf IAE} model, the profiles obtained with and without applying the smoothing on the temperature profiles are also consistent with each other. This can be also seen in the bottom panel where the ratio between reconstructed 3D temperature profiles with and without applying smoothing is seen to be less than 1\% over most of the radial range. Figure~\ref{fig_IAE_der_profiles} shows the logarithmic derivatives of the 3D temperature profiles of the clusters in the DR1 sample obtained with the  {\bf IAE} model. Also, in the bottom panel, we show the difference in logarithmic derivatives derived from  {\bf IAE} and parametric models ($\Delta$). We find that, although dispersion in the difference increases with the radius, the difference is consistent with zero throughout the radial range. While it is difficult to quantify this difference in the inner region, since logarithmic derivatives are close to zero, in the range [0.5-2]\,R$_{500}$ the difference in logarithmic derivatives between the {\bf IAE} and the parametric model can be more than as 20\%. The impact of this on the total mass estimate is not straightforward but 
is expected to be about 5\%-30\%.

\section{Discussion and conclusions}
Classical statistical modelling techniques can be sensitive to inaccuracies and may lead to poor performance if the data are complex (non-linear) and/or have a dynamic structure. Data-driven (model-agnostic) deep-learning techniques are now becoming increasingly popular. They make use of the topology to learn the underlying structure of the data, and often have been found to give superior performance in terms of accuracy and precision when the underlying structure of data are non-linear.  However, one typically requires a massive dataset and vast computational resources to train the neural network, limiting their applicability for some scenarios. In this paper, we demonstrate the first use of  deep learning techniques to build a model of galaxy cluster temperature profiles and apply this model to the problem of temperature profile deprojection. Using a non-linear interpolatory scheme with five anchor points (temperature profiles), allows us to have frugal learning with a sparse training set, and the neural network is able to uncover the lower dimensional non-linear manifold of data by way of mapping between latent space and real space. 

The resulting Interpolatory Auto-Encoder ({\bf IAE}) model is trained and evaluated in the radial range of [0.02-2]\,R$_{500}$ using a simulated dataset of 315 temperature profiles from the {\sc Three Hundred Project}. We then implement a new deconvolution scheme using efficient and cost-effective learning-based regularisation to achieve a stable and accurate reconstruction of the 3D temperature profiles by optimising the latent parameters (barycentric weights) of the anchor points using MCMC. Moreover, the deconvolution algorithm can be easily extended to include the instrumental PSF effect. We test the {\bf IAE} with a different set of deconvolution schemes with respect to the resolution, projection, and quality of the data. We find that, in general, the {\bf IAE} model can recover unbiased 3D temperature  profiles in the fitting range. The performance of the {\bf IAE} model to recover the true temperature profiles can be summarised as follows:
\begin{itemize}
\item We first considered the simplest case, where we tested the efficiency of the {\bf IAE} model in directly fitting the high resolution simulated 3D temperature profiles, defined in 48 fixed radial bins in the range  [0.02-2]\,R$_{500}$, the resolution with which the {\bf IAE} model is trained. We find that in this case, the reconstruction of temperature profiles from the {\bf IAE} model is robust, with the median fractional residuals centered around zero and a 1-$\sigma$ dispersion (determined by the 16th and 84th percentile range of fractional residuals) of about $\pm$5\% over most of the radial range. The dispersion in the outskirts is somewhat larger (about $\pm$10\%). This can be interpreted as being due to the complex nature of the ICM as a result of merging/accretion processes that are dominant there. Moreover, dispersion in the fractional residuals  for the sub-sample of 20 most relaxed clusters (MR20) and smooth temperature profiles (MS20) is about 35\% smaller compared to the sub-sample of 20 most disturbed clusters (MD20) and irregular temperature profiles (MI20). We find that the model fidelity can be further improved by increasing the  number of anchor points in the {\bf IAE} model. However, since observed temperature profiles are generally of much lower resolution, increasing the complexity of the model is undesirable as it could lead to overfitting.\newline
\item We then considered a case where we fitted the  high resolution simulated 2D temperature profiles to the {\bf IAE} model using classical emission measure weights. Here too we find the median fractional residual is centered around zero with a 1-$\sigma$ dispersion of about $\pm$5\% over most of the radial range. In the first few innermost bins, however, we find that the dispersion is increased to about $\pm$10\%. This is understandable since the projection operation introduces a degeneracy in the 3D temperature profiles which is significant in the inner regions i.e the mapping between input 2D temperature profiles and {\bf IAE} reconstructed 3D temperature profiles is not as strong as compared to a mapping between input 3D temperature profiles to {\bf IAE} reconstructed 3D temperature profiles. However, this degeneracy can be mitigated to a large extent in the observational-like cases since the 2D temperature profiles in the inner bins have relatively smaller errors associated with them as compared to the rest of the  radial bins.  Moreover, as in the previous case, the distribution of the fractional residuals over all radii for the MR20 (MS20) sub-sample is narrowly peaked compared to the MD20 (MI20) sub-sample.\newline

\item We next considered observation-like fitting cases, with typical temperature profile data quality such as would be obtained from the {\it XMM-Newton} or {\it Chandra} satellites. We first considered a case where we fit 2D temperature profiles defined at twelve radial points and up to R$_{500}$ only, mimicking the profile expected from the moderately deep X-ray exposures. We find that in the 2D fitting range i.e. [0.02-1]\,R$_{500}$, with the relatively low resolution input 2D temperature profiles, the performance of the {\bf IAE} model is negligibly degraded. However, beyond R$_{500}$, where we do not consider any 2D data in the fit, the 1-$\sigma$ dispersion in the 3D reconstruction increases with radius and becomes  about $\pm$20\% in the last bin. The 3D median fractional residual is found close to zero over most of the radial range, except beyond 1.5\,R$_{500}$ where it is underestimated by about 7\%. We also considered a case where we only use only six 2D temperature data points in the fit  and find that the {\bf IAE} is still able to provide an unbiased estimate of the reconstructed temperature profile, albeit with a slightly larger uncertainty. \newline
\item We considered a more realistic temperature-dependent spectroscopic-like weighting scheme \citep{Mazzotta2004} in the deprojection. We find that there is a small bias of about 4\% excess in the fractional residual in the innermost few bins, in addition to the underestimation in the outer regions as in the previous case. \newline

\item We also compared the {\bf IAE} model with a parametric temperature model. With the high resolution hydrodynamical simulated temperature profiles, the parametric model based on \cite{Vikhlinin2006} showed poor performance when the realistic error covariance matrix is ignored in the fit. Including the error covariance matrix improved the fit. The non-linearity and parameter degeneracy of the parametric model also contributed to sub-optimal performance, making the 3D reconstruction dependent on the choice of priors. In contrast, the {\bf IAE} model performed better, particularly in complex cases with ongoing merger shocks, demonstrating its superior adaptability to diverse data scenarios. 
 \newline
\item  Finally, in a first application to X-ray data, we built an augmented version of the {\bf IAE} model in the radial range [0.005-2]\,R$_{500}$. The data augmentation was necessary because the simulated profiles did not have sufficient resolution to probe the very core regions that are accessible to good quality X-ray data. The augmentation step was achieved by extrapolating the simulated profiles to lower radii (below $\approx$0.02\,R$_{500}$) by fitting them to the \cite{Vikhlinin2006} parametric model in the range $\approx$ [0.02-0.5]\,R$_{500}$. We then used this updated {\bf IAE} model to reconstruct the 3D temperature profiles and logarithmic derivative  of the representative (DR1) sample galaxy clusters drawn from the CHEX-MATE project. The resulting non-parametric {\bf IAE} profiles were compared to those derived from parametric deprojection and deconvolution. We find that, in such  observational cases where the typical number of annular data points is much fewer compared to the  simulations, the difference between the {\bf IAE} and parametric model is less than 10\% over most of the observed region. However, in the inner and outer regions, the difference between them can be as high as 25\%. Moreover, the results from the \cite{Vikhlinin2006} parametric model, especially inner and outer regions, depends on the priors chosen on the parameters as they are very poorly constrained during the fit.
\end{itemize}

It should be noted that the inner regions of the clusters, which involve processes such as AGN feeding/feedback, gas condensation, sloshing, etc., are complex and may not be accurately represented by current state-of-art cosmological simulations. Moreover, the augmentation of the central regions of the training set using the extrapolation of a parametric model could potentially introduce bias in the underlying model recovered from the {\bf IAE}. Despite these limitations, we believe that the {\bf IAE} model provides higher-fidelity  results compared to traditional parametric modelling, as demonstrated in this study. As the size and quality of both X-ray observations and simulations are set to improve in the coming years, the robustness of {\bf IAE} will also be enhanced resulting in a much lower scatter. Our future plan is to perform network training and testing on different sets of simulations so as to have a larger training and validation sample. This will potentially also help us to understand the systematics, if any, in the {\bf IAE} model inherited from the  particular set of numerical simulations used for training. For example, \cite{DeLuca2021} showed that the dynamical state of clusters in the \thethree clusters varies with  redshift: the relaxed clusters decrease in number from redshift $z=0$ to $z=1$. It remains to be seen if issues such as possible  redshift dependence have any impact on learning.  This effect, in principle, can be taken into account by training the model using simulated clusters across a large redshift range. 

Another important step in improving the deconvolution scheme will be to force the neural network model to learn the features shared between simulations and real data using transfer/adversarial learning \citep{ganin2016domain}. This will essentially mitigate the biases inherited by the neural network model from simulations. Moreover, we expect with an upgraded {\bf IAE} model, the reconstruction of 3D temperature profiles beyond the observational range of R$_{500}$ will be significantly improved due to an increase in the size of the training sample. We further plan to implement a more robust model extrapolation technique in future work.
 
 The usefulness of the {\bf IAE} is not only limited to the estimation of the temperature of the galaxy clusters. We further plan to use the {\bf IAE} interpolatory technique to recover the underlying density, pressure and hence dark matter profiles in the galaxy clusters. An important extension of this will be to train a neural network to estimate the total mass profiles of the galaxy clusters directly from the thermal profiles of the ICM without considering the hydrostatic equation. Another interesting prospect for  our work will be to implement the deconvolution technique in SZ and lensing data, to recover the robust model of the galaxy clusters. This will further help us to understand the biases introduced in calibrating the mass and scaling relations for cosmological studies. Such studies might be also used to assess more robustly relative density/temperature fluctuations, hence constraining turbulence and relative parameters (Mach number, injection scale, etc). Our methodology can also be implemented in other areas of astrophysics and cosmology. In fact, the {\bf IAE} scheme has already been implemented in the source separation algorithm to tackle physical hyper-spectral data \citep{CARLONIGERTOSIO2023108776}. 

 One of our immediate plans is to implement the proposed deconvolution technique to the most recent high quality CHEX-MATE X-ray sample of clusters \citep{2021A&A...650A.104C}, and compare to other approaches 
 such as those used in \cite{Bartalucci2018} (semi-parametric reconstruction) and \cite{2022A&A...662A.123E} (multi-scale non-parametric reconstruction). The comparison of the estimated logarithmic derivatives will be instructive  since these are highly related to the shape of mass profiles of clusters. Our ultimate goal will be to test the $\Lambda$CDM predictions on the total mass distribution in galaxy clusters using a new and sophisticated fully non-parametric approach.
\begin{acknowledgements}
The work of AI was supported by CNES, the French space agency. 
SE, LL and FG acknowledge the financial contribution from the contracts ASI-INAF Athena 2019-27-HH.0, ``Attivit\`a di Studio per la comunit\`a scientifica di Astrofisica delle Alte Energie e Fisica Astroparticellare'' (Accordo Attuativo ASI-INAF n. 2017-14-H.0), and from the European Union’s Horizon 2020 Programme under the AHEAD2020 project (grant agreement n. 871158).
This research was supported by the International Space Science Institute (ISSI) in Bern, through ISSI International Team project \#565 ({\it Multi-Wavelength Studies of the Culmination of Structure Formation in the Universe}).
MS acknowledges the financial contribution from contract ASI-INAF n.2017-14-H.0. and from contract INAF mainstream project 1.05.01.86.10.
EP acknowledges the financial support of CNRS/INSU and of CNES, the French Space Agency.
MED acknowledges partial financial support from a NASA ADAP award/SAO subaward SV9-89010.
MDP and AF acknowledge financial contribution from Sapienza Università di Roma, thanks to Progetti di Ricerca Medi 2020, RM120172B32D5BE2.
AF thanks financial support from Universidad de La Laguna (ULL), NextGenerationEU/PRTR, and Ministerio de Universidades (MIU) (UNI/511/2021) through grant "Margarita Salas".
HB, DdL, and PM acknowledge support from the Spoke 3 Astrophysics and Cosmos Observations. National Recovery and Resilience Plan (Piano Nazionale di Ripresa e Resilienza, PNRR) Project ID CN\_00000013 "Italian Research Center on High-Performance Computing, Big Data and Quantum Computing"  funded by MUR Missione 4 Componente 2 Investimento 1.4: Potenziamento strutture di ricerca e creazione di "campioni nazionali di R\&S (M4C2-19)" - Next Generation EU (NGEU)  and from the European Union’s Horizon 2020 Programme under the AHEAD2020 project (grant agreement n. 871158).  The authors would like to thank the reviewer for his/her careful, constructive and insightful comments in relation to this work.
\end{acknowledgements}

%
   \bibliographystyle{aa} 
   \bibliography{biblography} 
%
\appendix
\section{Supplementary Material}
\subsection{Correlation between $\chi_{_D}$  and $\chi_{_S}$ }
\begin{figure}
   \centering
    \includegraphics[width=0.49\textwidth]{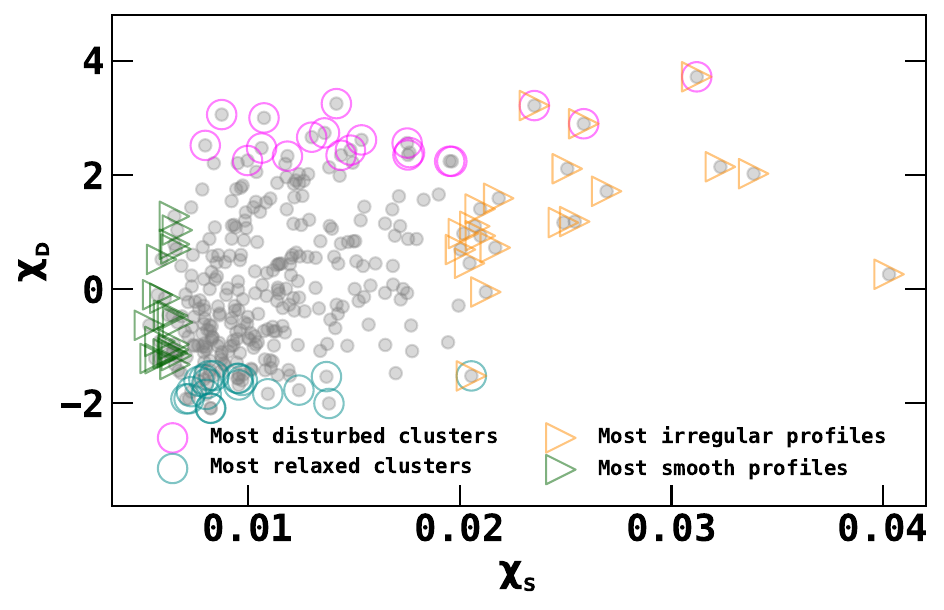}
    \caption{\footnotesize Correlation between $\chi_{_D}$  and $\chi_{_S}$ for the simulated clusters in {\sc The Three Hundred Project}. Cyan circles and green triangles represent the 20 most relaxed clusters and smooth profiles respectively. Magenta circles and orange triangles represent the 20 most disturbed clusters  and irregular profiles respectively. }
    \label{appx1}
\end{figure}

In Fig~\ref{appx1}, we present the correlation analysis between $\chi_{_D}$ and $\chi_{_S}$ for the clusters in the {\sc Three Hundred Project}. These parameters serve to classify the clusters based on their intrinsic dynamical state and the smoothness of their temperature profiles. Specifically, small values of $\chi_{_D}$ indicate relaxed clusters, while small $\chi_{_S}$ values suggest clusters with smooth temperature profiles. Our findings reveal a noteworthy correlation between $\chi_{_D}$ and $\chi_{_S}$, as evidenced by the calculated Spearman’s correlation coefficient of 0.42 and a $P$ value of $5 \times 10^{-15}$. 

\subsection{Temperature profile reconstruction with {\bf IAE} model for the fine binning cases}
\label{app:app1}

\begin{figure*}
\rotatebox{90} {
		\includegraphics[width=0.63\textwidth]{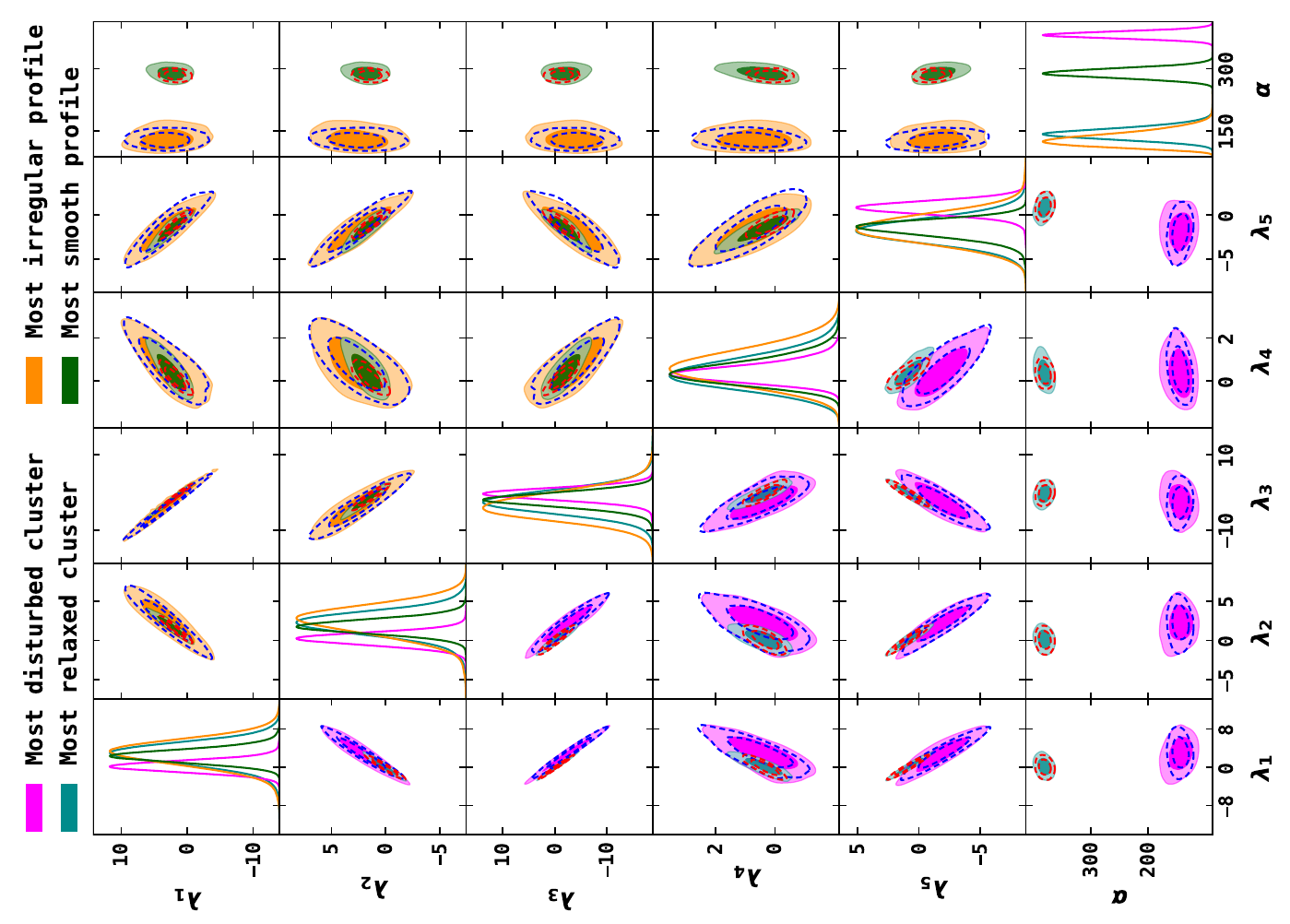}

	\includegraphics[width=0.63\textwidth]{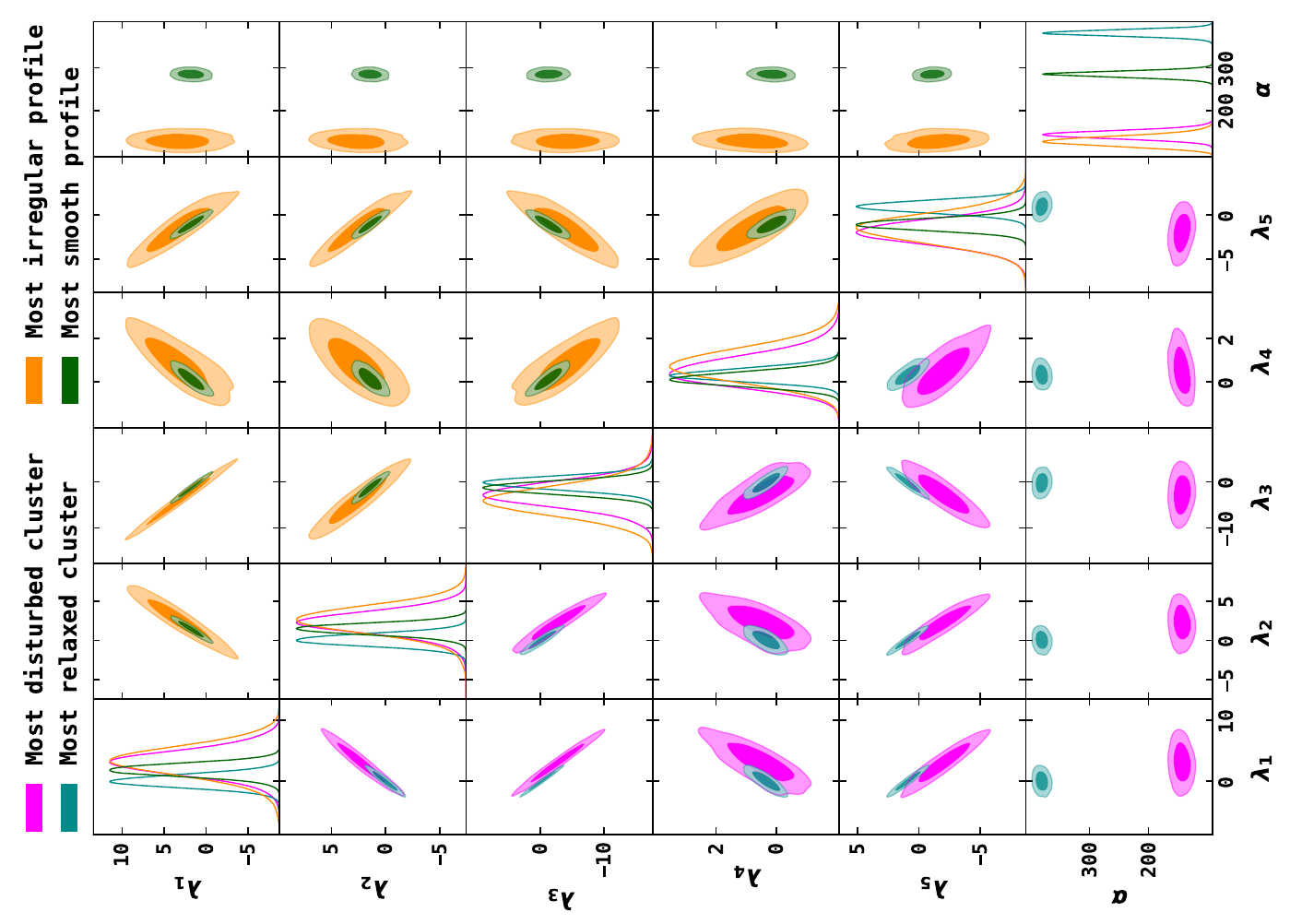}
} 
	\caption{\footnotesize Two-dimensional joint posterior probability distributions and one-dimensional marginal posterior probability distribution of {\bf IAE} model parameters with 3D-3D fine binning (top panel) and 2D-3D fine (bottom panel) binning cases for the most relaxed cluster (smooth profile) and most disturbed cluster (irregular profile). The shaded contours represent the 68\% and 95\% confidence regions. For comparison, 2D contours  in the 3D-3D fit case  are shown with the dashed red and blue lines for the  2D-3D fine binning case.  }
	\label{L1}
\end{figure*}

Figure~\ref{L1} shows  the posterior distribution of the parameters of the IAE model obtained using MCMC for the 3D-3D and 2D-3D fine binning cases for the most relaxed / disturbed clusters and of the most regular / irregular profiles in the validation sample. The parameters are found to be well-constrained, with the relaxed cluster profile having tighter constraints than the disturbed cluster profile, which has larger contour levels.

\begin{figure*}
\centering
{\includegraphics[width=0.95\textwidth]{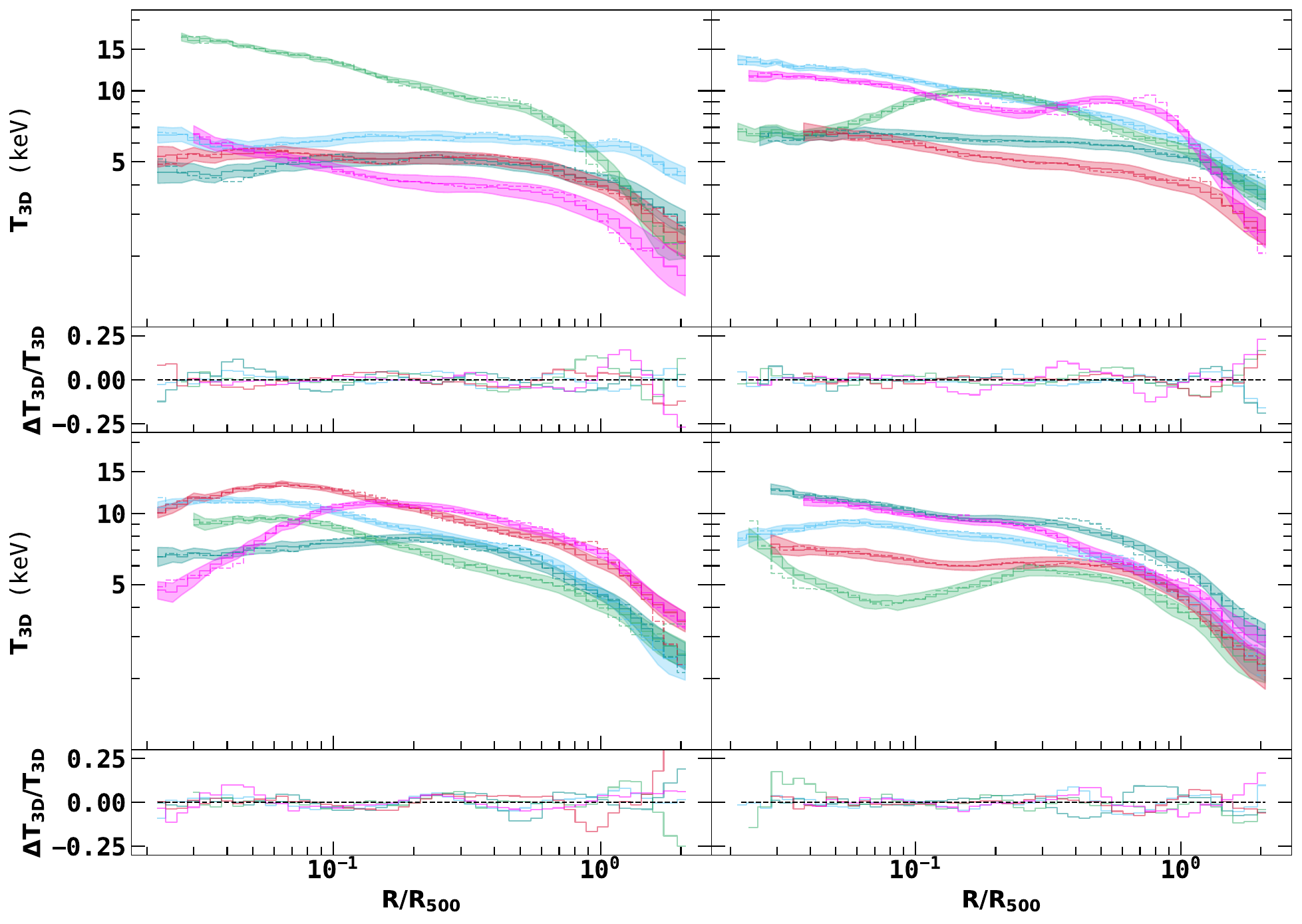}}
\caption{\footnotesize Comparison of  the 20 3D mass-weighted temperature profiles in the validation sample (dashed lines) and reconstructed 3D median temperature profiles obtained from the {\bf IAE} model (solid lines). The shaded regions represent the 1-$\sigma$ dispersion (16th–84th percentile range) of the recovered profile. Also shown, in the smaller subplots, are the residuals of the fit.} 
\label{appx2}
\end{figure*}

Figure~\ref{appx2} illustrates the comparison of 20 randomly selected 3D mass-weighted temperature profiles from the {\sc Three Hundred Project} validation sample with the corresponding reconstructed 3D median temperature profiles obtained from the {\bf IAE} model (3D-3D fine binning case),. The analysis covers the radial range of [0.02-2]\,R$_{500}$ using 48 radial bins. As shown, the discrepancy between the true and reconstructed profiles remains below 5\% across most of the cluster radial range. However, it is important to note that in certain cases, particularly in the outer regions (R > 1.5\,R$_{500}$), we observe larger (generally less than 20\%) discrepancies between the true and reconstructed temperature profiles. This behaviour can be attributed to the intricate interplay of complex physical processes within the cluster, such as merger events, shocks, and interactions between the ICM and accreting matter in the outskirts. These processes can influence the temperature distribution, leading to local variations that the {\bf IAE} model might encounter challenges in accurately capturing.

\begin{figure*}
	\includegraphics[width=0.49\textwidth]{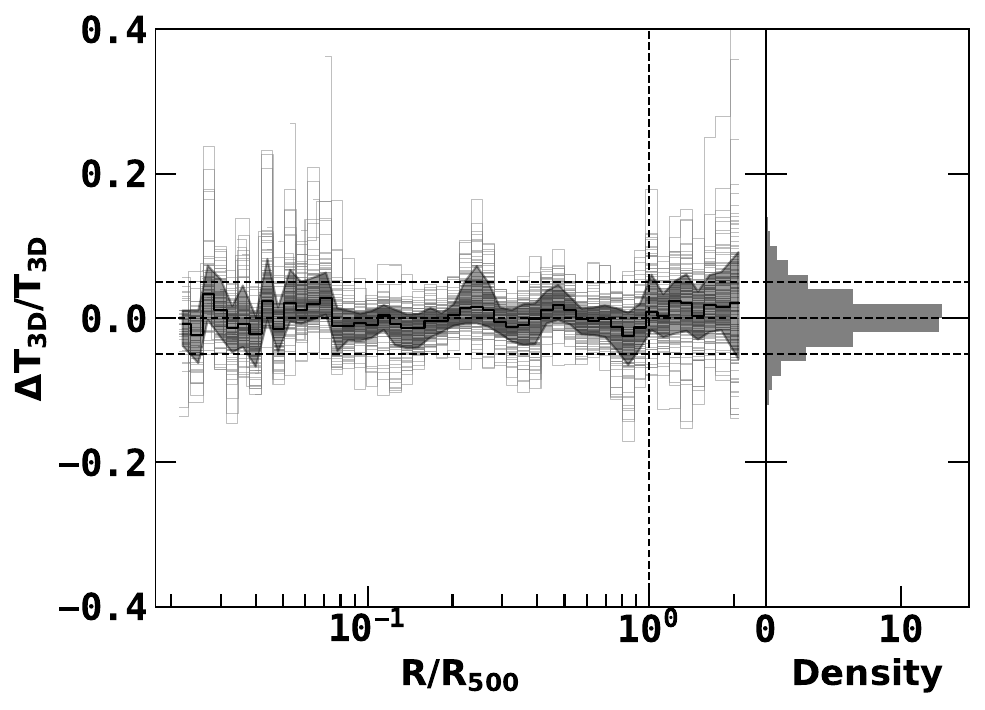}
		\includegraphics[width=0.49\textwidth]{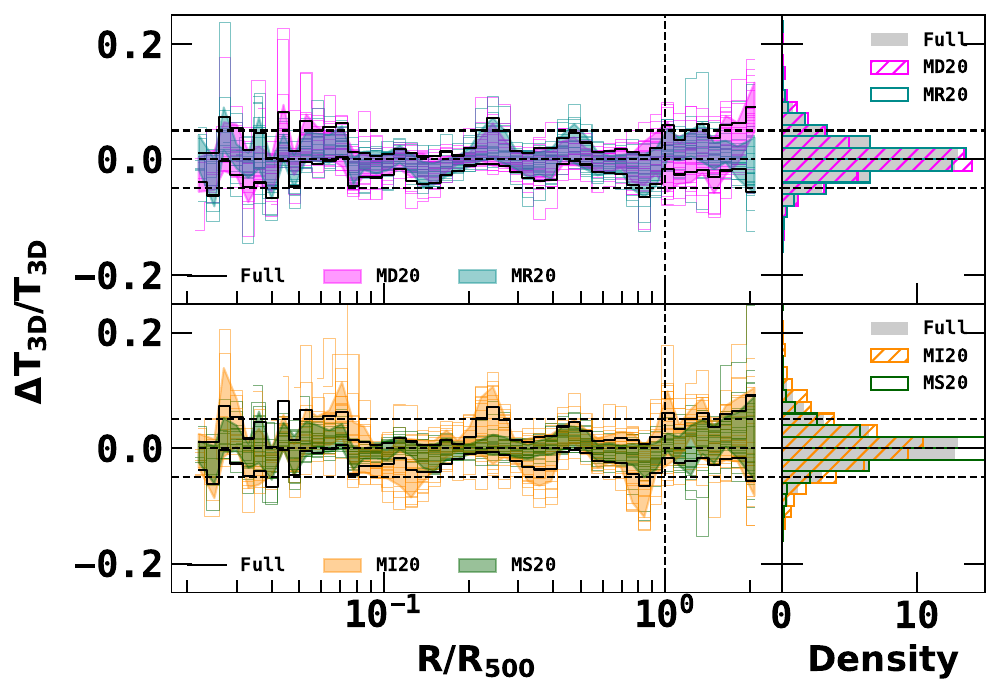}

	\caption{\footnotesize Fractional residuals obtained with {\bf IAE} model having 20 anchor points for 3D-3D fine binning case. Colour coding is the same as in Fig.~(\ref{fig6}).  There is a significant improvement of 25\% in the average fractional residual compared to the {\bf IAE} model with 5 anchor points. }
	\label{appx2b}

\end{figure*}

Figure~\ref{appx2b} shows the fractional residuals  between the true and reconstructed temperature profiles with {\bf IAE} having 20 anchor points for all the individual clusters in the validation sample considering 3D-3D fine binning case.  The median fractional residual profile is found to be close to zero throughout the radial range: at radii,  0.02\,R$_{500}$, R$_{500}$, and 2\,R$_{500}$, the values are $-0.008\pm0.025$, $0.008\pm0.037$ and $-0.020\pm0.074$ respectively.

\subsection{Temperature profile reconstruction with {\bf IAE} model for the observational-like binning cases}
\label{appx:appx4}
\begin{figure*}
		\includegraphics[width=0.49\textwidth]{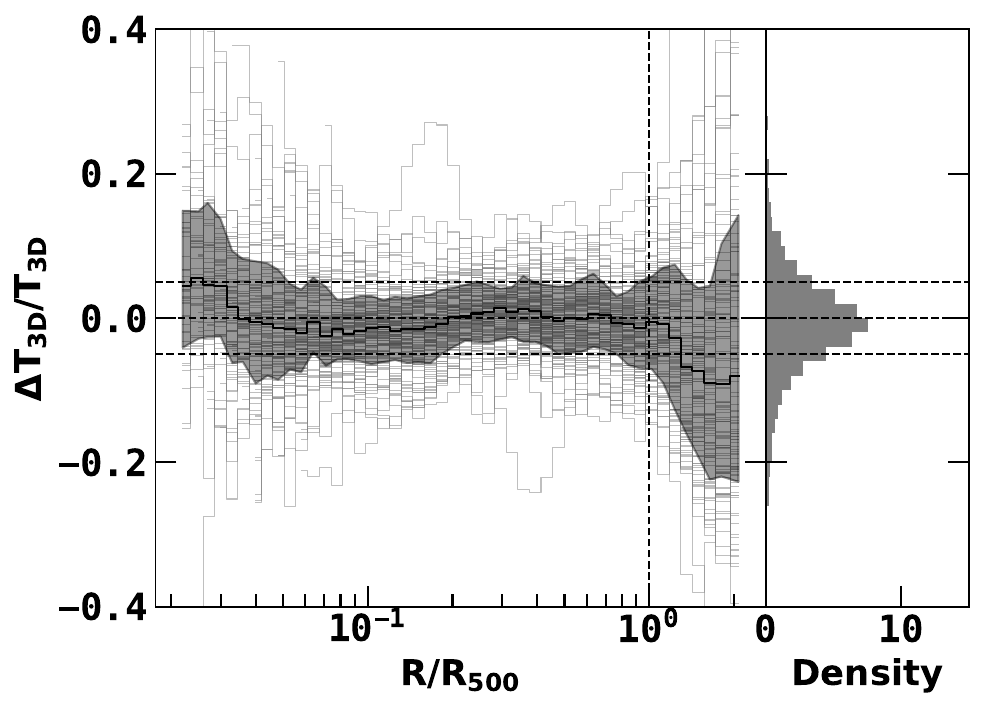}	\includegraphics[width=0.49\textwidth]{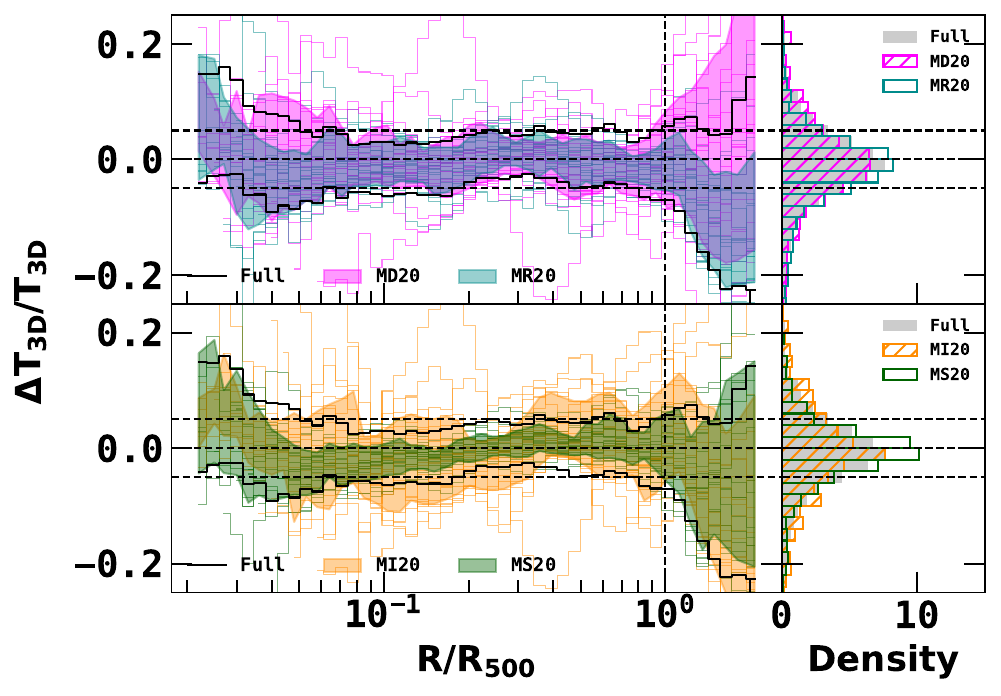}
	\caption{\footnotesize The fractional residuals for 115 clusters in the validation sample with {\bf IAE} for the 2D-3D fit (coarse binning) using 2D temperature profiles defined at twelve radial bins up to R$_{500}$, without considering errors on the 2D temperature profiles. Colour coding is the same as in Fig.~(\ref{fig9}). For simplicity, only the 3D temperature reconstruction is plotted.}
	\label{fig:appx4}
\end{figure*}
Figure~\ref{fig:appx4} illustrates the 3D fractional residuals obtained when error bars (i.e. error covariance matrix, see Eqn.~\ref{eq:b}) are not considered to fit the 2D temperature profiles defined at twelve coarse bins with {\bf IAE} model. In this scenario, it is evident that the scatter is amplified in the inner regions compared to the where the error covariance matrix is taken into account.   
The median 3D fractional residuals at radii 0.02\,R$_{500}$, R$_{500}$, and 2\,R$_{500}$ are determined as $0.048\pm0.090$, $-0.005\pm0.063$, and $-0.080\pm0.175$ respectively. The median of fractional residuals across the entire radial range for the complete sample is calculated to be $-0.006\pm-0.062$.

\begin{figure*}
\rotatebox{90} {
		\includegraphics[width=0.63\textwidth]{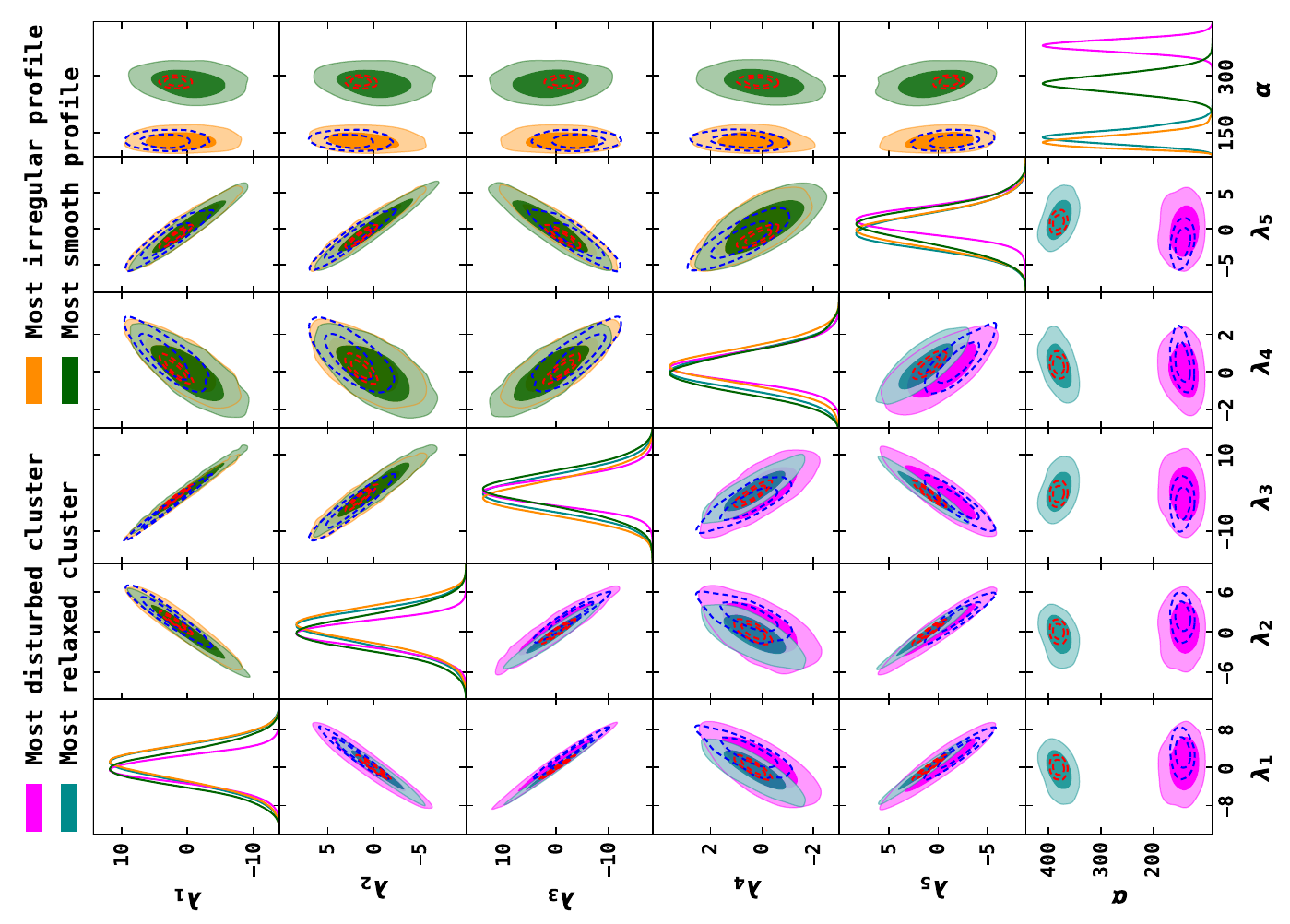}

	\includegraphics[width=0.63\textwidth]{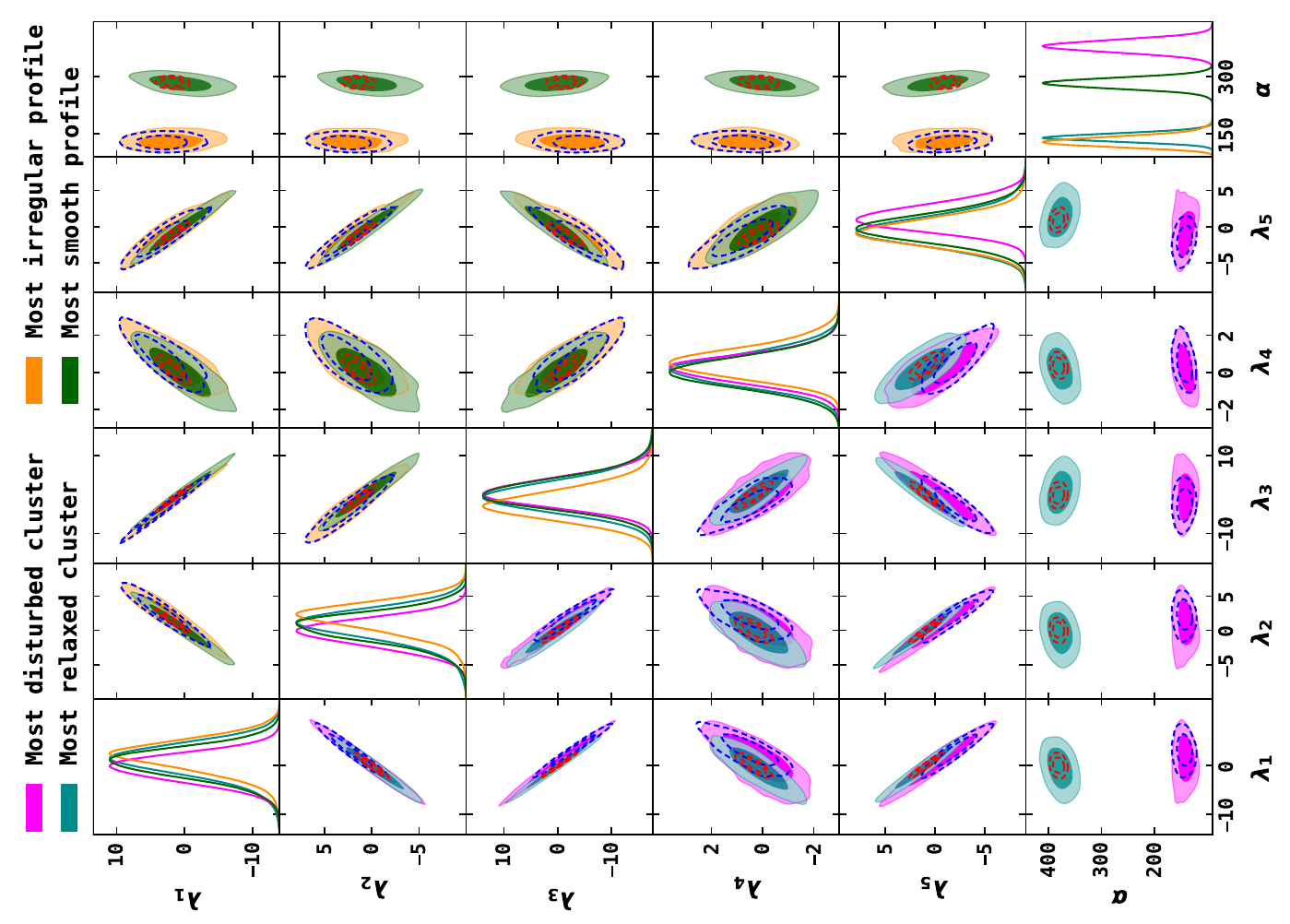}
} 
	\caption{\footnotesize Two-dimensional joint posterior probability distributions and one-dimensional marginal posterior probability distribution of {\bf IAE} model parameters with 2D-3D coarse binning of twelve (top panel) six (bottom panel) radial bins for the most relaxed cluster (smooth profile) and most disturbed cluster (irregular profile). The shaded contours represent the 68\% and 95\% confidence regions. For comparison, 2D contours for the 3D-3D binning case  are shown with the dashed red and blue lines for the bottom panel.  }
	\label{L2}
\end{figure*}

Figure~\ref{L2} illustrates the posterior distribution of the IAE model parameters obtained via MCMC for 2D-3D cases binning cases (both twelve as well as six bins). The analysis covers the most relaxed and disturbed clusters, along with the most regular and irregular profiles in the validation sample. While the parameters are well-constrained, the constraints are slightly weaker compared to the fine binning cases.

\begin{figure*}
{\includegraphics[width=0.90\textwidth]{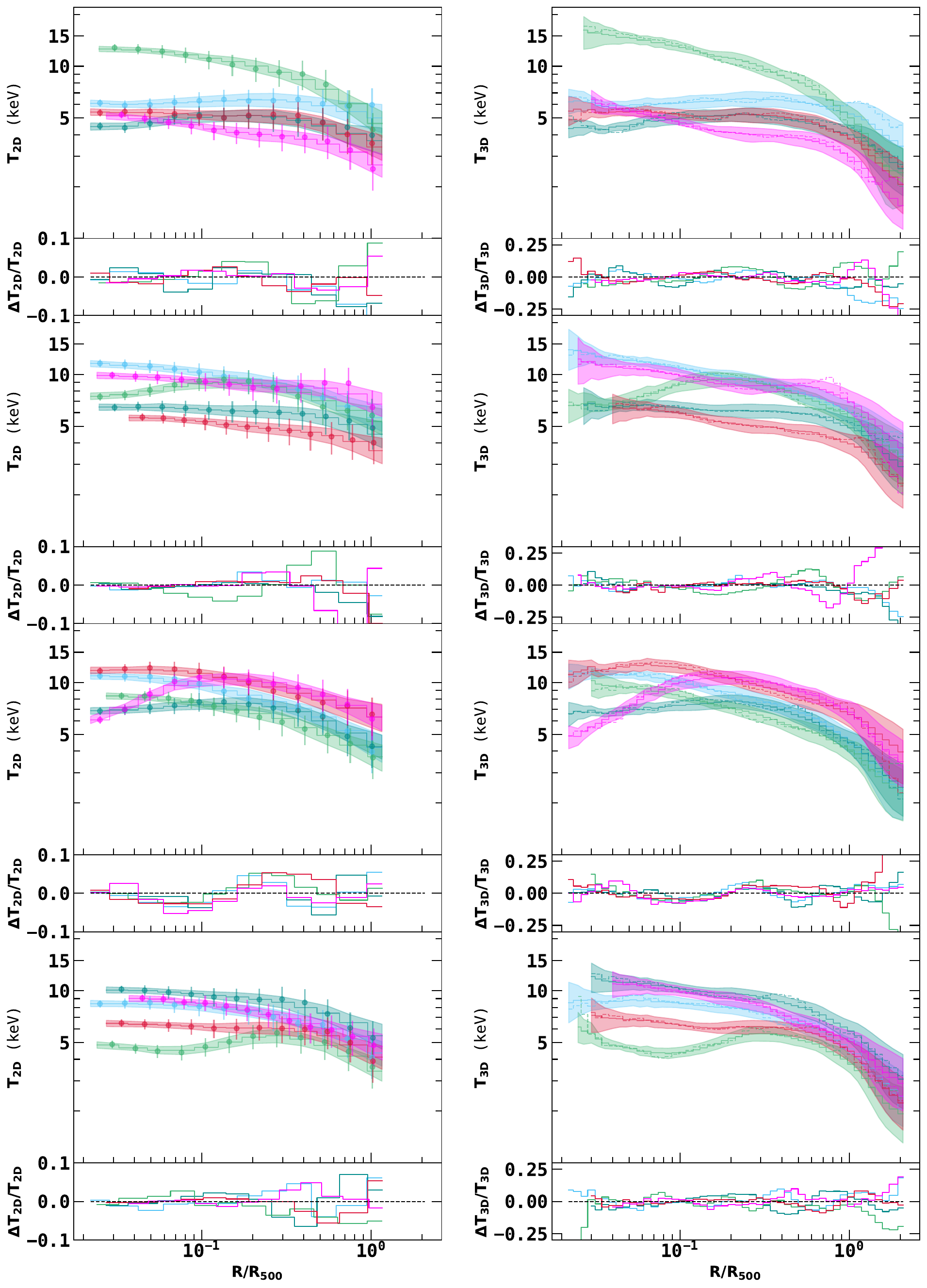}}
\caption{\footnotesize Left panel: Comparison of the 20 simulated 2D temperature profiles (solid points with errors) and reconstructed 2D temperature profiles obtained using {\bf IAE} model (solid lines), the fitting being performed in the range [0.02-1]\,R$_{500}$ considering twelve 2D temperature bins. The shaded regions represent the 1-$\sigma$ dispersion of the reconstructed 2D temperature profiles. The smaller subplots show the residuals of the fit. Right panel: Solid lines and the shaded regions show the corresponding reconstructed 3D temperature profiles and the 1-$\sigma$ dispersion respectively. Also shown in the dashed lines are the true 3D mass-weighted temperature profiles.}
\label{appx5}
\end{figure*}

\begin{figure*}
{\includegraphics[width=0.90\textwidth]{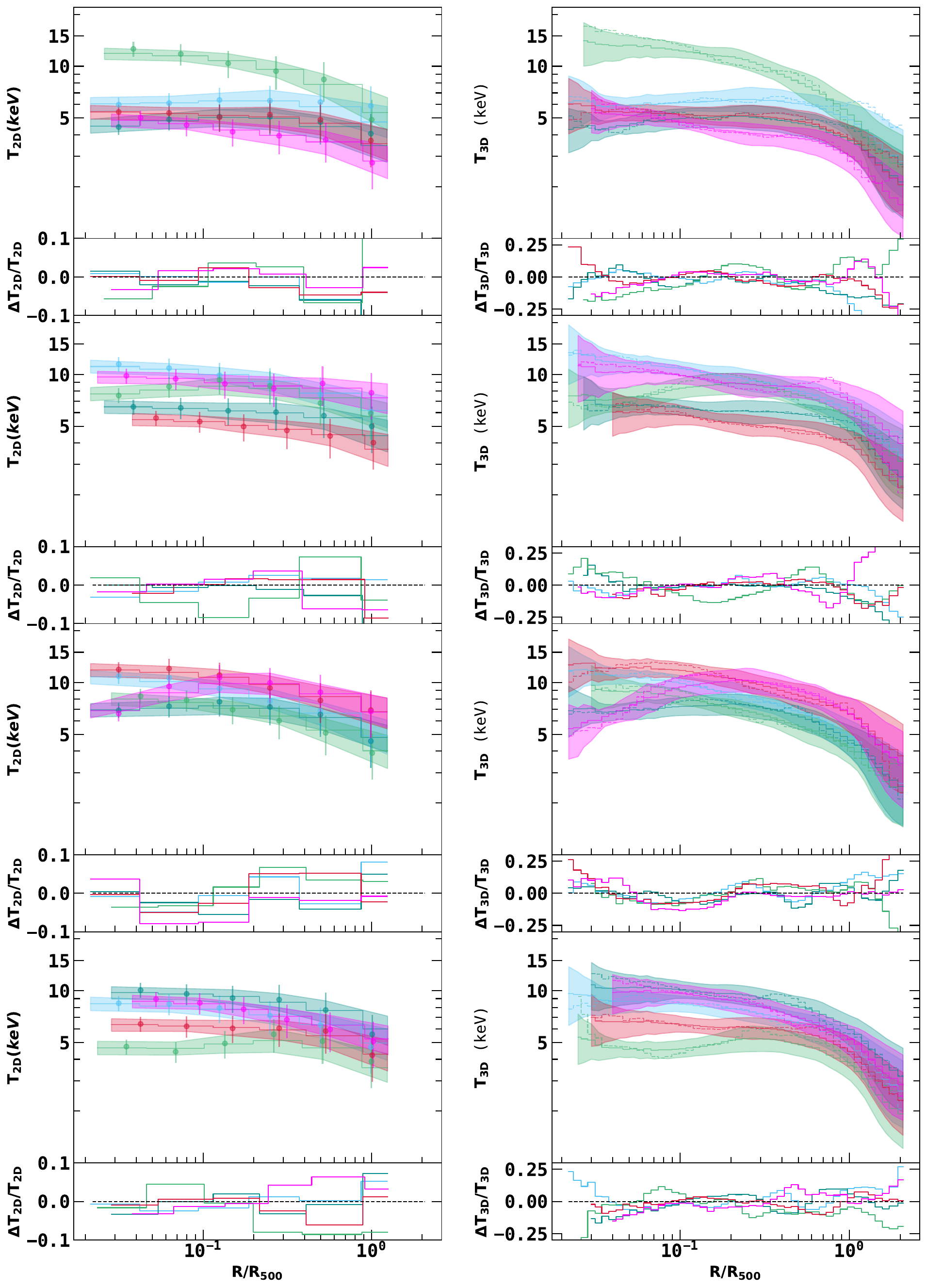}}
\caption{\footnotesize Left panel: Comparison of the 20 simulated 2D temperature profiles (solid points with errors) and reconstructed 2D temperature profiles obtained using {\bf IAE} model (solid lines), the fitting being performed in the range [0.02-1]\,R$_{500}$ considering six 2D temperature bins. The shaded regions represent the 1-$\sigma$ dispersion of the reconstructed 2D temperature profiles. The smaller subplots show the residuals of the fit. Right panel: Solid lines and the shaded regions show the corresponding reconstructed 3D temperature profiles and the 1-$\sigma$ dispersion respectively. Also shown in the dashed lines are the true 3D mass-weighted temperature profiles.}
\label{appx5a}
\end{figure*}

Figures~\ref{appx5} and \ref{appx5a} present a comparison between 20 randomly selected 3D and 2D profiles from {\sc The Three Hundred Project} validation sample with reconstructed 2D and 3D median temperature profiles obtained from the {\bf IAE} model using observational binning of twelve and six respectively, which is typical of X-ray observations. The analysis focuses on fitting simulated 2D temperature profiles in the  radial range of [0.02-1]\,R$_{500}$ with the convolved {\bf IAE} model assuming errors in temperature profiles increase linearly with radius. The results indicate that the discrepancy between the true and reconstructed temperature profiles remains around 5\% in the 2D fitting range from [0.02-1]\,R$_{500}$.

\subsection{DR1 sample used in this work}
\label{appx:final}
Table~\ref{allclusters} provides comprehensive details for all the clusters included in the DR1 sample used in this work. The table encompasses information on cluster names and redshifts, and other relevant properties, allowing for a comprehensive examination and analysis of each cluster characteristics.  The compilation comprises 30 clusters, reflecting the mass, redshift, and {\it Planck} S/N distribution akin to the properties observed in the CHEX-MATE parent sample.

\begin{table*}[!h]
\caption{List of the clusters of the DR1 sample.}	
\scriptsize
 \begin{center}	
\begin{threeparttable}
\begin{tabular}{c c c c c c}
\toprule
\toprule
  \multicolumn{1}{c}{PSZ2 Name} &
  \multicolumn{1}{c}{Redshift} &
  \multicolumn{1}{c}{$M_{500}$} &
  \multicolumn{1}{c}{T$_{x}$} &
  \multicolumn{1}{c}{Tier\tnote{*}} &
  \multicolumn{1}{c}{XMM obsid} \\
  \multicolumn{1}{c}{ } &
  \multicolumn{1}{c}{ } &
  \multicolumn{1}{c}{$10^{14} M_\odot$} &
  \multicolumn{1}{c}{ (keV)} &
  \multicolumn{1}{c}{ } &
  \multicolumn{1}{c}{center/offset} \\
\midrule
  PSZ2 G008.31-64.74 & 0.312 & 7.42$_{-0.39}^{+0.41}$ & $6.77 \pm  0.170$  & 2 & 0827010901\\
  PSZ2 G041.45+29.10 & 0.178 & 5.41$_{-0.35}^{+0.35}$ & $5.61 \pm  0.088$  & 1 & 0601080101\\
  PSZ2 G042.81+56.61 & 0.072 & 4.22$_{-0.19}^{+0.18}$ &  -              & 1 & 0202080201/0827361101\tnote{**}\\
  PSZ2 G046.88+56.48 & 0.115 & 5.10$_{-0.23}^{+0.22}$ & $5.08 \pm  0.094$  & 1 & 0827010601\\
  PSZ2 G050.40+31.17 & 0.164 & 4.22$_{-0.35}^{+0.34}$ & $4.65 \pm  0.056$  & 1 & 0827040101\\
  PSZ2 G056.77+36.32 & 0.095 & 4.34$_{-0.21}^{+0.18}$ &$ 5.02 \pm  0.063$  & 1 & 0740900101\\
  PSZ2 G056.93-55.08 & 0.447 & 9.49$_{-0.43}^{+0.43}$ & $8.04 \pm  0.121$  & 2 & 0503490201\\
  PSZ2 G057.78+52.32 & 0.065 & 2.32$_{-0.21}^{+0.21}$ & -              & 1 & 0827040301/0827041801\tnote{**}\\
  PSZ2 G057.92+27.64 & 0.076 & 2.66$_{-0.21}^{+0.21}$ &$ 3.59 \pm  0.054$  & 1 & 0827030301\\
  PSZ2 G066.41+27.03 & 0.575 & 7.69$_{-0.53}^{+0.51}$ &$ 8.86 \pm  0.215$  & 2 & 0827320601\\
  PSZ2 G072.62+41.46 & 0.228 &11.43$_{-0.27}^{+0.25}$ &$ 9.00 \pm  0.231$  & 2 & 0605000501\\
  PSZ2 G077.90-26.63 & 0.147 & 4.99$_{-0.25}^{+0.26}$ &$ 5.14 \pm  0.086$  & 1 & 0827020101\\
  PSZ2 G083.86+85.09 & 0.183 & 4.74$_{-0.33}^{+0.32}$ & $5.12 \pm  0.113$  & 1 & 0827030701\\
  PSZ2 G113.29-29.69 & 0.107 & 3.57$_{-0.26}^{+0.26}$ & $4.05 \pm  0.069$  & 1 & 0827021201\\
  PSZ2 G113.91-37.01 & 0.371 & 7.58$_{-0.55}^{+0.53}$ & $6.67 \pm  0.165$  & 2 & 0827021001\\
  PSZ2 G114.79-33.71 & 0.094 & 3.79$_{-0.22}^{+0.22}$ &$4.40 \pm  0.075$  & 1 & 0827320401\\
  PSZ2 G149.39-36.84 & 0.170 & 5.35$_{-0.50}^{+0.44}$ & $5.75 \pm  0.101$  & 1 & 0827030601\\
  PSZ2 G195.75-24.32 & 0.203 & 7.80$_{-0.41}^{+0.40}$ & $8.01 \pm  0.232$  & 2 & 0201510101\\
  PSZ2 G207.88+81.31 & 0.353 & 7.44$_{-0.43}^{+0.43}$ & $6.32 \pm  0.184$  & 2 & 0827020301\\
  PSZ2 G224.00+69.33 & 0.190 & 5.11$_{-0.34}^{+0.35}$ & $5.23 \pm  0.098$  & 1 & 0827020901\\
  PSZ2 G238.69+63.26 & 0.169 & 4.17$_{-0.40}^{+0.35}$ & $4.80 \pm  0.093$  & 1 & 0500760101\\
  PSZ2 G243.15-73.84 & 0.410 & 8.09$_{-0.50}^{+0.48}$ & $7.11 \pm  0.151$  & 2 & 0827011301\\
  PSZ2 G243.64+67.74 & 0.083 & 3.62$_{-0.20}^{+0.20}$ &$4.51 \pm  0.053$  & 1 & 0827010801\\
  PSZ2 G277.76-51.74 & 0.438 & 8.65$_{-0.38}^{+0.36}$ & $7.40 \pm  0.183$  & 2 & 0674380301\\
  PSZ2 G287.46+81.12 & 0.073 & 2.56$_{-0.24}^{+0.22}$ &$ 3.86 \pm  0.105$  & 1 & 0149900301\\
  PSZ2 G313.33+61.13 & 0.183 & 8.77$_{-0.33}^{+0.33}$ & $8.15 \pm  0.129$  & 2 & 0093030101\\
  PSZ2 G313.88-17.11 & 0.153 & 7.86$_{-0.27}^{+0.26}$ & $8.52 \pm  0.210$  & 2 & 0692932001\\
  PSZ2 G324.04+48.79 & 0.452 &10.58$_{-0.60}^{+0.56}$ & $11.5 \pm  0.295$  & 2 & 0112960101\\
  PSZ2 G340.94+35.07 & 0.236 & 7.80$_{-0.48}^{+0.45}$ & $6.28 \pm  0.153$  & 2 & 0827311201\\
  PSZ2 G349.46-59.95 & 0.347 &11.36$_{-0.34}^{+0.34}$ & $10.7 \pm  0.257$  & 2 & 0504630101\\
\bottomrule
\end{tabular}
   \smallskip
    \footnotesize
    \begin{tablenotes}
    \item[*] Tier 1 comprises a low redshift sample of clusters with a redshift range of [0.05-0.2] and mass values in the range of $M_{500}=[2-9]\times 10^{14}M_\odot$. Tier 2 represents a sample of the massive clusters with $M_{500}>7.5\times 10^{14}M_\odot$ in the [0.2-0.6] redshift range.
    \item[**] Due to the need for background treatments using off-set observations, these two clusters were excluded from the analysis in this work.
    \end{tablenotes}
    \end{threeparttable}
\end{center}
\footnotesize{\textbf{Notes}:  We quote: the PSZ2 name, the redshift, the nominal integrated mass from {\it Planck} data as derived from the mass with Eq. 9 in \cite{2013A&A...550A.131P} ($M_{500}$), the average  spectroscopic temperature measured in the [0.15-0.75]\,R$_{500}$ range (T$_{x}$), the Tier to which each cluster belongs (see \cite{2021A&A...650A.104C}  for details) and the {\it XMM-Newton} observations used in the analysis.}
\label{allclusters}
\end{table*}

\end{document}